\documentclass[12pt,a4paper,jhep]{article}
\pdfoutput=1
\usepackage{jheppub}
\usepackage{textcomp}

\usepackage{graphicx,subfigure,color,array,slashed
}
\usepackage{caption}
\usepackage{dcolumn}
\usepackage{bm}
\usepackage{hyperref}
\usepackage{amsmath,amssymb,amsbsy,amsfonts,latexsym,graphicx,hyperref}
\usepackage{bbold}
\newmuskip\pFqmuskip

\newcommand\fH[1]{\sbox0{#1}\dimen0=\ht0 \advance\dimen0 -1ex
	\sbox2{\'{}}\sbox2{\raise\dimen0\box2}%
	{\ooalign{\hidewidth\kern.1em\copy2\kern-.5\wd2\box2\hidewidth\cr\box0\crcr}}}

\newcommand*\pFq[6][8]{%
	\begingroup 
	\pFqmuskip=#1mu\relax
	\mathchardef\normalcomma=\mathcode`,
	\mathcode`\,=\string"8000
	\begingroup\lccode`\~=`\,
	\lowercase{\endgroup\let~}\pFqcomma
	{}_{#2}F_{#3}{\left[\genfrac..{0pt}{}{#4}{#5};#6\right]}%
	\endgroup
}
\newcommand{\pFqcomma}{{\normalcomma}\mskip\pFqmuskip}

\newcommand{\be}{\begin{equation}}
	\newcommand{\ee}{\end{equation}} 
\newcommand{\ba}{\begin{array}}
	\newcommand{\ea}{\end{array}}
\newcommand{\bea}{\begin{eqnarray}}
	\newcommand{\eea}{\end{eqnarray}}

\newcommand{\vev}[1]{\langle #1 \rangle}

\renewcommand{\Im}{{ \rm Im}}

\usepackage{multirow}

\title{   Heun's equation and analytic structure of  the Gap in Holographic superconductivity} 
\author[a,b]{Yoon-Seok Choun,}
\author[c]{Wenhe Cai,} 
\author[a]{Sang-Jin Sin}
\emailAdd{ychoun@gmail.com}
\emailAdd{whcai@shu.edu.cn} 
\emailAdd{sangjin.sin@gmail.com}
\affiliation[a]{ Department of Physics, Hanyang University, Seoul 04763, South Korea }
\affiliation[b]{Asia Pacific Center for Theoretical Physics (APCTP), Pohang 790-784, South Korea}
\affiliation[c]{ Department of Physics, Shanghai University, Shanghai 200444,  China} 
\abstract{ We present the new method to calculate the critical temperature as a function of $\Delta$, conformal dimension of the cooper operator. 
	We find that, in  the regime $1/2\leq \Delta <1$ where the  AC conductivity  does not show  a gap, the critical temperature is not well defined. 
	We also got expression of AC conductivity for $\Delta=2$, which agrees with numerical result in the probe approximation. }

\subheader{\today }
\begin{document}
	\maketitle  
	\section{\bf Introduction }
	Recent progress in  the  holographic superconductivity \cite{Hart2008,Gubser:2008px,Hartnoll:2016apf},   based on the  gauge gravity duality \cite{Maldacena:1997re,Witten:1998qj,Gubser:1998bc},  made an essential contribution in understanding the symmetry broken phase of AdS/CFT by constructing a dynamical symmetry breaking mechanism. 
	While the symmetry breaking in the the Abelian Higgs model in flat space is  adhoc by assuming the presence of the potential having a  Maxican hat shape,    the symmetry breaking  of the abelian Higgs model in AdS can be done by the gravitational instability of near horizon geometry to create a haired black hole,  thereby the model is equipped with  a fully dynamical mechanism of the symmetry breaking.   
	The observables' dependence on $\Delta$  is interesting because $\Delta$  depends on the  strength of the interaction.

After initial stage of the the model building \cite{Gubser:2008px,Hart2008} where probe limit of the gravity background was used, full back reacted version \cite{Hartnoll:2016apf} worked out. 
It turns out that 	although   there are significant differences in the zero temperature limit between the probe limit and the full back reacted version,   
the  former captures    the physics\cite{Horo2009} correctly near the critical temperature $T_{c}$, which is expected because the back reaction cannot be large when the condensation just begin to appear. 
The analytic expressions  of observables within the probe approximation were also obtained in  \cite{Siop2010,siopsis2011holographic}. 
One problem is that  \cite{denef2009landscape,Siop2010,siopsis2011holographic} 
	  the critical temperature is divergent at the $\Delta=1/2$, which does not seems to make physical sense and it has not been understood as far as we know. 
 This  was also noticed as a problem   \cite{Horo2009} but the reason for it has not been cleared yet.    
 	
	In this paper, we consider the problem by recomputing $T_{c}$ and physical observables analytically near the $T_{c}$, where the probe approximation is a good one.
	We apply Pincherle's theorem\cite{Leav1990}  to handle the Heun's equation which   appears in the computation of the critical temperature 
	in the blackhole background. 
	We   find   that the region of $1/2\leq \Delta <1$ for AdS$_{4}$  does not have a well defined eigenvalue and therefore does not have well defined 
	critical temperature either. See figure \ref{Lambda_big4}.
	We will also see that,  in this same regime, the AC conductivity gap $\omega_{g}$ does not exist either, giving us another  confidence in concluding the absence of the superconductivity in this regime. 
The situation remind us the physics of the pseudo gap regime where one can find Cooper pairs but not superconductivity due to the absence of the phase alignment of the pairs.    Similar phenomena exists for $1<\Delta<3/2$ for AdS$_{5}$, which is described in the appendix B. 
    	
	
	\section{Set up}
	We start with the action \cite{Hart2008}, 
	\begin{small}
		\be  {
			S=\int d^{d+1}x \sqrt{-|g|}\left( -\frac14 F_{\mu\nu}^{2}- |D_{\mu}\Phi|^{2}-m^{2}|\Phi|^{2} \right),}
		\ee
	\end{small}
	where $|g|=\det g_{ij}$,   $D_{\mu}\Phi=\partial_{\mu}-igA_{\mu}$ and $F=dA$. 
	Following the ref.\cite{Hart2008}, we use the fixed metric of    AdS$_{d+1}$  blackhole,  
	\begin{small}
		\begin{equation}
			ds^2= -f(r)dt^2+\frac{dr^2}{f(r)}+r^2d\vec{x}^2,  \;\;  f(r)=r^2\left(1-\frac{r_h^d}{r^d}\right).\label{eq:1}
		\end{equation}
	\end{small}
	The AdS radius is set to be $1$ and $r_h$ is the radius of the horizon.  The Hawking temperature is
	$T=\frac{d}{4\pi} r_h. $
	In the coordinate $z= r_h/r$, the field equations are %
	\begin{small}
		\bea
		&&  \frac{d^2 \Psi }{d z^2} -\frac{d-1+z^d}{z(1-z^d)}\frac{d \Psi }{d z}+\left( \frac{g^2 \Phi^2}{r_h^2(1-z^d)^2}-\frac{m^2}{z^2(1-z^d)}\right)\Psi =0, \nonumber \\
		&& \frac{d^2 \Phi}{d z^2}-  \frac{d-3}{z}\frac{d \Phi}{d z } -\frac{2g^2\Psi^2}{z^2(1-z^d)}\Phi =0.
		\label{eq:3}
		\eea
	\end{small}
	Here, $\Psi(z)$  is the scalar field  and $\Phi(z)$ is an electrostatic scalar potential $A_{t}$.
	Near the boundary $z=0$, we have  
	\bea
	\Psi(z)&=& z^{\Delta_{-}}\Psi^{(-)}(z)+z^{\Delta_{+}}\Psi^{(+)}(z),
	\nonumber \\ \Phi(z) &=& \mu- ({\rho}/{r_h^{d-2}})z^{d-2} +\cdots
	\label{eq:4}
	\eea
	where $\Delta_{\pm} =  {d/2\pm \sqrt{(d/2)^2+m^2}} $,
	$\mu$ is the chemical potential  and  $\rho$ is the charge density. By $\Delta$ we mean $\Delta_{+}$. 

	We  examine the range 
	$\frac{d-1}{2} \leq  \Delta < d$ 
	only,  because  the regime $0< \Delta <\frac{d-1}{2} $ is not physical. 
	Notice also that $\Delta=d/2$ is the value for which $\Delta_{+}=\Delta_{-}$ and   $\Delta=d$ is the value where $m^{2}=0$.  We request the boundary conditions at the horizon $z=1$: $\Phi(1)=0$ 
	and the finiteness of  $\Psi(1)$. Then  the condensate of the  Cooper pair
	operator $\mathcal{O}_{\Delta}$   dual to the field $\Psi$ is given   by  
	$
	\left< \mathcal{O}_{\Delta}\right>=\lim_{{r\to \infty}}\sqrt{2}r^{\Delta}\Psi(r)
	$
	under the assumption that the source  is zero. 
	
	\vskip.2cm
	\section{     Critical temperature $T_{c}$   in AdS4}\label{Tc}    
	At $T= T_c$, $\Psi =0$,  and Eq.(\ref{eq:3}) is integrated \cite{Siop2010} to give 
	\begin{equation}
		\Phi(z)= \lambda_{d} r_c (1-z^{d-2})  { }\mbox{ with } \lambda_{d}= {\rho}/{r_c^{d-1}}, 
	\end{equation}
	where $r_c$ is the horizon radius at $ T_c$. As $T\rightarrow T_c$,   the field equation of  $\Psi$ becomes
	\begin{equation}
		-\frac{d^2 \Psi }{d z^2} +\frac{d-1+z^d}{z(1-z^d)}\frac{d \Psi }{d z}+  \frac{m^2}{z^2(1-z^d)} \Psi =\frac{\lambda_{g,d}^2(1-z^{d-2})^2}{(1-z^d)^2}\Psi
	\end{equation}
	where $\lambda_{g,d}= g\lambda_d$. 
	Our   result  for the critical temperature is given by  
	\be
	T_c 
	=\frac{d}{4\pi} \left(\frac{g \rho }{\lambda_{g,d}}\right)^{\frac{1}{d-1}},
	\label{si:1}
	\ee
	which is  a part of the first line of table 1. 
	Details of deriving this result  is in sections \ref{Tcp} \& \ref{unphysical1}.  
	For $\Delta=1$ and 2 in   AdS$_{4}$, we have 
	$T_c/(g^{1/2}\sqrt{\rho } )=0.2256 $  and 0.1184  respectively. 
	If we set our  coupling  $g=1$, these   are in good agreement with the    numerical data of \cite{Hart2008} confirming the validity of our method.

	To find the $\Delta$-dependence of the $T_{c}$,  we first calculate $\lambda_{g,d}$.   The procedures are rather involved both analytically and numerically.  
	Here, we display  the analytic structure   of  the calculated data of $\lambda_{g,d}$ leaving the details to the section \ref{Tcp} and appendix \ref{hahaha}: 
	\begin{small}
		\bea
		\lambda_{g,3} &&= 1.96 \Delta^{4/3} - 0.87  \;\;\mbox{at}\;\; 1\leq \Delta \leq 3, \nonumber\\
		\lambda_{g,4} &&=  1.18 \Delta^{4/3} - 0.97 \;\;\mbox{at}\;\; 3/2\leq \Delta \leq 4.  
		\label{lambdafit}
		\eea    
	\end{small} 	
	Here, we used 	 the Pincherle’s Theorem with   matrix-eigenvalue algorithm\cite{Leav1990}.  
	Notice that the variational  method used in  \cite{Siop2010} is not applicable near the singularity $\Delta=(d-2)/2$. 
	\subsection{   Matrix algorithm and Pincherle’s Theorem}\label{Tcp}
	
	At the critical temperature $T_c$, $\Psi =0$, so  Eq.(\ref{eq:3}) tells us $\Phi^{''}=0$. Then, we can set
	\begin{equation}
		\Phi(z)= \lambda_3 r_c (1-z) \hspace{1cm}\mbox{where}\;\;\lambda_3 =\frac{\rho}{r_c^2}
		\label{eq:9}
	\end{equation}
	here, $r_c$ is the radius of the horizon at $T=T_c$. As $T\rightarrow T_c$, the field equation $\Psi$ approaches to  
	\begin{equation}
		-\frac{d^2 \Psi }{d z^2} +\frac{2+z^3}{z(1-z^3)}\frac{d \Psi }{d z}+  \frac{m^2}{z^2(1-z^3)} \Psi =\frac{\lambda_{g,3}^2}{(z^2+z+1)^2}\Psi
		\label{eq:10}
	\end{equation}
	where $\lambda_{g,3}= g\lambda_3$. Factoring out the behavior near the boundary $z=0$ and the horizon, we define
	\begin{equation}
		\Psi(z)= \frac{\left< \mathcal{O}_{\Delta}\right>}{\sqrt{2}r_h^{\Delta}}z^{\Delta}F(z)  \hspace{1cm}\mbox{where}\;\;F(z)=
		(z^2+z+1)^{-\lambda_{g,3}/\sqrt{3}}y(z)
		\label{eq:11}
	\end{equation}   
	Then, $F$ is normalized as $F(0)=1$ and we obtain
	\begin{eqnarray}
		&& \frac{d^2 y }{d z^2} +\frac{(1-\frac{4}{\sqrt{3}} \lambda_{g,3} +2\Delta)z^3+\frac{2\lambda_{g,3}}{\sqrt{3}}z^2 +\frac{2\lambda_{g,3}}{\sqrt{3}}z+2(1-\Delta)}{z( z^3-1)}\frac{d y}{d z}   	\label{eq:12}  \\
		&&+\frac{(3\Delta^2-4\sqrt{3}\Delta\lambda_{g,3}+4\lambda_{g,3}^2)z^2-(4\lambda_{g,3}^2-2\sqrt{3}\Delta\lambda_{g,3}+\sqrt{3}\lambda_{g,3})z-2\sqrt{3}(1-\Delta)\lambda_{g,3}}{3z(
			z^3-1)}y=0. 
		\nonumber
	\end{eqnarray}
	Notice that this is the generalized Heun's equation\cite{hounkonnou2009generalized} that has five regular singular points at $z=0,1,\frac{-1\pm\sqrt{3}i}{2},\infty$.
	Substituting $y(z)= \sum_{n=0}^{\infty } d_n z^{n}$ into (\ref{eq:12}), we obtain the following four term  recurrence relation:
	\begin{equation}
		\alpha_n\; d_{n+1}+ \beta_n \;d_n + \gamma_n \;d_{n-1}+\delta_n\;d_{n-2}=0  \quad
		\hbox{  for  } n \geq 2, \label{eq:13}
	\end{equation}
	with
	\begin{equation}
		\begin{cases} \alpha_n=-3(n+1)(n+2\Delta-2) \cr
			\beta_n=2\sqrt{3}\lambda_{g,3}(n+\Delta-1) \cr
			\gamma_n=\sqrt{3}(2n+2\Delta-3)\lambda_{g,3}-4\lambda_{g,3}^2 \cr
			\delta_n=3(n-\frac{2}{\sqrt{3}}\lambda_{g,3}+\Delta-2)^2
		\end{cases}
		\label{eq:14}
	\end{equation}
	The first four $d_{n}$'s are given by   $\alpha_0 d_1+ \beta_0 d_0=0$, $\alpha_1 d_2+ \beta_1 d_1+ \gamma_1 d_0=0 $, $d_{-1}=0$ and $d_{-2}=0$.
	Eq.(\ref{eq:11}), Eq.(\ref{eq:13}) and Eq.(\ref{eq:14})  give us the following boundary condition
	\begin{equation}
		F^{'}(0)=0.
		\label{eq:15}
	\end{equation}

	Since the 4 term relation can be reduced to the 3 term  relation, 
	we first   review for a minimal solution  of the three term recurrence relation \begin{equation}
		\alpha_n\; d_{n+1}+ \beta_n \;d_n + \gamma_n \;d_{n-1} =0  \quad
		\hbox{  for  } n \geq 1, \label{eq:18}
	\end{equation}
	with  $\alpha_0 d_1+ \beta_0 d_0=0$ and $d_{-1}=0$.
	Eq.(\ref{eq:18}) has two linearly independent solutions $X(n)$, $Y(n)$. We recall that $\{X(n)\}$ is a minimal solution of  Eq.(\ref{eq:18}) if not all
	$X(n)=0$ and if there exists another solution $Y(n)$ such that $\lim_{n\rightarrow\infty}X(n)/Y(n)=0$.
	Now $(d_n)_{n\in \mathbb{N}}$ is the minimal solution  if  $\alpha_0 \ne 0$ and
	\begin{equation}
		\beta_0 +\cfrac{-\alpha_0 \gamma_1}{\beta_1 -\cfrac{\alpha_1 \gamma_2}{\beta_2-\cfrac{\alpha_2
					\gamma_3}{\beta_3 -\cdots}}}=0. \label{eq:19}
	\end{equation}
	One should remember that $\alpha_{n},\beta_{n},\gamma_{n}$'s are functions of $\lambda$
	so that above equation should be read as equation for $\lambda$. 
	
	As we mentioned above, we can  transform the four term recurrence relations into three-term recurrence relations by the Gaussian elimination steps.  More explicitly,
	the transformed recurrence relation is
	\begin{equation}
		\alpha_n^{'}\; d_{n+1}+ \beta_n^{'} \;d_n + \gamma_n^{'} \;d_{n-1} =0  \quad
		\hbox{  for  } n \geq 1, \label{eq:21}
	\end{equation}
	where 
	\begin{equation}
		\alpha_n^{'}=\alpha_n,\quad  \beta_n^{'}=\beta_n,\quad  \gamma_n^{'}= \gamma_n  \quad
		\hbox{  for  } n =0,1
		\nonumber\\
	\end{equation}
	and
	\begin{equation}
		\begin{cases}  \delta_n^{'} =0\cr
			\alpha_n^{'}=\alpha_n \cr
			\beta_n^{'}= \beta_n-\frac{\alpha_{n-1}^{'}\delta_n}{\gamma_{n-1}^{'}} \cr
			\gamma_n^{'}= \gamma_n-\frac{\beta_{n-1}^{'}\delta_n}{\gamma_{n-1}^{'}}\quad
			\hbox{  for  } n \geq 2
		\end{cases}
		\label{eq:22}
	\end{equation}
	and  $\alpha_0^{'} d_1+ \beta_0^{'} d_0=0$ and $d_{-1}=0$. 
	Now the minimal solution is determined by 
	\begin{equation}
		\beta_0^{'} +\cfrac{-\alpha_0^{'} \gamma_1^{'}}{\beta_1^{'} -\cfrac{\alpha_1^{'} \gamma_2^{'}}{\beta_2^{'}-\cfrac{\alpha_2^{'} \gamma_3^{'}}{\beta_3^{'}
					-\cdots}}} =0
		\label{eq:23}
	\end{equation}
	which, in terms of the unprimed parameters,  is equivalent to
	\begin{equation}
		det\left(M_{N\times N}\right)= \begin{vmatrix}
			\beta_0& \alpha_0&  &  &  &  &  &     \\
			\gamma_1 & \beta_1  &  \alpha_1 &  &  &  &  &    \\
			\delta_2  & \gamma_2  & \beta_2  & \alpha_2  &  &  &  &     \\
			& \delta_3 & \gamma_3   & \beta_3 &   \alpha_3 &  &  &    \\
			&  & \delta_4 &  \gamma_4 & \beta_4 &  \alpha_4 &  &     \\
			&  &  & \ddots & \ddots & \ddots & \ddots &     \\
			&  &  &  &\delta_{N-1}  &\gamma_{N-1}  &\beta_{N-1}  &\alpha_{N-1}   \\
			&  &  &  &  & \delta_{N} & \gamma_{N} & \beta_{N}
			
		\end{vmatrix} =0, 
		\label{eq:24}
	\end{equation}
	or
	\begin{equation}
		d_N =0
		\label{eq:25}
	\end{equation}
	in the limit   $N\rightarrow \infty$.
	
	We now show why $y(z)$ is convergent at $z=1$ if $d_n$ in Eq.(\ref{eq:13}) is a minimal solution.
	We rewrite Eq.(\ref{eq:13}) as
	\begin{equation}
		d_{n+1}+ A_n\;d_n + B_n\;d_{n-1}+ C_n\;d_{n-2}=0,
		\label{eq:26}
	\end{equation}
	where $A_n$, $B_n$ and $C_n$ have asymptotic expansions of the form
	\begin{equation}
		\begin{cases} A_n= \frac{\beta_n}{\alpha_n} \sim \sum_{j=0}^{\infty}\frac{a_j}{n^j}   \cr
			B_n= \frac{\gamma_n}{\alpha_n} \sim  \sum_{j=0}^{\infty}\frac{b_j}{n^j}   \cr
			C_n= \frac{\delta_n}{\alpha_n}  \sim  \sum_{j=0}^{\infty}\frac{c_j}{n^j}
		\end{cases}
		\label{eq:27}
	\end{equation}
	with
	\begin{equation}
		\begin{cases} a_0= 0, \quad a_1= -\frac{2\lambda_{g,3}}{\sqrt{3}},\quad a_2= \frac{2\Delta\lambda_{g,3} }{\sqrt{3} } \cr
			b_0= 0, \quad b_1= -\frac{2\lambda_{g,3}}{\sqrt{3}},\quad b_2= \frac{(2\Delta +1)\lambda_{g,3} }{\sqrt{3} }  \cr
			c_0= -1, \quad c_1= 3+\frac{4\lambda_{g,3}}{\sqrt{3}},\quad c_2= -3-\frac{4\lambda_{g,3}}{\sqrt{3}}-(\frac{2\lambda_{g,3}}{\sqrt{3}}+\Delta)^2 .
		\end{cases}
		\label{eq:28}
	\end{equation}
	The radius of convergence, $\rho$, satisfies characteristic equation associated with Eq.(\ref{eq:26})   \cite{Miln1933,Perr1921,Poin1885} :  
	\be 
	\rho^3+ a_0\rho^2+b_0\rho^1+c_0\rho =\rho^3 -1= 0,
	\ee
	whose roots  are given by 
	\begin{equation}
		\rho_1 = 1,\hspace{1cm} \rho_2 = \frac{-1+\sqrt{3}i}{2},\hspace{1cm} \rho_3 = \frac{-1-\sqrt{3}i}{2}.   \label{eq:17}
	\end{equation}
	So   for a four-term recurrence relation in Eq.(\ref{eq:13}), the radius of convergence is 1 for all three cases.   Since the solutions should converge at the horizon, $y(z)$ should be convergent at
	$|z|\leq1$.   According to Pincherle’s Theorem  \cite{Jone1980},
	we have a convergent solution of $y(z)$ at $|z|=1$ if only if the four term recurrence relation Eq.(\ref{eq:13}) has a minimal solution. 
	Since  we have three different roots $\rho_i$'s, so Eq.(\ref{eq:26}) has three linearly independent solutions $d_1(n)$, $d_2(n)$, $d_3(n)$. 
	One can show that  \cite{Jone1980} for the large $n$, 
	\begin{equation}
		d_i(n)\sim \rho_i^n n^{\alpha_i}\sum_{r=0}^{\infty} \frac{\tau_i(r)}{n^r},\quad i=1,2,3
		\label{eq:29}
	\end{equation}
	with
	\begin{equation}
		\alpha_i=\frac{a_1\rho_i^2+b_1\rho_i+c_1}{a_0\rho_i^2+2b_0\rho_i+c_0 },\quad i=1,2,3
		\label{eq:30}
	\end{equation}
	and $\tau_i(0)=1$.
	In particular, we obtain
	\begin{equation}
		\tau_i(1)= \frac{(-a_0 \rho_i^2+3c_0)\alpha_i^2+(a_0\rho_i^2-2b_1\rho_i-3c_0-4c_1)\alpha_i+2(a_2\rho_i^2+b_2\rho_i+c_2)}
		{2(a_0\rho_i^2+2b_0\rho_i+3c_0)\alpha_i-2((a_0+a_1)\rho_i^2+(b_1+2b_0)\rho_i+c_1+3c_0)},\quad i=1,2,3
		\label{eq:31}
	\end{equation}
	Substituting Eq.(\ref{eq:17}) and Eq.(\ref{eq:28}) into Eq.(\ref{eq:29})--Eq.(\ref{eq:31}), we obtain
	\begin{equation}
		\begin{cases}  d_1(n)\sim n^{-1}\left( 1+\frac{-3\Delta^2+(3\sqrt{3}-4\lambda_{g,3})\lambda_{g,3}}{9n} \right)  \cr
			d_2(n)\sim  \left( \frac{-1+\sqrt{3}i}{2} \right)^n  n^{-1-\frac{2\lambda_{g,3}}{\sqrt{3}}}\left( 1+\frac{\varpi-\chi i}{n} \right)  \cr
			d_3(n)\sim  \left( \frac{-1-\sqrt{3}i}{2} \right)^n n^{-1-\frac{2\lambda_{g,3}}{\sqrt{3}}}\left( 1+\frac{\varpi+\chi i}{n} \right)
		\end{cases}
		\label{eq:32}
	\end{equation}
	with
	\begin{equation}
		\begin{cases} \varpi =\frac{-6\Delta^2-12\sqrt{3}\Delta\lambda_{g,3}+(15\sqrt{3}+16\lambda_{g,3})\lambda_{g,3}}{18}  \cr
			\chi = \frac{(3+4\sqrt{3})\lambda_{g,3}}{18}\end{cases}
		\label{eq:33}
	\end{equation}
	Since   $\lambda_{g,3}>0$, 
	\begin{equation}
		\lim_{n\rightarrow\infty}\frac{ d_2(n)}{ d_1(n)}=0, \quad \lim_{n\rightarrow\infty}\frac{ d_3(n)}{ d_1(n)}=0. 
		\label{eq:34} 
	\end{equation}
	Therefore
	$d_2(n)$ and $d_3(n)$ are minimal solutions.
	Also,
	\begin{equation}
		\begin{cases} \sum|d_1(n)|\sim \sum \frac{1}{n}\rightarrow\infty  \cr
			\sum|d_2(n)|\sim \sum n^{-1-\frac{2\lambda_{g,3}}{\sqrt{3}}}<\infty  \cr
			\sum|d_3(n)|\sim \sum n^{-1-\frac{2\lambda_{g,3}}{\sqrt{3}}} <\infty
		\end{cases}
		\label{eq:35}
	\end{equation}
	Therefore, $y(z)=\sum_{n=0}^{\infty}d_n z^n$ is convergent at $z=1$ if only if  we take $d_2$ and $d_{3}$ which are  minimal solutions.
	
	\vskip .5cm 
	Eq.(\ref{eq:24}) becomes a polynomial of degree $N$ with respect to $\lambda_{g,3}$.
	The algorithm to find $\lambda_{g,3}$ for a given $\Delta$ is as follows: 
	\begin{enumerate}
		\item  Choose an $N$. 
		\item
		Define a function returning the determinant of system Eq.(\ref{eq:24}).
		\item
		Find the roots of interest of this function. 
		\item
		Increase $N$ until those
		roots become constant to within the desired precision \cite{Leav1990}. 
	\end{enumerate}
	\subsection{ Presence of unphysical regime:   $\frac12<\Delta<1$}\label{unphysical1}
\begin{figure}[h]
		\centering
		\subfigure[  $\lambda_{g,3}$  vs $\Delta$ overall. ]
		{\includegraphics[width=5.8cm]{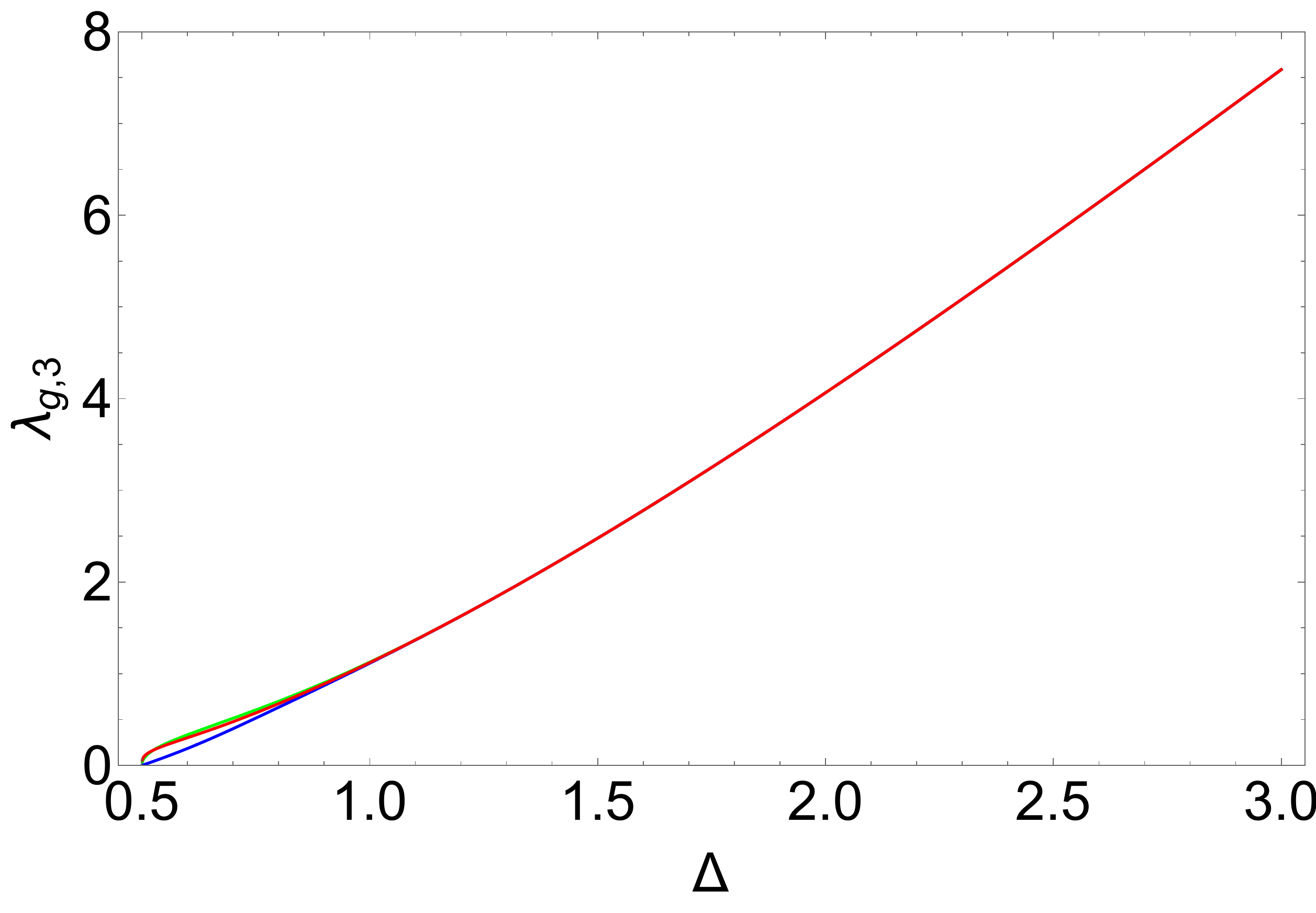}}
		\subfigure[  $\lambda_{g,3}$  vs $\Delta$ zoomed]
		{\includegraphics[width=6cm]{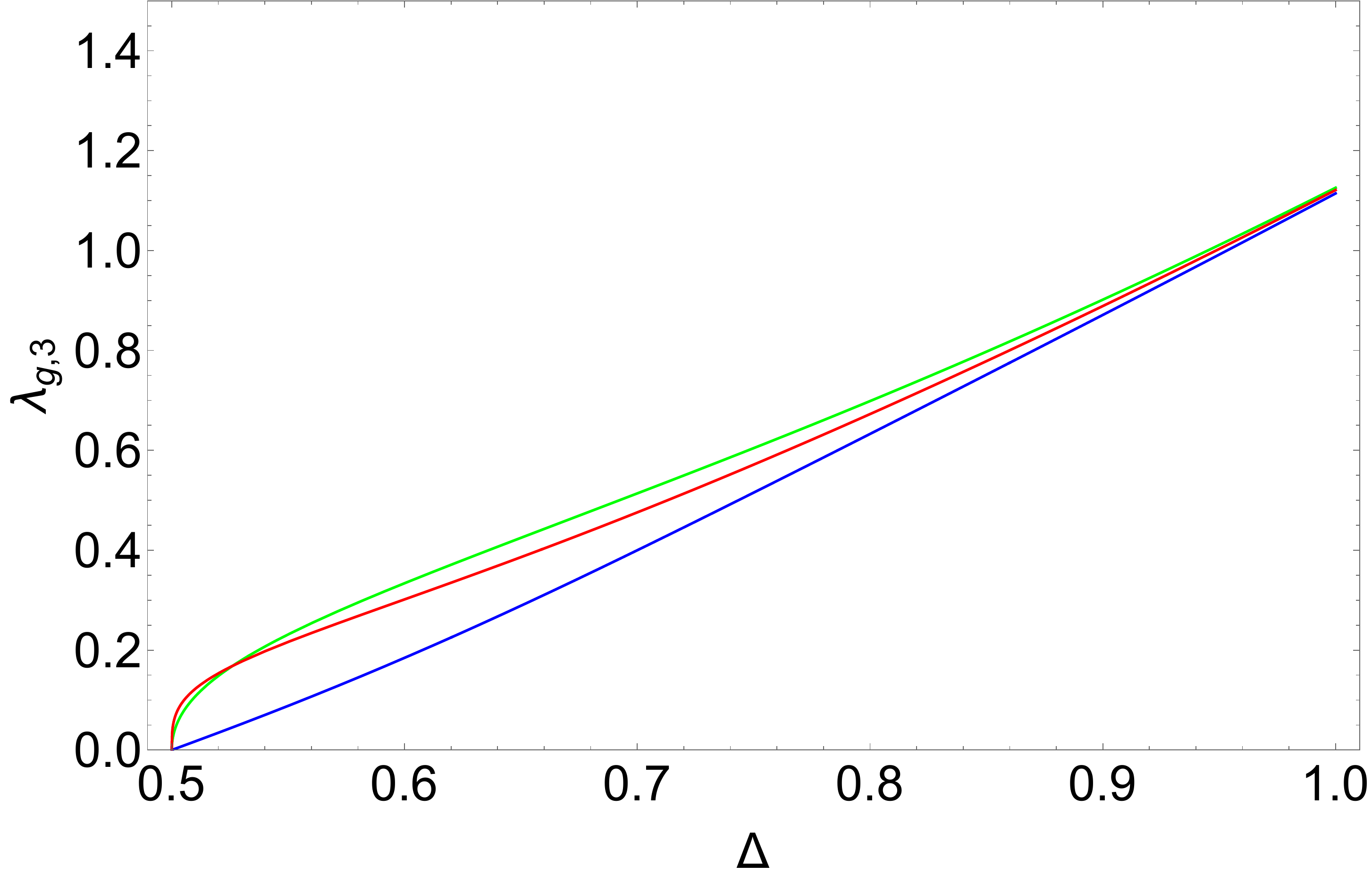}} 
		\caption{\small
			(a)  $\lambda_{g,3}$  vs $\Delta$: Green, blue and  red  colored curves for $\lambda_{g,3}$ are obtained by letting  $d_N =0$, with $N=30, 31, 32 $ respectively.  There are three branches in  $1/2<\Delta<1$. And such  brances merge for $\Delta\geq1$.  
			(b) Zoom  in of figure (a) for the regime $1/2<\Delta<1$.  
		}
		\label{Lambda_big4}
	\end{figure}

	We  numerically compute the determinant to locate  its roots. We are only interested in
	smallest positive real roots of $\lambda_{g,3}$. Taking $N=32$, we first compute the roots and then    find an approximate fitting function, which turns out to be given by
	\be
	\lambda_{g,3} \approx  1.96 \Delta^{4/3} - 0.87 \;\;\mbox{for}\;\; 1\leq \Delta\leq 3 \label{eq:36}
	\ee
However, for 	$1/2< \Delta<1$, we will see that there is no convergent solution, because there are three branches  so that it is impossible  to get an unique  value  $\lambda_{g,3}$ no matter how large $N$ is.  See  the	Fig.~\ref{Lambda_big4} (b).  Notice, however, that 	these three branches merge to the single value $\lambda_{g,3} \approx 1$ as $N$ increases as Fig.~\ref{Lambda_big4} (b) shows.

We now want to understand analytically  why three branches occur near $\Delta =1/2$ regardless of the size of $N$. Eq.(\ref{eq:24}) can be simplified using the formula for 
	the determinant of a block matirix,
		\begin{equation}
		\det \begin{pmatrix}
			A & B \\
			C & D 
		\end{pmatrix}= \det(A)\det(D-C A^{-1} B), \quad {\rm with }
	\end{equation}
%
	$ \begin{footnotesize}
		A= \begin{pmatrix}
			\beta_0 & \alpha_0 \\
			\gamma_1 & \beta_1 
		\end{pmatrix}  , 
		\quad B=\begin{pmatrix}
			0 & 0 &     \cdots & 0 \\
			\alpha_1 &     0 & \cdots & 0 
		\end{pmatrix}  , 
		\quad C=\begin{pmatrix}
			\delta_2 & \gamma_2 \\
			0 & \delta_3 \\
			0 & 0 \\
			0 & 0 \\
			\vdots & \vdots \\
			0 & 0 
		\end{pmatrix} ,
		\quad D=\begin{pmatrix}
			\beta_2 & \alpha_2 &  &  &  &  \\
			\gamma_3 & \beta_3 & \alpha_3 &  &  &  \\
			\delta_4 & \gamma_4 & \beta_4 & \alpha_4 &  &  \\
			& \ddots & \ddots & \ddots & \ddots &  \\
			&  & \delta_{N-1} & \gamma_{N-1} & \beta_{N-1} & \alpha_{N-1} \\
			&  &  & \delta_{N} & \gamma_{N} & \beta_{N} 
		\end{pmatrix}. 
	\end{footnotesize} $
By explicit computation,  we can see  the factor $\det(A)=9 \lambda^2$ at $\Delta =1/2$ so that the minimal real root is  $\lambda_{g,3}=0$.  
	Near $\Delta =1/2$, we can expand the determinant as a series in $ \varepsilon=\Delta- 1/2 \ll 1$ and  $0< \lambda_{g,3}\ll 1$. After some calculations, we found that    
	$d_N =0$  gives     following results: 
	\begin{enumerate}
		\item For $N=3m$ with positive interger $m $, 
		\be
		\lambda_{g,3}^3 \sum_{n=0}^{N-2}  \alpha_{1,n} \lambda_{g,3}^n +\varepsilon \lambda_{g,3}  \sum_{n=0}^{N}  \beta_{1,n} \lambda_{g,3}^n + \mathcal{O}(\varepsilon^2) =0. 
		\ee
		This leads us   $\lambda_{g,3} \sim \varepsilon^{1/2} \sim (\Delta-1/2)^{1/2}$ as far as $\alpha_{1,0}\beta_{1,0}\neq 0$, which can be confirmed by explicit computation.
		 This result does not depends on the size of  $N$.	Similarly, 	
		\item For $N=3m+1$,  
		\be
		\lambda_{g,3}^2 \sum_{n=0}^{N-1}  \alpha_{2,n} \lambda_{g,3}^n +\varepsilon \lambda_{g,3}  \sum_{n=0}^{N}  \beta_{2,n} \lambda_{g,3}^n + \mathcal{O}(\varepsilon^2) =0,
		\ee
		  giving  us   $\lambda_{g,3} \sim (\Delta-1/2)$.
		  		\item For $N=3m+2$, 
		\be
		\lambda_{g,3}^3 \sum_{n=0}^{N-2}  \alpha_{3,n} \lambda_{g,3}^n +\varepsilon    \sum_{n=0}^{N+1}  \beta_{3,n} \lambda_{g,3}^n + \mathcal{O}(\varepsilon^2) =0, 
		\ee
		leading to   $\lambda_{g,3} \sim (\Delta-1/2)^{1/3}$.
	\end{enumerate}
	These   results   prove the presence of  three branches     near $\Delta=1/2$. 	
	
	We numerically calculated  101 different values of $\lambda_{g,3}$'s  at various $\Delta$ and the result is the red colored curve in Fig.~\ref{Tcsingular}.
	These data fits well by above formula. 
	
	
	\begin{figure}[h]
		\centering
		\subfigure[   $T_{c}$  vs $\Delta$  in $AdS_4$]	
	{\includegraphics[width=5.5cm]{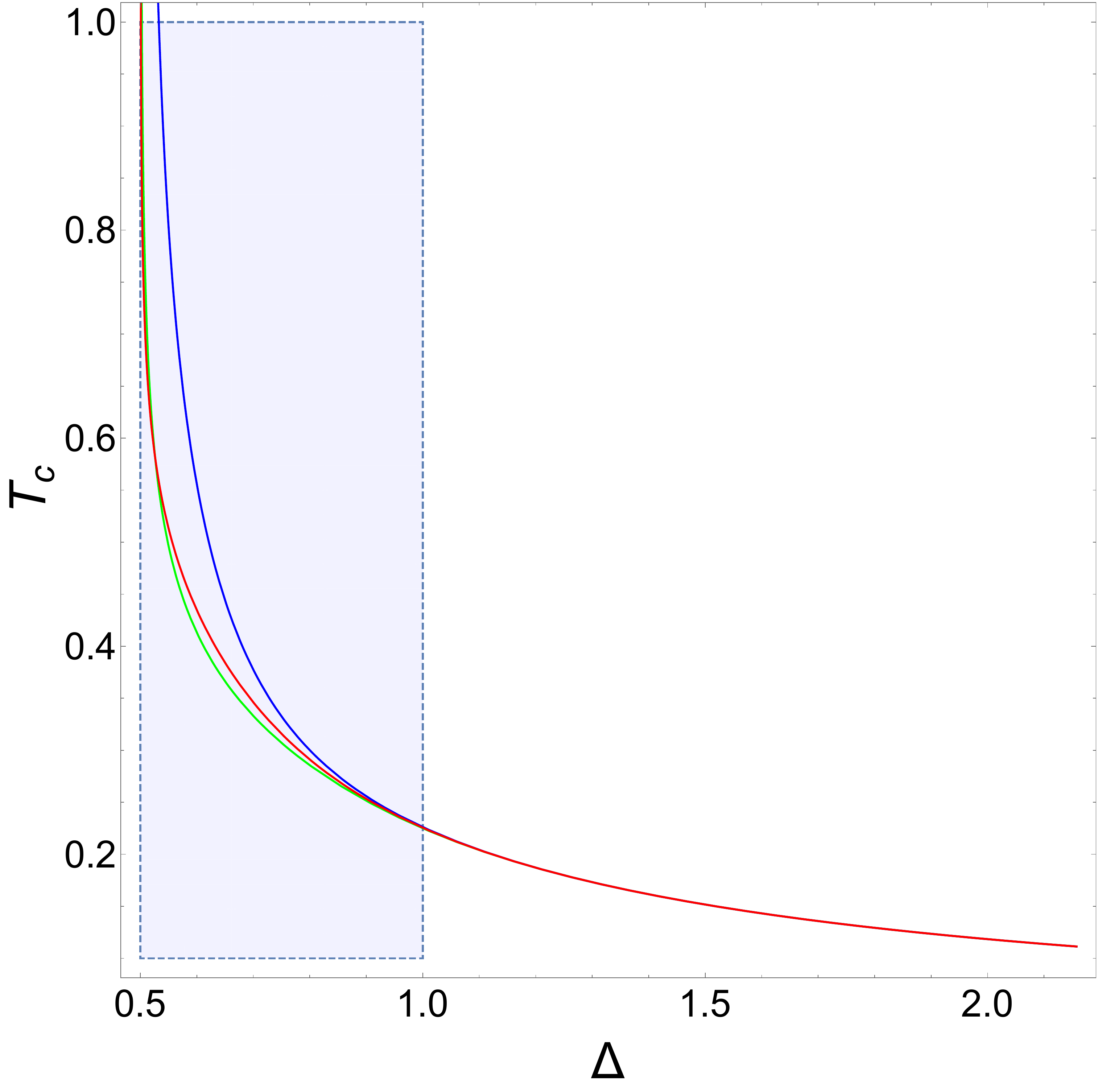}}
	\subfigure[   $T_{c}$  vs $\Delta$  in $AdS_5$]	
	{\includegraphics[width=5.5cm]{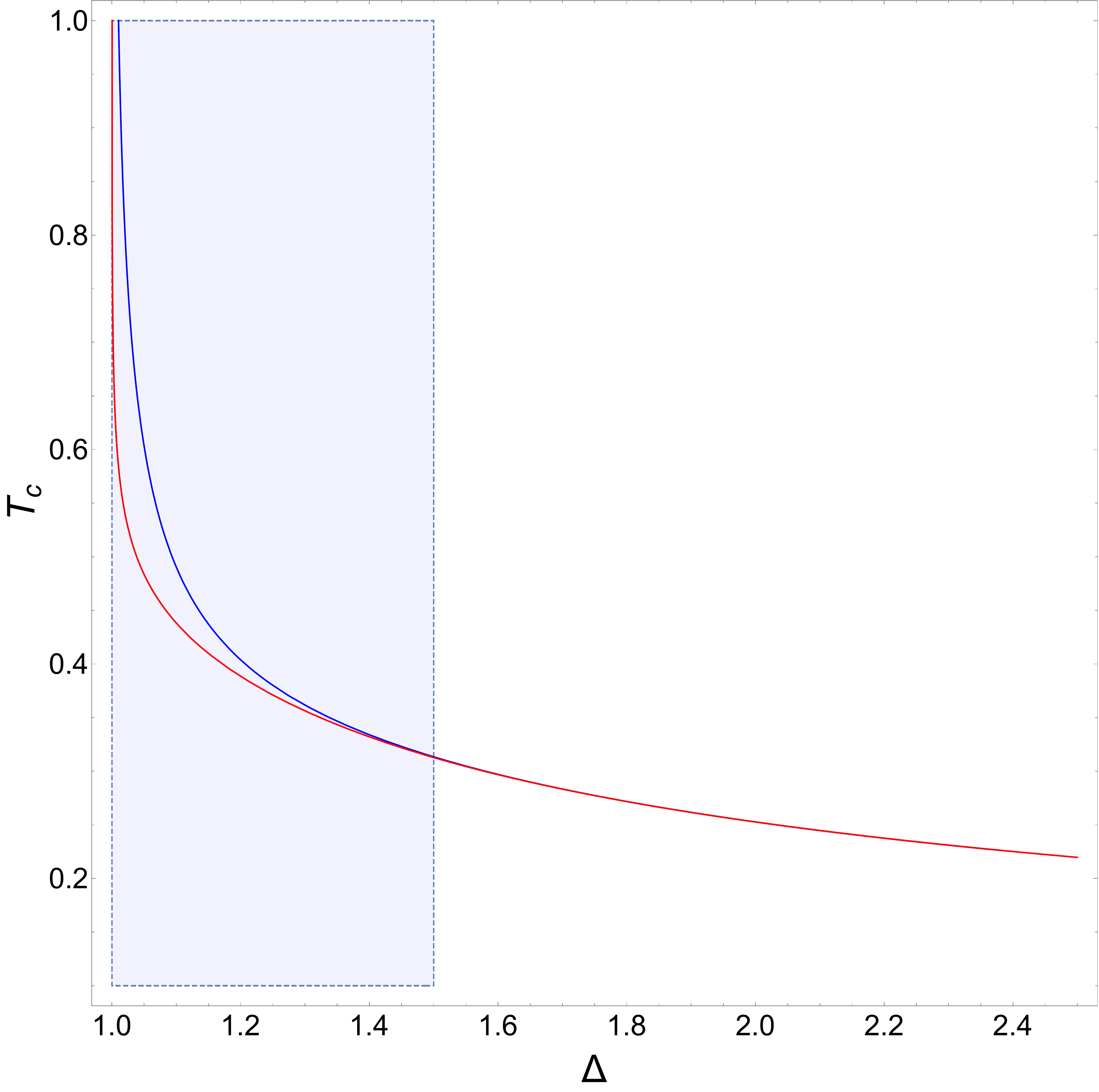}}
		\caption{\small
			(a)  Shaded area should be eliminated  due to the non-uniqueness of eigenvalues $\lambda_{g,3}$   shown in (a,b) at Fig.~\ref{Lambda_big4}.	This  naturally elliminates $\Delta=1/2$ where $T_{c}$ diverges. 
			We have set $g=\rho=1$. 
			(b) Similarly, shaded area at $1 \leq  \Delta < 3/2$  should  be eliminated in $AdS_5$ due to the non-uniqueness of eigenvalues $\lambda_{g,4}$. 	This  naturally elliminates $\Delta=1$ where $T_{c}$ diverges. There are two branches in   $AdS_5$ unlike  $AdS_4$.   
		}
		\label{Tc_value}
	\end{figure}

	\begin{figure}[h] 
			\centering
		\subfigure[  $\lambda_{g,d}$  vs $\Delta$ for d=3,4]
		{\includegraphics[width=7cm]{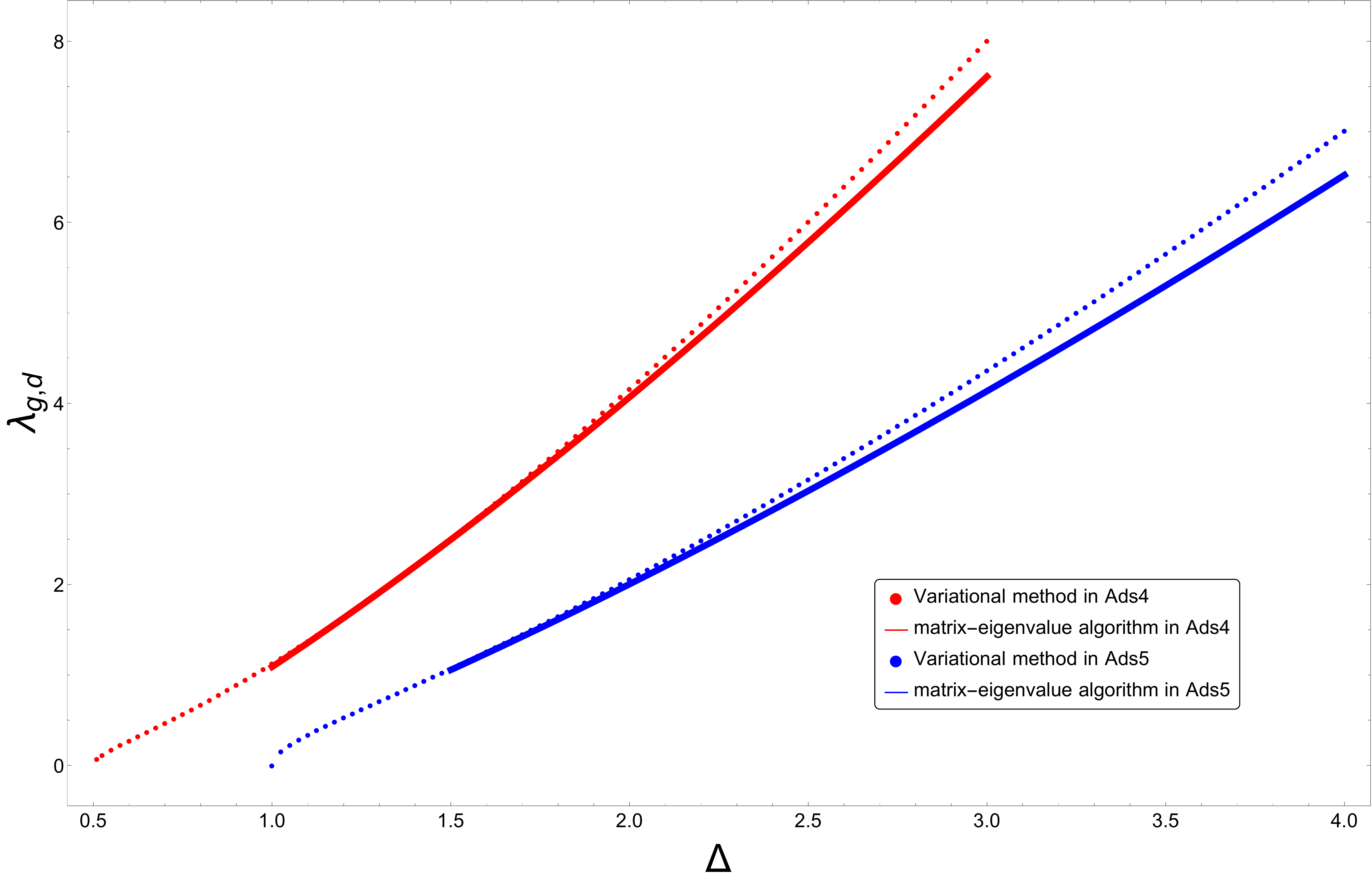}} 
		\caption{  $\lambda_{g,d}$  vs $\Delta$: red   line for $\lambda_{g,3}$ and blue line for $\lambda_{g,4}$. The results of variational  method in ref.  \cite{Siop2010} is denoted by red and blue dotted lines for AdS$_{4}$ and  AdS$_{5}$ respectively. 
		} 
		\label{Tcsingular}
	\end{figure}

	The authors of ref.\cite{Siop2010}  got  $\lambda_{g,3}$'s by using  variational method using the fact that the eigenvalue $\lambda_{g,3}$
	minimizes the expression
	\be
	\lambda_{g,3}^2 = \frac{\int_{0}^{1}dz\; z^{2\Delta-2}\left( (1-z^3)\left[F^{'}(z)\right]^2+\Delta^2 z \left[F(z)\right]^2\right)}{\int_{0}^{1}dz\;
		z^{2\Delta-2}\frac{1-z}{1+z+z^2}\left[F(z)\right]^2}
	\label{eq:37}.
	\ee
	for $\Delta>1/2$. The integral does not converge at $\Delta=1/2$ because of $\ln(z)$. The   trial function used is $F(z)=1-\alpha
	z^2$  where $\alpha$ is the variational parameter.  Their result is given by the  red   dotted line  in Fig.~\ref{Tcsingular}. While the variational method tells us that there are  numerical values of $\lambda_{g,3}$ for $1/2<\Delta<1$,    our method tells us that  this region  does not allow  well defined  value of  $\lambda_{g,3}$, hence $T_{c}$ is not defined there.

	
	
	The critical temperature  is given by 
	$
	T_c =\frac{3}{4\pi}\sqrt{\frac{\rho}{\lambda_3}}
	$, 
	so that it can be 
  calculated by   once $\lambda$ is given.    Notice that    $T_c$ is  a monotonically  decreasing function of $\Delta$. 

Similar statements are true for AdS5:  Depending on even-ness or odd-ness of $n$, there are two branches if  $1<\Delta<1.5$.  Two branches merge in $\Delta \geq1.5$  for AdS5. For more detail, see section \ref{unphysical2}.
	
	\section{The condensation   near    critical temperature}
	Substrituting Eq.(\ref{eq:11}) into  Eq.(\ref{eq:3}), the field equation $\Phi$ becomes
	\be
	\frac{d^2 \Phi}{d z^2} =   \frac{g^2\left< \mathcal{O}_{\Delta}\right>^2}{ r_h^{2\Delta}}\frac{z^{2(\Delta-1)} F^{2}(z)}{1-z^3 }  \Phi, 
	\label{eq:39}
	\ee
	where $ {g\left< \mathcal{O}_{\Delta}\right>^2}/{ r_h^{2\Delta}}$ is small because $T\approx T_c$.  The above equation    have the   expansion around  Eq.(\ref{eq:9}) with small correction \cite{Siop2010}:  
	\be
	\frac{ \Phi}{r_h} =   \lambda_3(1-z)+ \frac{g^2\left< \mathcal{O}_{\Delta}\right>^2}{ r_h^{2\Delta}}\chi_1(z)
	\label{eq:40}
	\ee
	We	have $\chi_1(1)=\chi_1^{'}(1)=0$  due  to  the boundary condition
	$\Phi(1)=0 $.   
	Taking the  derivative of Eq.(\ref{eq:40}) twice with respect to $z$
	and using the result in   Eq.(\ref{eq:39}),    
	\be
	\chi_1^{''}=\frac{z^{2(\Delta-1)}F^2(z)}{1-z^3}\left\{ \lambda_3(1-z)+ \frac{g^2\left< \mathcal{O}_{\Delta}\right>^2}{ r_h^{2\Delta}}\chi_1 \right\}\approx \frac{\lambda_3 z^{2(\Delta-1)}F^2(z)}{z^2+z+1}.  
	\label{eq:41}
	\ee
	Integrating Eq.(\ref{eq:41})  gives us 
	\be
	\chi_1^{'}(0) = -\lambda_3 \mathcal{C}_3 \quad
	\hbox{  for  }\mathcal{C}_3 = \int_{0}^{1}dz\;\frac{  z^{2(\Delta-1)}F^2(z)}{z^2+z+1}. 
	\label{eq:42}   
	\ee
	Eq.(\ref{eq:11}) with Eq.(\ref{eq:14}) shows
	\begin{equation}
		F(z)= (z^2+z+1)^{-\lambda_{g,3}/\sqrt{3}}y(z)\approx \left(z^2+z+1\right)^{-\frac{\lambda_{g,3}}{\sqrt{3}}} \sum_{n=0}^{15}d_n z^n .
		\label{eq:43}
	\end{equation}
	Here, we ignore $d_n z^n$ terms if  $n\geq16$ because  $0<|d_n|\ll1$ numerically and $y(z)$ converges for $0\leq z\leq1$.

\begin{figure}[h]
		\centering
		\includegraphics[scale=.28]{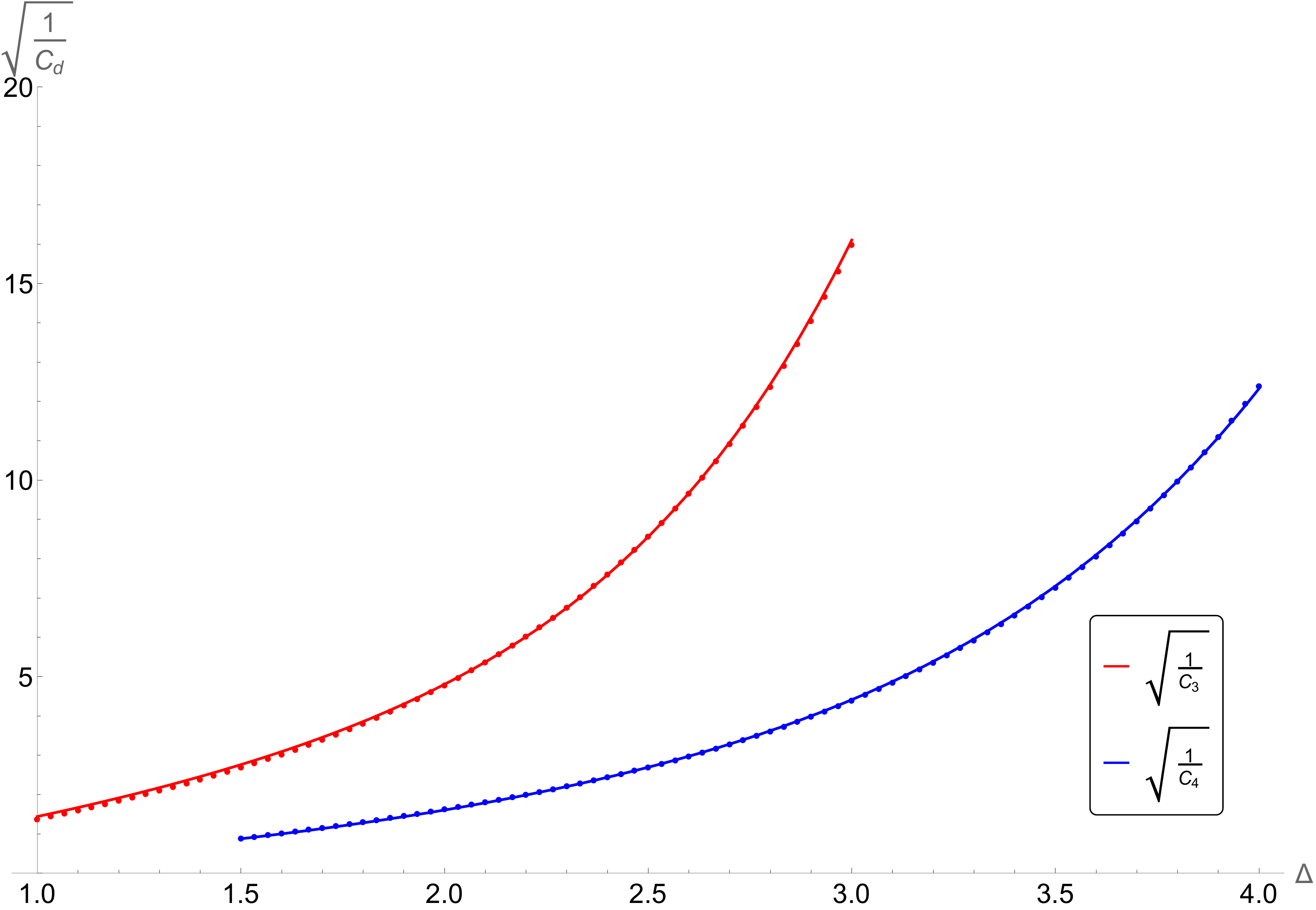}
		\caption{ (a) $\sqrt{1/\mathcal{C}_3}$ data by eq.(\ref{eq:42}) and  eq.(\ref{eq:43}) with eq.(\ref{eq:36}), as functions of $\Delta$. Red colored curve is the plot of eq.(\ref{eq:45}). (b) $\sqrt{1/\mathcal{C}_4}$ data by eq.(\ref{qq:42}) and  eq.(\ref{qq:43}) with eq.(\ref{qq:36}), as functions of $\Delta$. Blue colored curve is the plot of eq.(\ref{qq:45}).}
		\label{M.eigen}
	\end{figure}

\begin{figure}[!htb]
		\centering
		\includegraphics[width=0.6\linewidth]{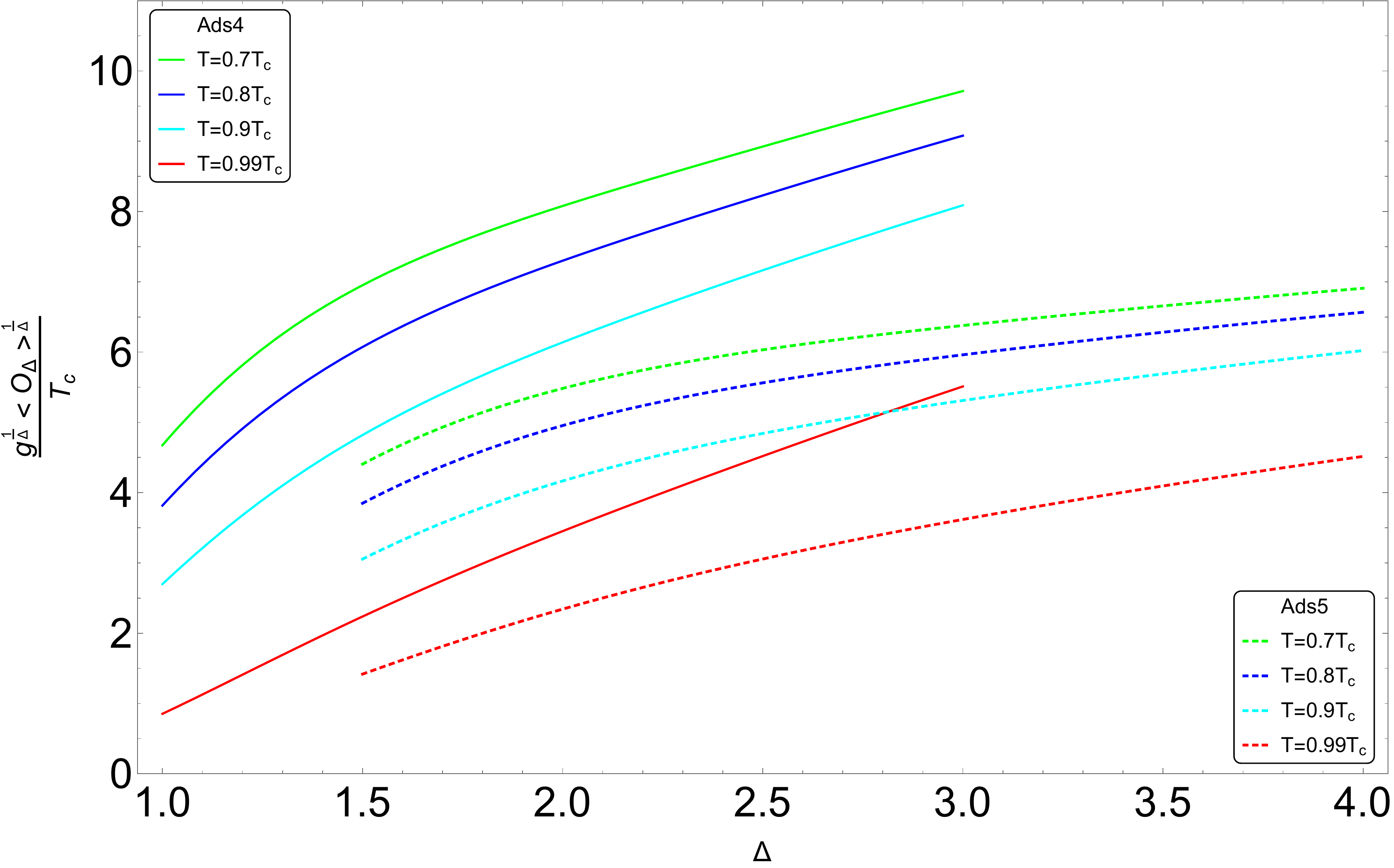}
		\caption{$ {g^{\frac{1}{ \Delta }}\vev{O_{\Delta }}^{\frac{1}{\Delta }}}/{T_c}$ vs  $\Delta$  for a few $T$'s  near $T_{c}$.   Sold curves are for  AdS$_{4}$ and dotted ones   are for  AdS$_{5}$.
		}
		\label{Odel_crit}
	\end{figure}

	We can calculate the numerical value of $\sqrt{1/\mathcal{C}_3}$
	by putting Eq.(\ref{eq:36}) and Eq.(\ref{eq:14}) into eq.(\ref{eq:42}) and  eq.(\ref{eq:43}). 
	We calculated  102 different values of $\sqrt{1/\mathcal{C}_3}$'s  at various $\Delta$,  which is drawn as  dots in Fig.~\ref{M.eigen}.
	Then we tried to find an approximate fitting function. 
	The result is given as follows, 
	\be
	\sqrt{ \frac{1}{\mathcal{C}_3} } \approx  \frac{\Delta ^6 +120 \Delta ^{1.5}}{84}\label{eq:45}   
	\ee
	Fig. \ref{M.eigen} shows how the data fits by above formula.
	From the Eq.(\ref{eq:40}) and Eq.(\ref{eq:4}),  we have 
	\be
	\frac{\rho}{r_h^2}=\lambda_3\left( 1+\frac{\mathcal{C}_3 g^2 \left< \mathcal{O}_{\Delta}\right>^2}{r_h^{2\Delta}}\right)
	\label{eq:46}
	\ee
	Putting $T=\frac{3}{4\pi} r_h  $ with $\lambda_3=\frac{\rho}{r_c^2}$ into Eq.(\ref{eq:46}), we obtain the condensate near $T_c$:  
	\be
	g\frac{\left< \mathcal{O}_{\Delta}\right>}{T_c^{\Delta}}\approx  \mathcal{M}_3 \;  \sqrt{1-\frac{T}{T_c}} \quad
	\hbox{  for  } \mathcal{M}_3 = \left(\frac{4\pi}{3}\right)^{\Delta}\sqrt{\frac{2}{\mathcal{C}_3}}
	\label{eq:47}
	\ee
	In ref. \cite{Siop2010}  it was argued that  $\lim_{\Delta \rightarrow d}\mathcal{C}_d=0$,  which would lead to the  divergence of the condensation in  eq. (\ref{eq:42}).  
	However,  our result shows that  $\lim_{\Delta \rightarrow d}\mathcal{C}_d =finite $ so that  eq. (\ref{eq:42}) is finite, which can be confirmed in the FIG. \ref{Odel_crit}.   
	The condensate is an increasing function of the $\Delta$ but it decreases with increasing $T$.  
	
	As we substitute  eq. (\ref{si:1}) into  eq. (\ref{eq:47}), we obtain
	\be
	g\frac{\left< \mathcal{O}_{\Delta}\right>}{(g \rho)^{\frac{\Delta}{2}} }\approx \lambda_{g,3}^{-\frac{\Delta}{2}} \sqrt{\frac{2}{\mathcal{C}_3}} \;  \sqrt{1-\frac{T}{T_c}}  
	\label{www:1}
	\ee

	The square root temperature dependence   is typical of a mean field theory   \cite{Hart2008,Herzog:2010vz,Siop2010}.    Our main interest here  is the $\Delta$ dependence of  the    ${\cal M}_{3}$, especially  the singular dependence through ${\cal C}_{3}$ whose values for some particular value of $\Delta$ was obtained before: for $\Delta=1$, we have $\mathcal{M}_3 =8.53$ which is in good agreement with the ${\mathcal M}_3 =9.3$ \cite{Hart2008}. For $\Delta=2$, we   have ${\mathcal M}_3 =119.17$  which roughly agrees with   the   results ${\mathcal M}_3  =119$ of ref. \cite{Jing2020}  and  ${\mathcal M}_3 =144$  of ref. \cite{Hart2008}. We  obtained  the  approximate results for general   $\mathcal{C}_d$.  See   eq.(\ref{eq:45}) and Eq.(\ref{qq:45})  in   the appendix.  
	For large $\Delta$, $\mathcal{C}_d \sim \Delta^{-(d+9)}$. 
	We conclude that     we do not have a singular dependence of the condensation  anywhere  for the s-wave holographic superconductivity, which is different from the result of  ref. \cite{Siop2010}.  
	See the FIG. \ref{Odel_crit}.

	\section{The AC Conductivity for $\Delta=1,2$ in 2+1} 
\begin{figure}[h]
		\subfigure[ $Re$$(\sigma (\omega))$ for   $\Delta=1$. ]
		{\includegraphics[width=0.5\linewidth]{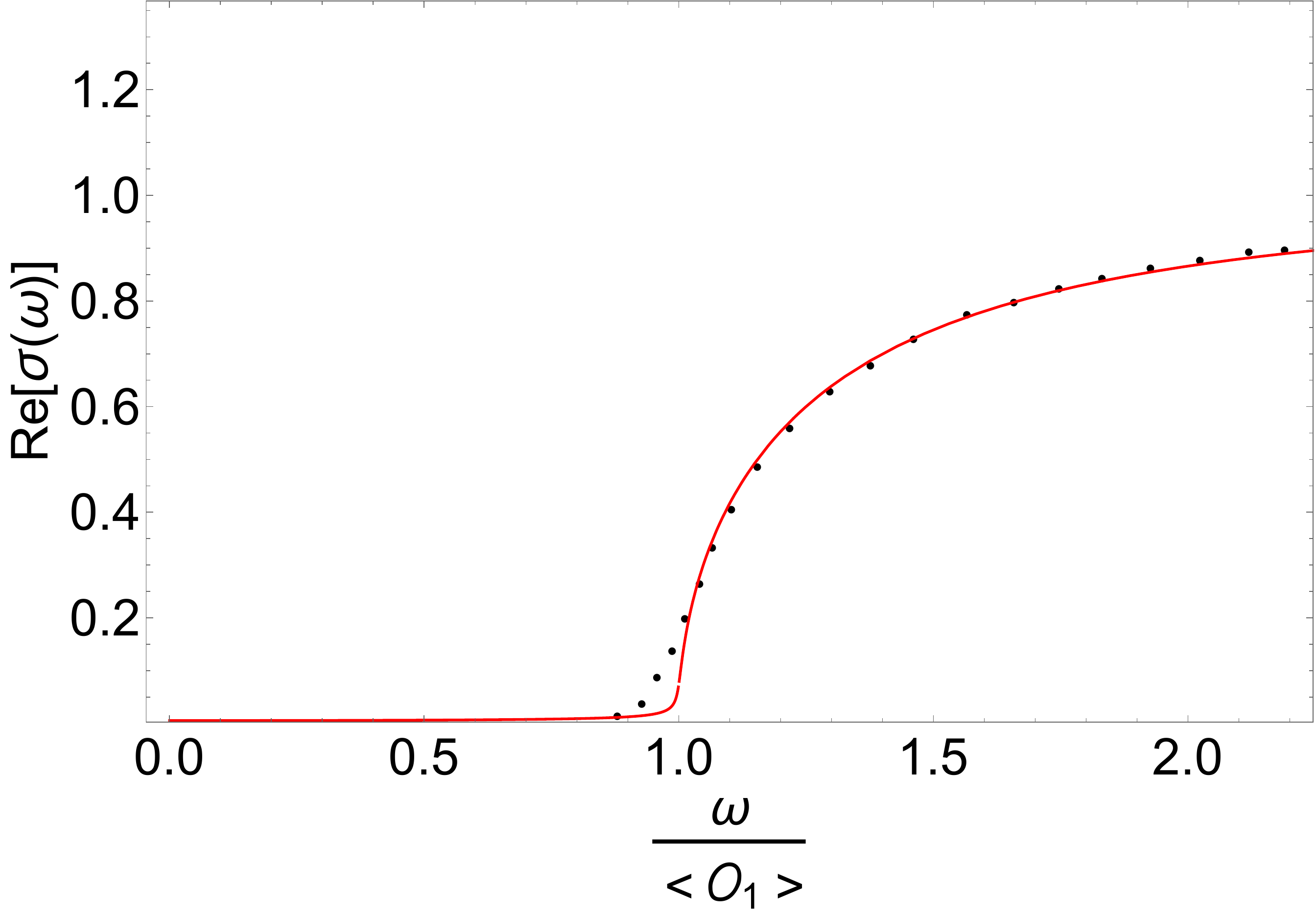}}
		\subfigure[  $Re$$(\sigma (\omega))$ for   $\Delta=2$. ]
		{\includegraphics[width=0.5\linewidth]{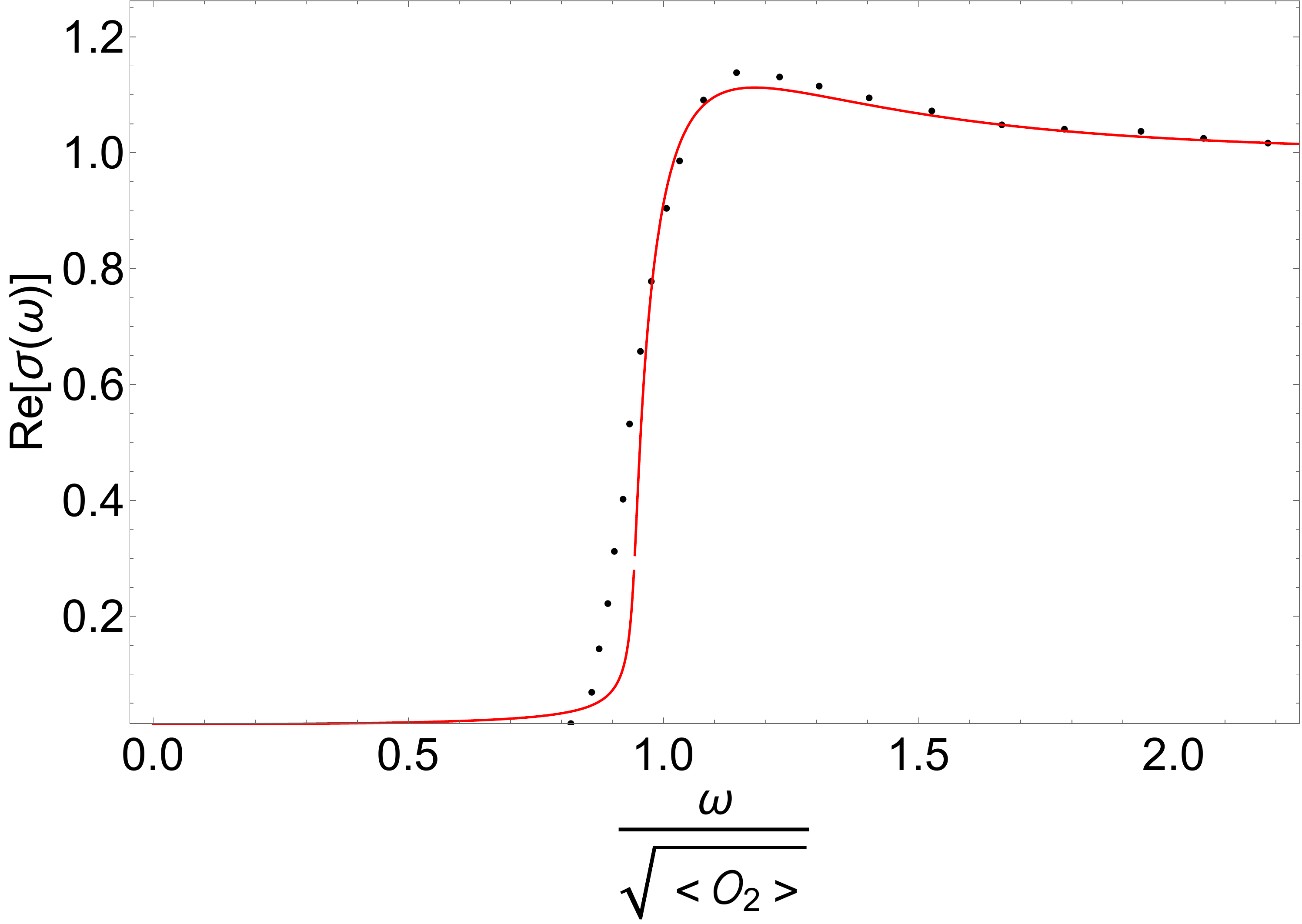}}
		\hfill
		\subfigure[   $Im$$(\sigma (\omega))$ for   $\Delta=1$. ]
		{\includegraphics[width=0.5\linewidth]{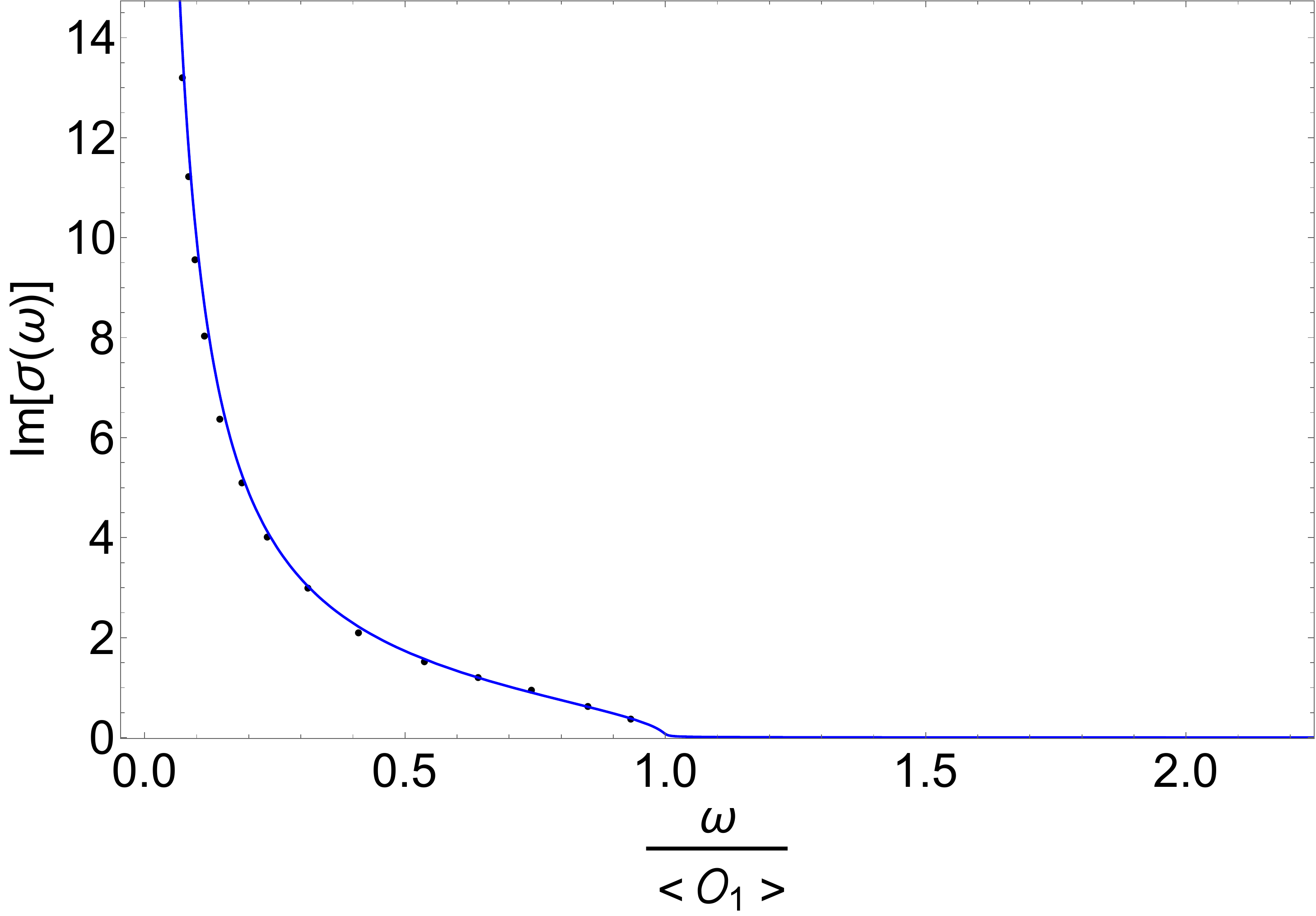}}
		\subfigure[  $Im$$(\sigma (\omega))$ for   $\Delta=2$. ]
		{\includegraphics[width=0.5\linewidth]{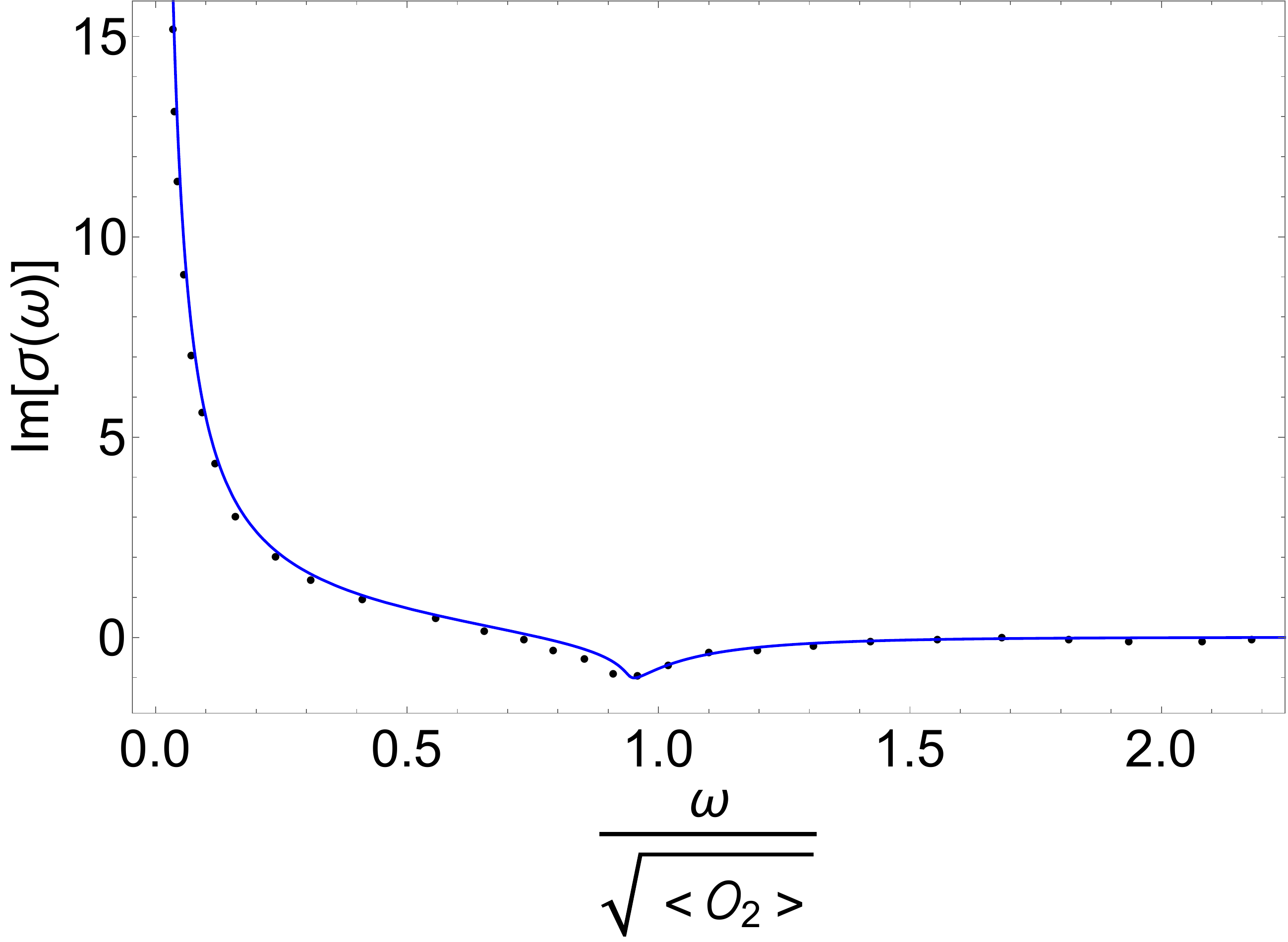}}
		\caption{\small   Analytic  results (real lines) vs numerical results(dotted lines).  
			(a,b)  Plots of  $Re$$(\sigma (\omega))$ in Eq.(\ref{codd:12}) and Eq.(\ref{codd:16}).  Dot points are numerical results in Fig. 3. of ref. \cite{Hart2008}.       (c,d)  Plots of   $Im$$(\sigma (\omega))$.   Dot points are  from Fig. 4. of ref. \cite{Hart2008}.   In all cases $T/T_c=0.1$.}.    
		\label{sigma3}
	\end{figure}

	The Maxwell equation for the planar wave solution with  zero spatial momentum and frequency   $ \omega$ is 	
	\begin{equation} 
		r_{+}^2 (1-z^3)^2 \frac{d^2 A_x }{d z^2} -3 r_{+}^2 z^2 (1-z^3)\frac{d A_x }{d z}+\left( \omega^2-V(z) \right) A_x=0, 
		\label{cod:1}
	\end{equation}	
	where $A_x$ is    the perturbing electromagnetic potential    and 
	$$V(z)= \frac{g^2 \left< \mathcal{O}_{\Delta}\right>^2}{r_{+}^{2\Delta-2}}(1-z^3) z^{2\Delta-2} F(z)^2,$$
	with $F$ defined as before. 	
	To request the ingoing   boundary conditions at the horizon,  $z=1$,  we introduce $G(z)$ by  $A_x(z)=(1-z)^{-\frac{i}{3} \hat{\omega}} G(z)$ where $\hat{\omega}=\omega/r_{+}$.      Then the wave equation  Eq.(\ref{cod:1}) reads
	{\small 
		\begin{eqnarray}  
			&&(1-z^3)  \frac{d^2 G }{d z^2} +  \left( -3 z^2+\frac{2i \hat{\omega}}{3} (1+z+z^2) \right) \frac{d G }{d z}\label{cod:2}\\
			&&+\left( \frac{(2+z)(4+z+z^2)}{9(1+z+z^2)}\hat{\omega}^2 + \frac{i\hat{\omega}}{3} (1+2z) - \frac{g^2\left< \mathcal{O}_{\Delta}\right>^2}{r_{+}^{2\Delta}} z^{2\Delta-2} F^2(z)  \right)G=0 
			\nonumber
	\end{eqnarray}	}
	If the asymptotic behaviour of the Maxwell field at large  $r$ is  given by 
	\begin{equation}
		A_x = A_x^{(0)}+\frac{ A_x^{(1)}}{r}+\cdots, 
		\label{cod:8}
	\end{equation}
	then the  conductivity is given by 	 
	\begin{equation}
		\sigma(\omega) =\frac{1}{i \omega} \frac{A_x^{(1)}}{A_x^{(0)}} = \frac{1}{i \hat{\omega}} \frac{\frac{d G(0) }{d z}+\frac{i\hat{\omega}}{3}G(0)}{G(0)}  .
		\label{cod:9}
	\end{equation}	   
	Near the $T=0$,  the equation  Eq.(\ref{cod:2})   is simplied to 	 
	\begin{equation} 
		\frac{d^2 G }{d z^2} +\frac{2i \hat{\omega} }{3}\frac{d G }{d z} +\left(\frac{8}{9} \hat{\omega}^2 +\frac{i \hat{\omega}}{3} - \frac{g^2\left< \mathcal{O}_{\Delta}\right>^2}{r_{+}^{2\Delta}} z^{2\Delta-2} F(z)^2 \right)G=0.
		\label{cod:10}
	\end{equation}	
	For $\Delta=1$,  $F(z)\approx 1$ so that   the solution of  Eq.(\ref{cod:10}) is  
	\begin{footnotesize}
		\bea
		G(z)=   \exp \left( i z\left(-\frac{\omega}{3}+\sqrt{ \omega^2+\frac{i \omega}{3}-\frac{g^2\left< \mathcal{O}_{1}\right>^2}{r_{+}^{2}}}\right)\right)  +R\;  \exp \left( i z\left(-\frac{\omega}{3}-\sqrt{ \omega^2+\frac{i \omega}{3}-\frac{g^2\left< \mathcal{O}_{1}\right>^2}{r_{+}^{2}}}\right)\right) 
		\nonumber
		\eea
	\end{footnotesize}  
	Here, $R$ is a  constant  called reflection coefficient.  Taking the zero temperature limit $T \rightarrow 0$  is equivalent to sending the horizon to  infinity. Then the in-falling boundary condition corresponds to $R=0$.   Then  it gives  the   conductivities,
	\begin{equation}
		\sigma(\omega) =\frac{g\left< \mathcal{O}_{1}\right>}{\omega} \sqrt{\left(1+\frac{i r_{+}}{3 \omega} \right)\left(\frac{\omega}{g\left< \mathcal{O}_{1}\right> } \right)^2-1}  . 
		\label{codd:12}
	\end{equation} 
	Compare  Figure \ref{sigma3}(a)  with   Figure \ref{sigma3}(c).  
	Similary, for $\Delta=2$, we  can obtain	the   conductivity  given as follow,  
	\begin{footnotesize}
		\begin{equation} 
			\sigma(\omega) =\frac{3i\sqrt{g\left< \mathcal{O}_{2}\right>}}{  \sqrt{2}\omega } \frac{\Gamma\left(   0.24-\frac{4i}{9}\sqrt{P(\omega)} \right) }{\Gamma\left(  -0.26-\frac{4i}{9}\sqrt{P(\omega)} \right) }  \frac{\Gamma\left(   1.26-\frac{4i}{9}\sqrt{P(\omega)} \right)}{\Gamma\left(   0.76-\frac{4i}{9}\sqrt{P(\omega)} \right)} 
			\label{codd:16}
		\end{equation}
	\end{footnotesize} 
	where
	\begin{equation} 
		P(\omega) =     \frac{9}{8}\left(1+\frac{i r_{+}}{3 \omega} \right)\left(\frac{\omega}{\sqrt{g\left< \mathcal{O}_{2}\right>}}\right)^2  -1
		\nonumber
	\end{equation} 
	This  result fits the numerical data almost exactly as one can 
	see in Figure \ref{sigma3}(b).  And it is consistent with the result of ref. \cite{Horo2009}; compare  Figure \ref{sigma3}(b)  with   Figure \ref{sigma3}(d). 
	For derivation of these results, see the appendix \ref{appendix3}.	
	To request the ingoing   boundary conditions at the horizon,  $z=1$,  we introduce $H(z)$ by  $A_x(z)=(1-z^3)^{-\frac{i}{3} \hat{\omega}} H(z)$ where $\hat{\omega}=\omega/r_{+}$.      Then the wave equation  Eq.(\ref{cod:1}) reads
	{\small 
		\begin{eqnarray}  
			&&(1-z^3)  \frac{d^2 H }{d z^2} -3 \left( 1-\frac{2i \hat{\omega}}{3} \right) z^2 \frac{d H }{d z}\label{codd:2}\\
			&&+\left( \frac{\hat{\omega}^2(1+z)(1+z^2)}{1+z+z^2} +2i\hat{\omega}z - \frac{g^2\left< \mathcal{O}_{\Delta}\right>^2}{r_{+}^{2\Delta}} z^{2\Delta-2} F^2(z)  \right)H=0 
			\nonumber
	\end{eqnarray}	}
	The boundary conditions   at the horizon are \cite{Horo2014}
	$$H(1)=1,  \quad \lim_{z\rightarrow 1}(1 -z^3)^{-\frac{i}{3}\hat{\omega}}H^{\prime}(z)=0. $$
	
	To evaluate the   conductivities at low frequency, it is enough to obtain $H(z)$ up to first order in  $\omega$, 
	\begin{equation} 
		H(z)= H_0(z) + \omega H_1(z) + \mathcal{O}(\omega^2) .
		\label{cod:18}
	\end{equation}	 
	Inserting this into Eq.(\ref{codd:2}), $H_0(z)$ and $H_1(z)$ satisfy	
	\begin{small}
		\bea
		&&(1-z^3) H_0^{\prime\prime}-3 z^2 H_0^{\prime}-  \Delta^2 b^{2\Delta} z^{2\Delta-2} F(z)^2 H_0  =0,  \label{cod:19} \\	 
		&&(1-z^3) H_1^{\prime\prime}-3 z^2 H_1^{\prime}-  \Delta^2 b^{2\Delta} z^{2\Delta-2} F(z)^2 H_1 
		= -2iz(z H_0^{\prime}+ H_0 )  .
		\label{cod:20}
	\end{eqnarray}	
\end{small}
where $b^{\Delta} = \frac{g\left< \mathcal{O}_{\Delta}\right> }{\Delta r_{+}^{ \Delta}}$. Near the $T=0$ we can simplify  two coupled equations  Eq.(\ref{cod:19}) and Eq.(\ref{cod:20}) as 	
\begin{eqnarray}   
	H_0^{\prime\prime} -  \Delta^2 b^{2\Delta} z^{2\Delta-2}  H_0  &=&0,  \label{cod:21}\\  
	H_1^{\prime\prime} -  \Delta^2 b^{2\Delta} z^{2\Delta-2}  H_1 &=& -2i z H_0 . 
	\label{cod:22}
\end{eqnarray}	
The  conductivity is given by 	 
\begin{equation}
	\sigma(\omega) =\frac{1}{i \omega} \frac{A_x^{(1)}}{A_x^{(0)}} = \frac{1}{i \hat{\omega}} \frac{\frac{d H(0) }{d z} }{H(0)}  .
	\label{coddd:9}
\end{equation}	  
The solution of Eq.(\ref{coddd:9}) is given in Eq.(\ref{codd:26}).  
Here, $n_s$ is   the coefficient of the pole in the imaginary part $\Im{\sigma(\omega)}\sim n_s/\omega $ as
$\omega \rightarrow 0$. For derivation of these results, see the appendix \ref{appendix3}.
For  the      $\Delta$ values other than 1 or 2, there is no analytic result  available  at this moment. 
 
\section{Discussion}

One problem is that  \cite{denef2009landscape,Siop2010,siopsis2011holographic} 
	  the critical temperature is divergent at the $\Delta=1/2$, which does not seems to make physical sense and it has not been understood as far as we know. 
 This  was also noticed as a problem   \cite{Horo2009} but the reason for it has not been cleared yet.    
 	
	In this paper, we consider the problem of divergence of the critical temperature at  $\Delta=1/2$
	 by recalculating  $T_{c}$ 
	using  Pincherle's theorem\cite{Leav1990}  to handle the Heun's equation. 
	We   find   that the region of $1/2\leq \Delta <1$ for AdS$_{4}$  does not have  well defined 
	critical temperature. Similar phenomena also  occur in AdS5. 
	We  also computed the AC conductivity gap $\omega_{g}$ and  in this same regime, it does not exist either. 
 The situation  is similar to  the physics of the pseudo gap where   Cooper pairs are formed  but   the phase alignment of the pairs are absent. 
 In the future work, we will work out the same phenomena in other background and also for non s-wave situation, to confirmed the universality of the phenomena.

%
%
\newpage

\section*{Appendices} 

\appendix 
\section{ H\lowercase{olographic superconductors with } A{\lowercase{d}}S{$_{4}$}}

The theory of 
holographic superconductors are much studied. Some of the relevant papers  for the  analytical techniques can be found, for example, in refs. 
\cite{cai2015introduction,ge2010analytical,gangopadhyay2012analytic,pan2010general,cai2011analytical,jing2011holographic,roychowdhury2012effect,gangopadhyay2012analytics,li2011analytical,pan2011analytical,zeng2011analytical,flauger2011striped,zhao2013notes,liu2011holographic,zhao2013holographic,roychowdhury2013ads,kanno2011note,peng2011various,pan2012analytical,banerjee2013holographic,yao2013analytical,lu2014lifshitz,sheykhi2016analytical,ghorai2016higher,sheykhi2016analyticals,lai2015analytical,liu2010dynamical,erdmenger2013striped,kuang2013building,gangopadhyay2014holographic,zhang2015holographic,cai2011magnetic,kim2013holographic,yao2014holographic}.
	After initial stage of the the model building \cite{Hart2008,Gubser:2008px} where probe limit of the gravity background was used, full back reacted version \cite{Hartnoll:2016apf}. 
Although   there are a few differences in the zero temperature limit,   the probe limit captures most of the physics\cite{Horo2009}. 
Later on,  physical observables of the superconductivity are  numerically calculated \cite{Horo2009} as functions of  the conformal weight ($\Delta$) of the Cooper pair operator. These include ${\mathcal O_{\Delta}}$.   $T_{c}, \vev{{\mathcal O}_{\Delta}}, \sigma({\omega}), \omega_{g},  \omega_{i}, n_{s}, $ which are the critical temperature, the condensation of  the Cooper pair operator, the AC conductivity, the gap in the AC conductivity, the resonance frequencies, and the density of the cooper pairs respectively. 

Since the parametric dependences of observables are   crutial in understanding the underlying  physics, it would be nice to have an analytic expressions  within the probe approximation, while it would be senseless to try  to replace  the fully back reacted numerical solution. 
Works in this direction had been initiated 
in \cite{Siop2010,siopsis2011holographic}. 
In this paper, we reconsider the problem since  many of the result could not be reproduced.  We got the analytic results which also agree with the numerical results of the original paper \cite{Horo2009}. 
Since the details are  rather long, we  summarize our results here.  

\begin{figure}[h]
	\centering
	\includegraphics[width=0.6\linewidth]{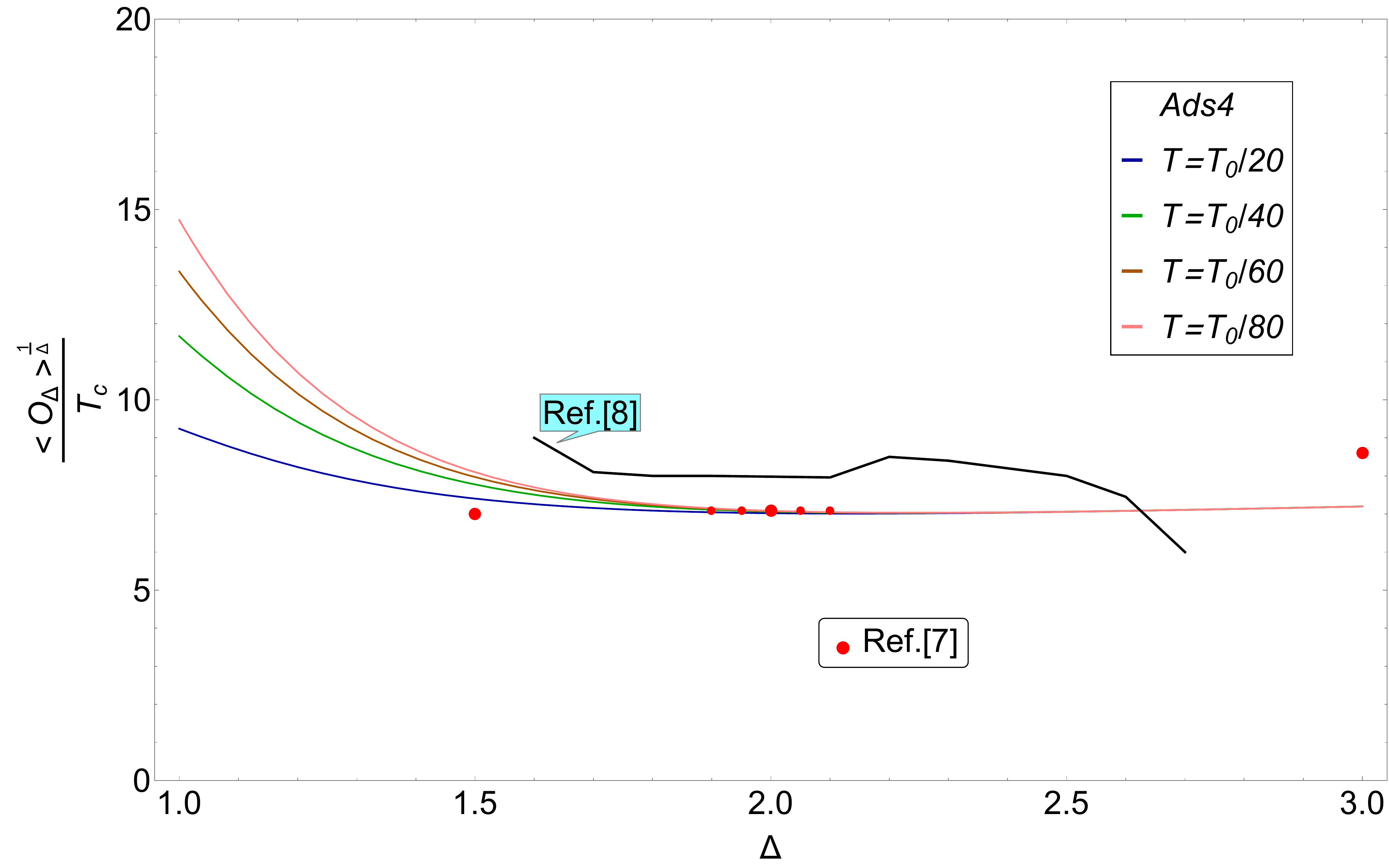}
	\caption{ $  {\left< \mathcal{O}_{\Delta}\right>^{\frac{1}{\Delta}}}/{T_c}$  vs    $\Delta$ for 
		$d=3$. 
		Smooth colored lines are our results. 
		Here, $T_{0}=\rho^\frac{1}{2}$ with $g=1$.
		For large $\Delta>3$,  the graph  does not saturate to a constant but increases slowly.} 
	\label{Odelta_b}
\end{figure}

	We  also  calculated  the  the Cooper pair condensation    $\left< \mathcal{O}_{\Delta}\right>$ as an analytic function of $\Delta$, which is plotted in   FIG. \ref{Odelta_b} where   we compared our results (real colored lines) with those of ref. \cite{Horo2009} (a few red dotted data ) and  \cite{Siop2010} (black broken line). 
Noticed that   the condensation does not change much for a  region around $\Delta=2$ and slowly increasing as $\Delta\to 3$.  Our analytic formula 
reproduces the values of of   ref. \cite{Horo2009} near $\Delta=2$ 
and gives a finite value of the condensation near $\Delta=3$ unlike  ref. \cite{Siop2010}. 
Notice also  that the condensation is almost independent of   $T$ and $\Delta$ over $3/2 < \Delta <3$ region.  
Interestingly, we will see that the flatness of the graph over the region $3/2 < \Delta <3$ comes  as a consequence of the remarkable cancellation of singularities of two functions at $\Delta=3/2$. Similar result  holds in three spatial dimension as well as in two dimension.   
Our results for $T_{c}$ , $\left< \mathcal{O}_{\Delta}\right>/T_c^{\Delta}$ and $\left< \mathcal{O}_{\Delta}\right>$ for both 
near $T=T_{c}$ and $T=0$  are summarized  in the Table \ref{d5}  and  \ref{d4}.
\begin{table}[!htb]
	\centering
	\begin{tabular}{|l|l|l|} 
		\hline
		{\small AdS}$_{d+1}$  & $\quad\; T_c\sim   \left({g}{\rho}\right)^{1/(d-1)}$
		for   $\frac{d-1}{2}\leq \Delta \leq d$              
		\\ \hline
		$T\approx T_c$ & $\left< \mathcal{O}_{\Delta}\right>\sim l^{-\Delta} g^{\gamma_g}\sqrt{1-\frac{T}{T_c}}$   for   $\frac{d-1}{2} \leq \Delta \leq d$                    \\ \hline
		$T\approx 0$ & $\left< \mathcal{O}_{\Delta}\right>\sim l^{-\Delta} g^{\gamma_g}  \left( \frac{T_0}{T}\right)^{\gamma_1}$  if  $\frac{d-2}{2} < \Delta\ll\frac{d}{2}$         \\	
		& $\left< \mathcal{O}_{\Delta}\right>\sim l^{-\Delta} g^{\gamma_g}  \left( \ln \frac{T_0}{T}\right)^{\gamma_2} $  if   $\Delta = \frac{d}{2} $   \\
		& $\left< \mathcal{O}_{\Delta}\right>\sim l^{-\Delta} g^{\gamma_g} $  if   $\frac{d}{2}\ll\Delta \leq d$  
		\\ 
		\hline
	\end{tabular}
	\caption{  $T_c$ and $\left< \mathcal{O}_{\Delta}\right>$  near   $T= T_c$  and  $T = 0$.  Here, 
		$\gamma_1 = \frac{\Delta (d-2\Delta)}{(d-2+2\Delta)}$, $\gamma_2 = \frac{\Delta}{2(d-1)} $ and $T_0 =\left( g \rho \right)^{\frac{1}{d-1}}$ respectively, and  
		$l={\rho}^{-\frac1{d-1}}$ is the   distance scale   given by the density $\rho$.    } \label{d5}
\end{table}  
\begin{table}[!htb]
	\centering
	\begin{tabular}{|l|l|l|} 
		\hline
		{\small AdS}$_{d+1}$  &         
		\\ \hline 
		$T\approx T_c$ & $\left<X\right> \sim l^{d-1}  \sqrt{1-\frac{T}{T_c}}$   for   $\frac{d-1}{2}\leq \Delta \leq d$                    \\ \hline
		$T\approx 0$ & $\left<X\right> \sim l^{d-1}    \left( \frac{T_c}{T}\right)^{\gamma_1}$  if  $\frac{d-1}{2} \leq \Delta\ll\frac{d}{2}$         \\	
		& $\left<X\right> \sim l^{d-1}    \left( \ln \frac{T_c}{T}\right)^{\gamma_2} $  if   $\Delta = \frac{d}{2} $   \\
		& $\left<X\right> \sim l^{d-1}   $  if   $\frac{d}{2}\ll\Delta \leq d$  
		\\ 
		\hline
	\end{tabular}
	\caption{    $\left<X\right>:=\frac{\left< \mathcal{O}_{\Delta}\right>}{T_c^{\Delta}} $  near   $T= T_c$  and  $T = 0$.      } \label{d4}
\end{table}  

\begin{figure}[h]   
	\centering
	\includegraphics[width=0.6\linewidth]{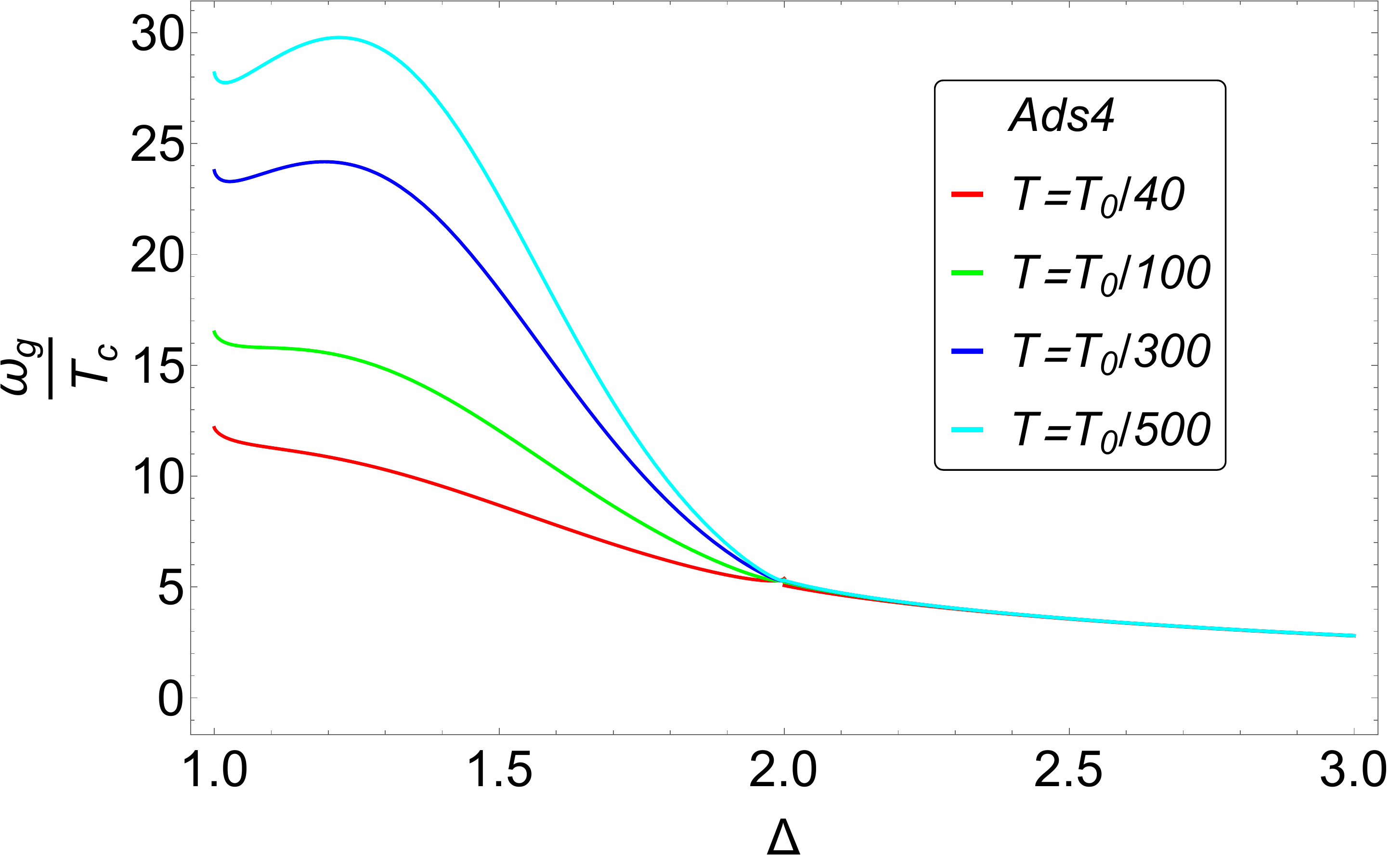}
	\caption{ $ \omega_g/T_c$  vs    $\Delta$ for  $d=3$. Here, $T_{0}=(g \rho)^\frac{1}{2}$.    For   $\Delta>3$,   it  decreases slowly.} 
	\label{omegag}
\end{figure}

The second   quantity calculated  is $\omega_g$, the 
gap in the  optical (AC)  conductivity.  
Notice that there is no solution for $\omega_g$ at $1/2 <\Delta <1$; see appendix \ref{appendix4}.  
The co-incidence of this regime with that of non-existence of the critical temperature gives us a confidence in concluding the absence of the superconductivity in this regime. 
Our  results for the $\omega_g$ is summarized in the Table \ref{omegag1}, which are   
plotted in   FIG. \ref{omegag}.  The size of the gap  is defined  by   $\omega_g =  \sqrt{V_{\mbox{max}}}$ \cite{Horo2009}; one should notice that ref. \cite{Horo2009} and ref. \cite{Horo2011} use slightly different definition of $\omega_g$. 
\begin{table}[!htb]
	\centering
	\begin{tabular}{|l|}
		\hline
		$\omega_g/T_c =c_1 X^{\Delta} \left(\frac{T_c}{T}\right)^{\Delta-1} $,   for   $1\leq \Delta \ll \frac{3}{2}$, {\scriptsize $c_1 = \left(\frac{3z_0}{4\pi} \right)^{\Delta-1} \left( 1-\left(\frac{\Delta}{3-\Delta} z_0^{3/2-\Delta}\right)^2\right)$}	\\ \hline
		$\omega_g/T_c = c_2 \frac{ {X^{3/2} \left(\frac{T_c}{T}\right)^{1/2}}}{\ln \left(X \frac{T_c}{T}\right)}$,  for   $ \Delta = \frac{3}{2}$,   ${ \scriptsize c_2 = \frac{7}{10} }$	\\ \hline 
		$\omega_g/T_c  = c_3   X^{3-\Delta} \left(\frac{T_c}{T}\right)^{2-\Delta}   $  for   $\frac{3}{2} \ll \Delta < 2$,	${ \scriptsize	c_3 = c_{30} \left[ 1-\left(\frac{\Delta}{3-\Delta} z_0^{\Delta-3/2}\right)^2\right]} $ 	\\ \hline
		$\omega_g/T_c  = c_4  X   $  for   $2 \leq \Delta \leq 3$, ${ \scriptsize c_4 =  \frac{1.1}{\mbox{Li}(\Delta^{1.2})} }$	\\ \hline
	\end{tabular} 
	\caption{  $\omega_g/T_c$  as function of $ { \scriptsize X=\frac{g^{1/\Delta}\left< \mathcal{O}_{\Delta}\right>^{1/\Delta}}{T_c}} $  near   $T \approx 0$.   
		{\scriptsize $  c_{30}= \frac{\sqrt{\pi} 
				\left( {3z_0}/{4\pi} \right)^{2-\Delta}
				\Gamma\left(\frac{3+\Delta}{2+\Delta}\right)\csc\left(\frac{\pi}{2\Delta}\right)}{\Delta^{3/\Delta}\Gamma\left(\frac{2}{\Delta}\right)\Gamma\left(\frac{3}{2\Delta}\right)\Gamma\left(1+\frac{1}{\Delta}\right)} $}. 
	} \label{omegag1}
\end{table}  
Notice that $\omega_g/T_c$  has the  slightly decreasing tendency as a function of $\Delta$ instead of the slowly increasing  behavior of ref. \cite{Horo2009}. So there is a small mismatch between the two.

\begin{figure}[h]
	\centering
	\includegraphics[width=7cm]{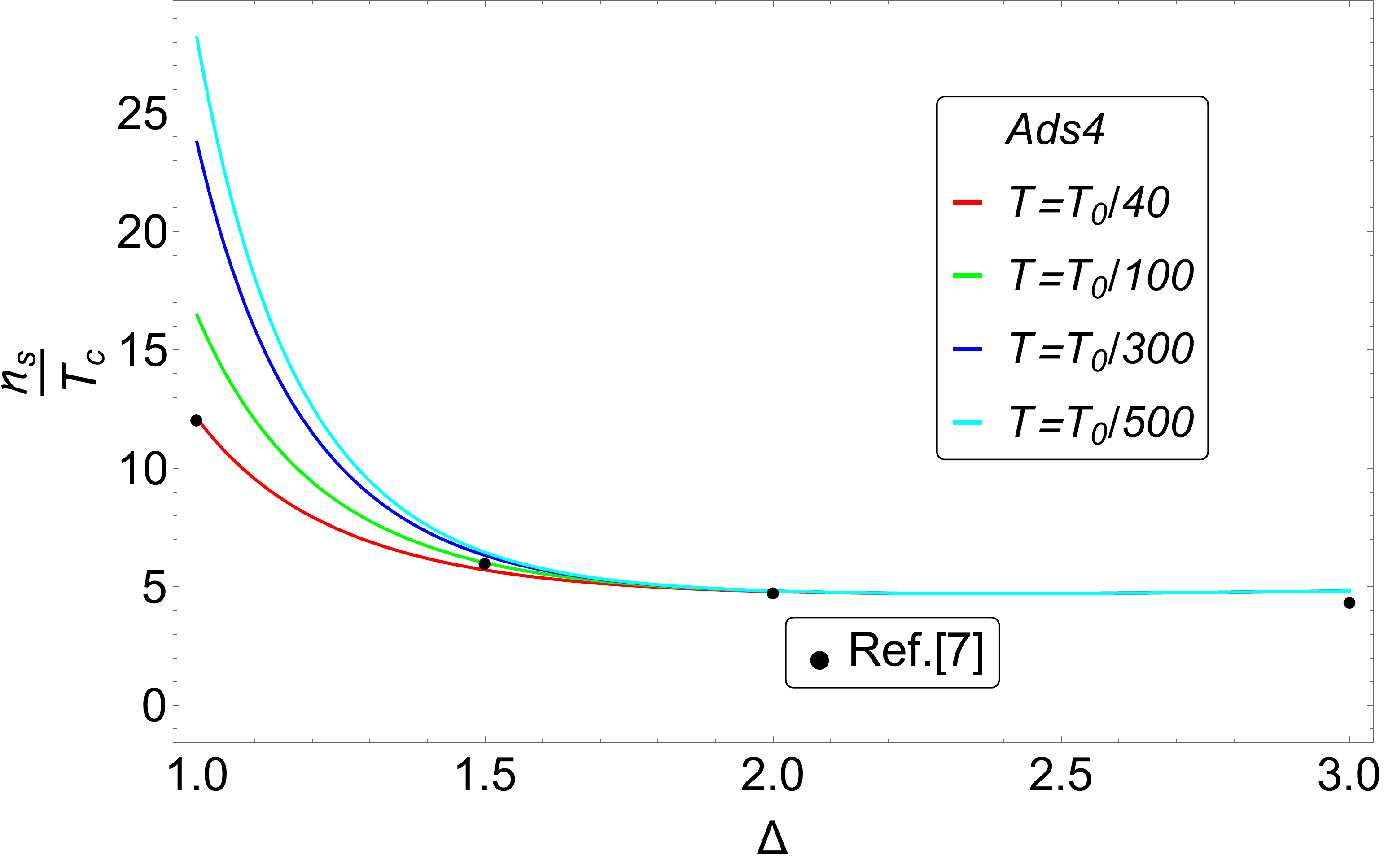}
	\caption{ $ n_s/T_c$  vs    $\Delta$ for  $d=3$.  Colored real lines are plots of  our analytic results  while the black   points are numerical data of  ref. \cite{Horo2009}. Notice that 
		it is    almost independent of   $T$  and   $\Delta$ over the  region  $2\leq \Delta \leq3$.}
	\label{superfluid}
\end{figure}

The third quantity we calculated is the superfluid density $n_s$, which appears as  the residue of the pole in the imaginary part of the optical conductivity  at  $\omega = 0$. 
We obtained it as an analytic function of $\Delta$   given below, 
\begin{equation}
	\frac{n_s}{T_c} = \frac{2\pi \Delta \csc\left( \frac{\pi}{2\Delta} \right)}{(2\Delta)^{1/\Delta} \left(\Gamma\left(\frac{1}{2\Delta}\right)\right)^2} \frac{g^{1/\Delta}\left< \mathcal{O}_{\Delta}\right>^{1/\Delta}}{T_c}, 
	\label{codd:26}
\end{equation}
which is plotted in   FIG. \ref{superfluid}.     By plotting  our result, we find that it agrees with the numerical result of ref. \cite{Horo2009} for all the  data points given there: See   FIG. \ref{superfluid}. 

\begin{figure}[!htb]
	\subfigure[$\left< \mathcal{O}_{\Delta}\right>^{1/\Delta}$  vs    $\Delta$.]
	{ \includegraphics[width=0.31\linewidth]{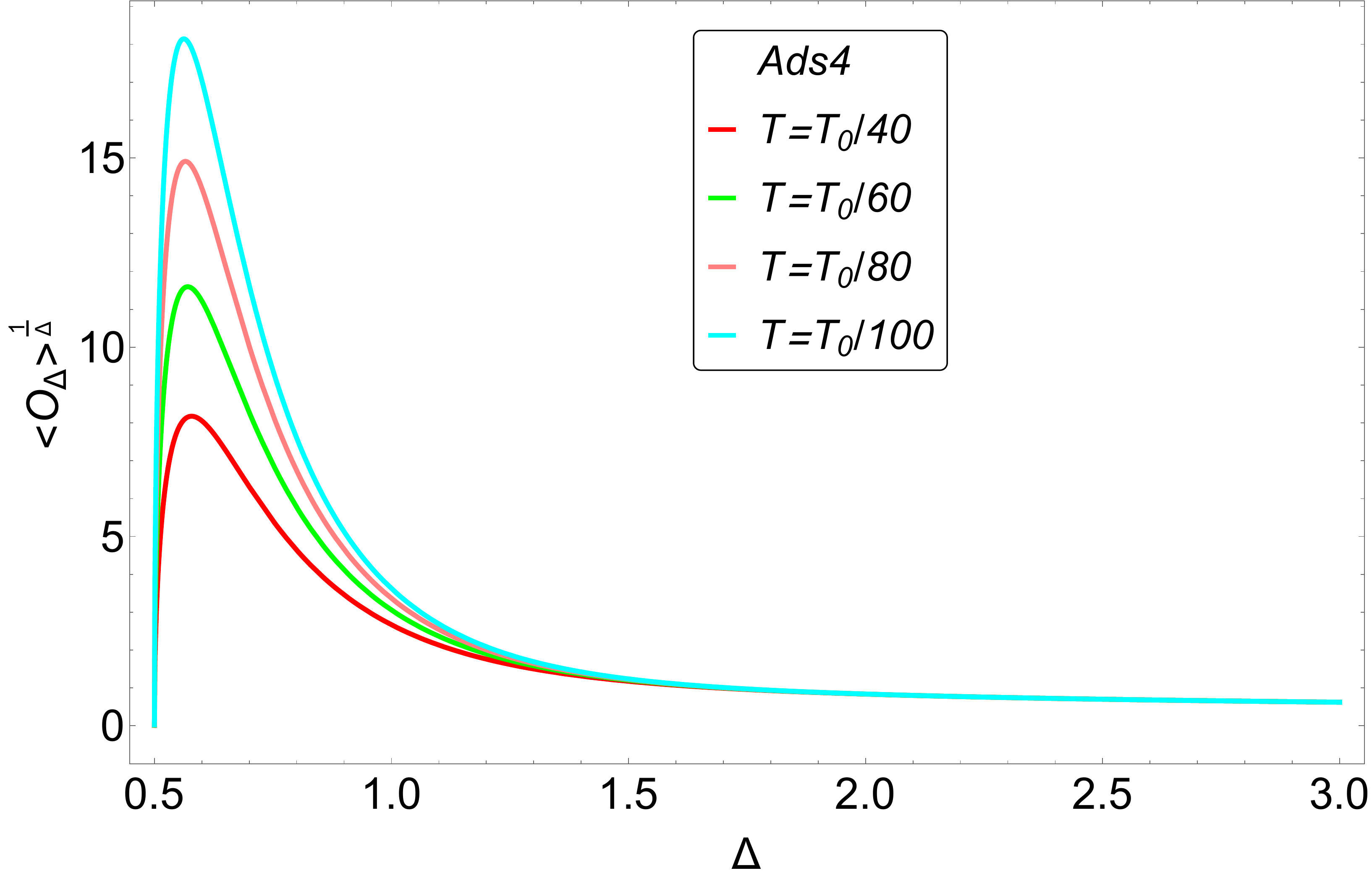}	}
	\subfigure[$T_c$  vs    $\Delta$.]
	{ \includegraphics[width=0.31\linewidth]{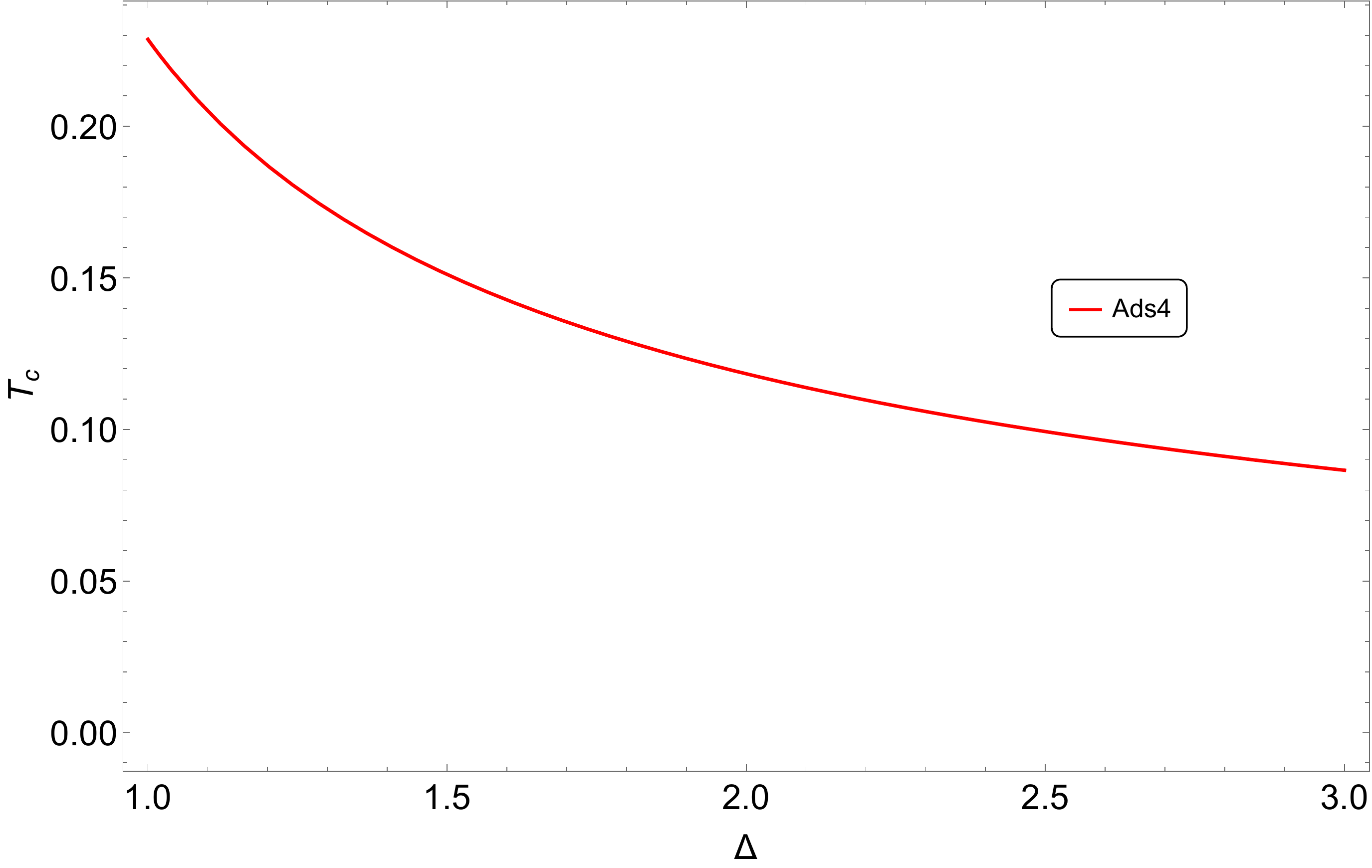}	}
	\subfigure[$\omega_g$  vs    $\Delta$.]
	{ \includegraphics[width=0.31\linewidth]{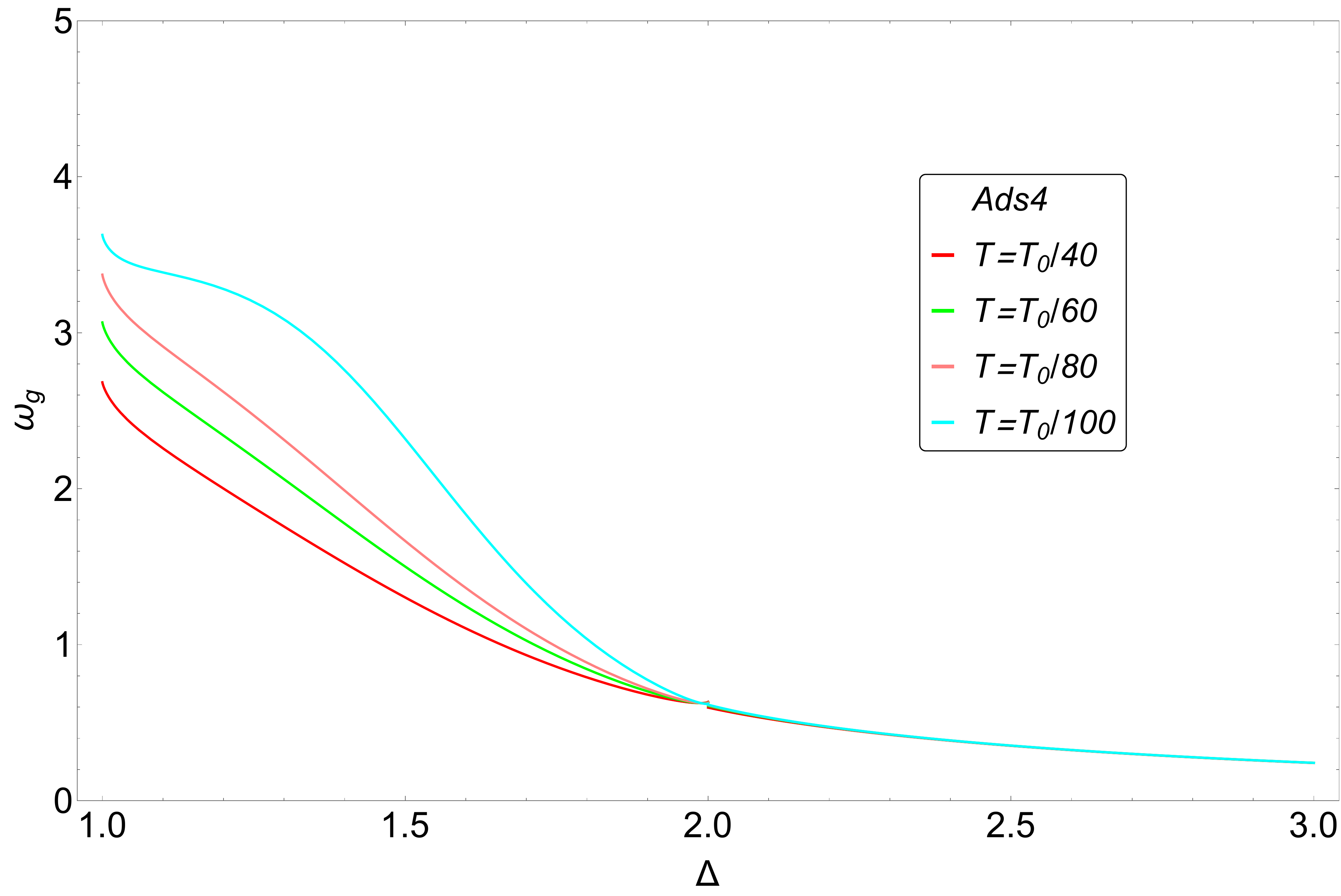}	}
	\subfigure[$\frac{\left< \mathcal{O}_{\Delta}\right>^{1/\Delta}}{T_c}$  vs    $\Delta$.]
	{ \includegraphics[width=0.31\linewidth]{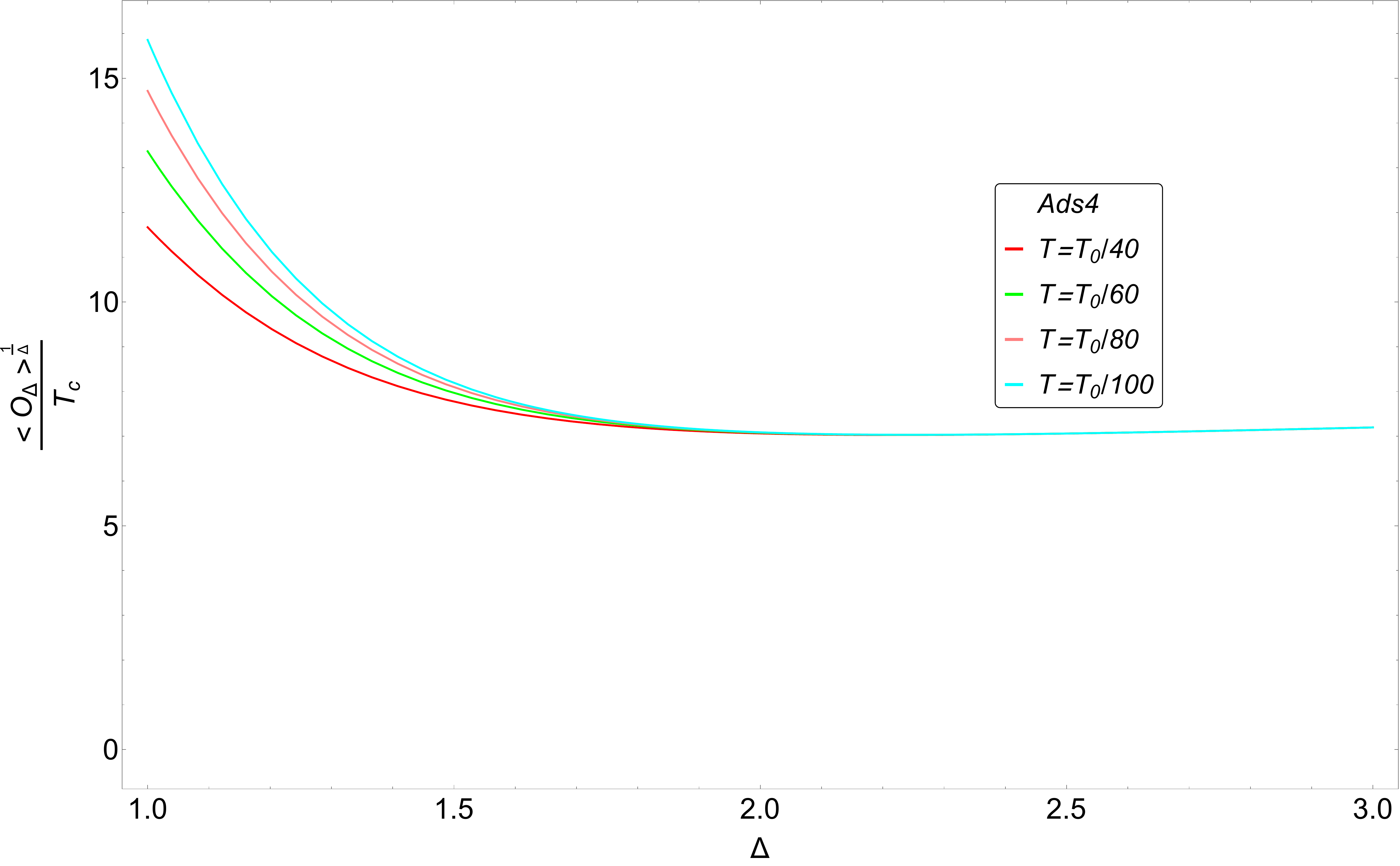}	}
	\subfigure[$\frac{\omega_g}{T_c}$  vs    $\Delta$.]
	{ \includegraphics[width=0.31\linewidth]{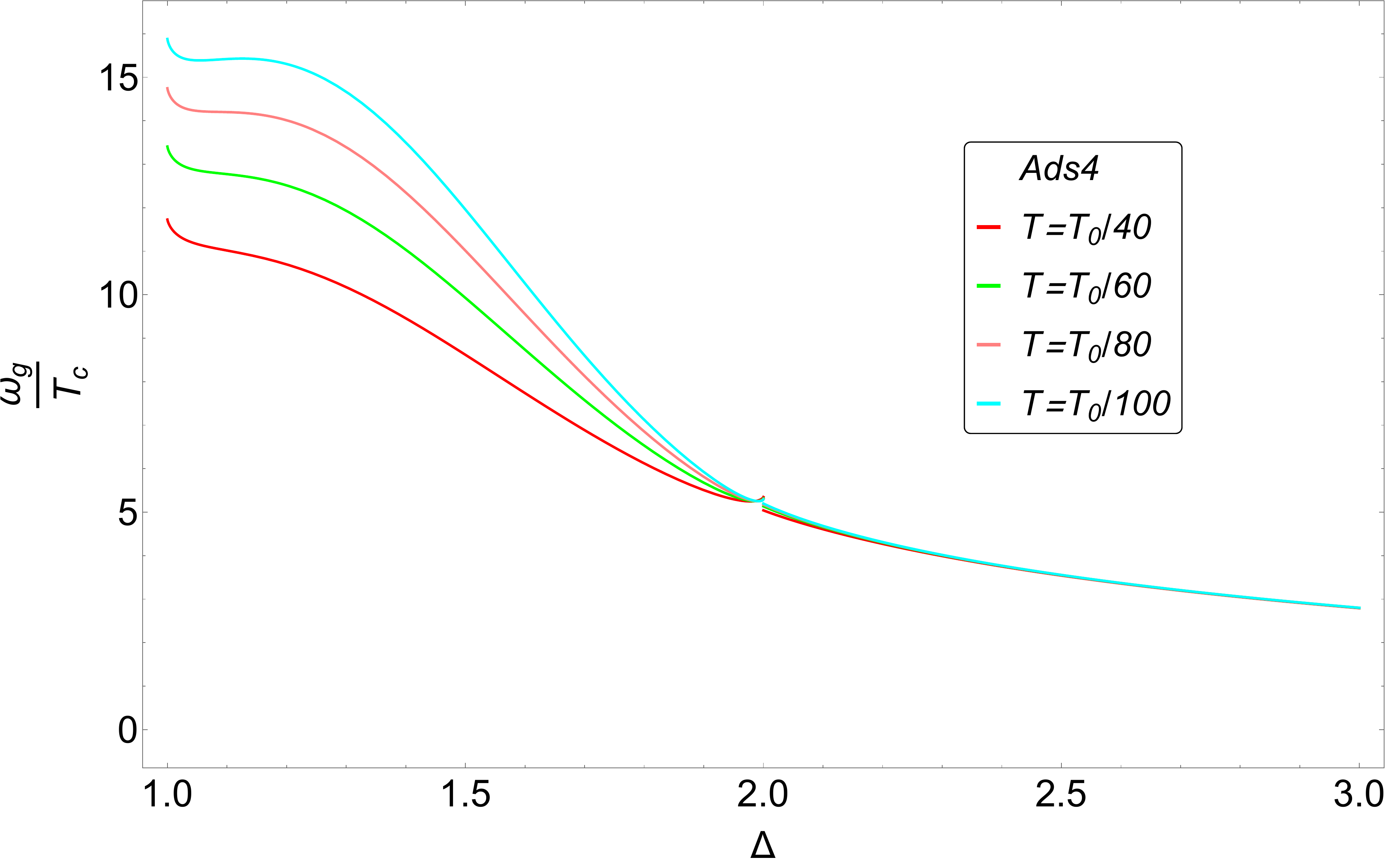}	}
	\subfigure[$\frac{\omega_g}{\left< \mathcal{O}_{\Delta}\right>^{1/\Delta}}$  vs    $\Delta$.]
	{ \includegraphics[width=0.31\linewidth]{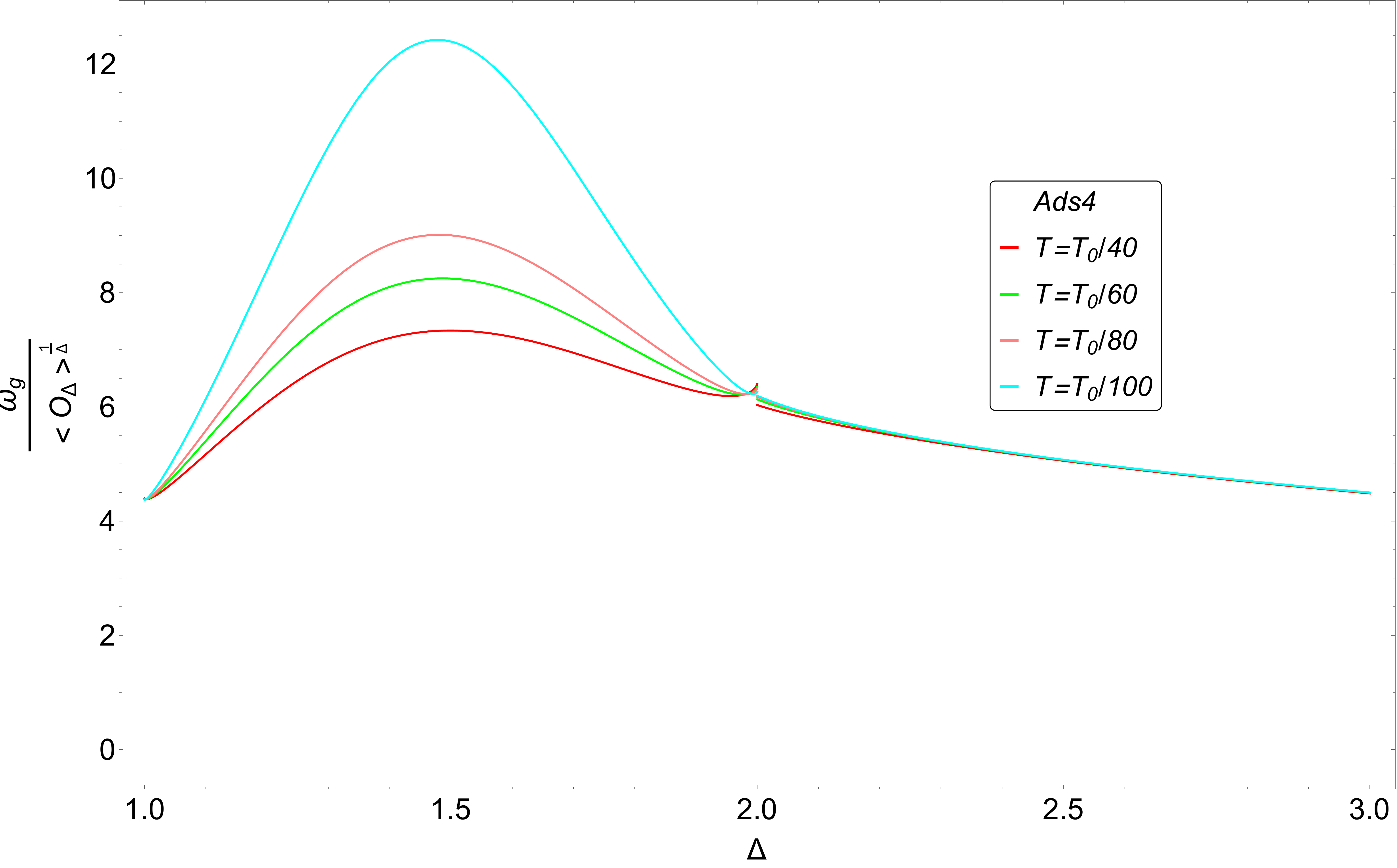}	}
	\caption{ (a-f)  Low temperature behavior of observables  we calculated in this paper.  Here, we set   $d=3$ and $g=\rho =1$.}
	\label{node4}
\end{figure}

It has been believed that  $\left< \mathcal{O}_{\Delta}\right>^{1/\Delta}$, $T_c$ and $\omega_g$ are the same quantity up to a numerical factor.  
This may the case if we look at them for  a given $\Delta$. However, as functions of $\Delta$,   they   are all different  ones, as we can see in 
figure \ref{node4}. The identification of these observables partially make sense in the relatively large $\Delta>2$ regime. 
It is also interesting to notice that $\frac{\omega_g}{\left< \mathcal{O}_{\Delta}\right>^{1/\Delta}}$  is the maximum at $\Delta=3/2$ as one can see in  figure \ref{node4}(f). 

\subsection{Condensate   near the zero temperature}
\label{appendix2B}
In general,  Eq.(\ref{eq:3}) shows us that $F(z)$ in Eq. (\ref{eq:11})  does not converge at $z=1$. But the previous section \ref{Tcp} says that it  is converged at the horizon with
specific value of $\left< \mathcal{O}_{\Delta}\right>$. Its means whether we can find eigenvalue of it at $z=1$ or not simply, satisfied for $F(1)<\infty$. Unlike $T\approx T_c$ case, it is really hard to find the eigenvalue $\left< \mathcal{O}_{\Delta}\right>$ at $T\approx 0$. Because
Eq.(\ref{eq:3})  are nonlinear coupled equations: $\Phi(z)$ cannot be described in a linear equation any longer unlike $T\approx
T_c$ case. Instead, we use the perturbation theory for the eigenvalue at $T\approx 0$.

We can simplify Eq.(\ref{eq:3})   in  $T\rightarrow 0$ limit by  defining   
\be 
\Psi(z)=  ({\left< \mathcal{O}_{\Delta}\right>}/{\sqrt{2}r_h^{\Delta}})z^{\Delta}F(z).\label{nomal}\ee
The equations of motion for F near the zero temperature becomes
\bea
&&	\frac{d^2 F}{d z^2} -\frac{2\Delta +1-d}{z }\frac{d F}{d z} +\frac{g^2 \Phi^2}{r_h^2}F=0,  \\ 
&&	\frac{d^2 \Phi}{d z^2}-  \frac{d-3}{z}\frac{d \Phi}{d z } -\frac{ g^2\left< \mathcal{O}_{\Delta}\right>^2 }{r_h^{2\Delta}}z^{2(\Delta-1)} F^2 \Phi =0 . \nonumber
\label{eq:sim2}
\eea
The boundary conditions (BC)  we should use are 
\be 
\frac{d \Phi(0)}{d z^{d-2}}=-\frac{\rho}{r_h^{d-2}}, \quad F(0)=1, 
\ee
and  
\be
\Phi(1)=0, \quad 3 \frac{d  F(1)}{d z}+\Delta^2 F(1)=0. 
\ee 
The latter is the horizon regularity conditions  at $z=1$, and from (A.39) one can derive $F'(0)=0$.  

We use $X$ to   denote $g^{\frac{1}{\Delta}}\frac{\left< \mathcal{O}_{\Delta}\right>^{\frac{1}{\Delta}}}{T_c}$ which appears often. Then,  $X$  satisfies 
\begin{equation}
	X^{2(d-1)}
	= G_d^{2(d-1)} \left(\alpha_{d} +\beta_d \tau_{d}^{d-2\Delta} X^{d-2\Delta} \right)
	\label{si:67}
\end{equation} 
\begin{eqnarray}
	\hbox{where }\quad 		G_d &=& \frac{4\pi \Delta^{1/\Delta}}{d}\left(\frac{-2^{1+\nu} \lambda_{g,d}}{\Gamma(-\nu)}\right)^{\frac{1}{d-1}}  \label{si:671a}\\  
	\alpha_{d} &=&-\frac{\sqrt{\pi}\Gamma\left(\frac{d-2\Delta}{2\Delta}\right)\Gamma\left(\frac{1}{\Delta}\right)\Gamma\left(\frac{d-1}{\Delta}\right)}{8\Delta^2 \Gamma\left(\frac{d+\Delta}{2\Delta}\right)} \label{si:671b}\\  
	\beta_d &=& \frac{\nu \pi  (d-\Delta)^2 \csc(\nu \pi)}{2\Delta^3 \left(d-2\Delta\right)}. 
	\label{si:671c}
\end{eqnarray}  	
with    $\nu=\frac{d-2}{2\Delta}$ and $\tau_{d}=\frac{d}{4\pi \Delta^{1/\Delta}}\frac{T_c}{T} $. 
For derivation of this result, see the appendix \ref{appendix2B}.
We can   get the solution of Eq.(\ref{si:67})    
according to the regimes of $\Delta$:  	\\
$		X = G_d^{\frac{2(d-1)}{d-2+2\Delta}} \beta_d^{\frac{1}{d-2+2\Delta}}\tau_{d}^{\frac{d-2\Delta}{d-2+2\Delta}} 
$ for  $(d-2)/2< \Delta \ll d/2$: \\
$X = G_d \;\alpha_d^{\frac{1}{2(d-1)}} $
for $d/2 \ll \Delta < d$. 
Especially,   for $ \Delta =\frac{d}{2}$
\begin{equation}
	X^{2(d-1)}
	= G_d^{2(d-1)} \left(\rho_d +\sigma_d \ln(\tau_d X) \right)
	\label{ii:10c}
\end{equation}
where $	\rho_d  =\frac{\sigma_d}{d}\left( 5-\frac{2}{d-2}-\pi  \cot \left(\frac{2 \pi }{d}\right)-2 \psi \left(\frac{2}{d}\right)-\log (4)\right)$ and  $\sigma_d =\frac{\pi  (d-2) \csc \left(\frac{2 \pi }{d}\right)}{d^2}  $. Here, $\psi\left(z\right)$ is  the digamma function. Details  are available in appendix \ref{big} and \ref{biga}.  
Numerical results tell us that $\rho_3 \approx 0.8$, $\rho_4 \approx 0.64 $ and $\sigma_3\approx 0.4 \approx \sigma_4$. 
Therefore, eq.(\ref{ii:10c})  becomes 
\begin{small}
	\begin{eqnarray} 
		X  &\approx&   6.76 \left( 1+0.45 \ln\left(\frac{T_c}{T}\right)\right)^{1/4} \;\;\;\mbox{for AdS$_4$}\nonumber\\
		X &\approx&  4.9 \left( 1+0.57 \ln\left(\frac{T_c}{T}\right)\right)^{1/6} \;\;\;\mbox{for AdS$_5$}. 
		\label{ii:11} 
	\end{eqnarray}
\end{small}
We can first  test our result with known results: 
For  $\Delta=1$  and $T=0.1 T_c$, our analytic    expression  with  
$g=1$ gives  
$ {\left< \mathcal{O}_{1}\right>}/{T_c}\approx   12.65 $,     which is comparable to  the numerical result 10.8  of ref. 
\cite{Hart2008}.  Our result, however,  is   different from that  of ref. \cite{Siop2010}   except at $\Delta=1$.

It is important to notice that  the temperature dependence of   the condensation $X=g^{\frac{1}{\Delta}}\frac{\left< \mathcal{O}_{\Delta}\right>^{\frac{1}{\Delta}}}{T_c}$  is very different depending on the regime of $\Delta$. It  diverges as  $T\to 0$  for   $ \Delta < \frac{d}{2}$, but it has   little dependence  on $T$ in 	$d/2 < \Delta <d$.   These results explains   the numerical features   of ref.  \cite{Hart2008}.   

Notice that  there are presence of  singularities  at $\Delta =d/2$ in both   $\alpha_d$ and $\beta_d$.	
Surprisingly, however, it turns out that there is no singularity in $X$. To understand this, notice that 
the   behaviors of $\alpha_d$   near $\Delta= d/2$ is  	
\begin{eqnarray} 
	\lim_{\Delta \rightarrow d/2}\alpha_{d} &=& \frac{(d-2)\pi \csc\left(\frac{2\pi}{d} \right)}{2 d^2} \frac{1}{\Delta-d/2} , \label{ii:1a}
\end{eqnarray} 
which   is exactly the same as  the behavior  of $-\beta_{d}$ near $\Delta=d/2$. 
Therefore, the singularity of $X$ of eq.(\ref{si:67})   disappears because 
at  $\Delta=d/2$, 
$X^{d-2\Delta}=1$ and   $\alpha_d+\beta_d$  is finite.   Such cancellation of two singularities  was rather  unexpected.  

\begin{figure}[!htb]
	\centering
	\subfigure[$\alpha_{3}$ vs $\Delta$]
	{ \includegraphics[width=0.3\linewidth]{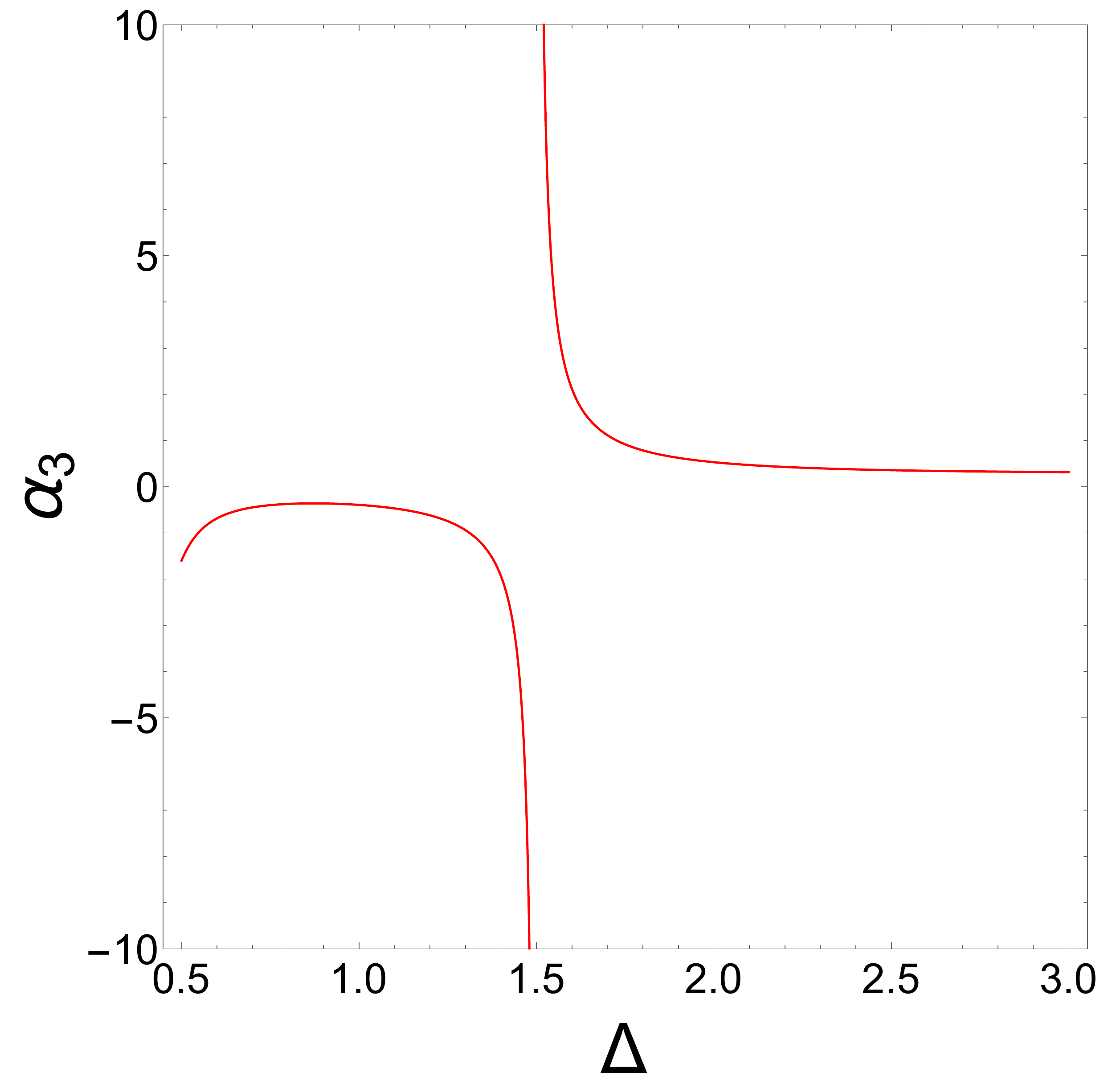}}
	\subfigure[$\beta_{3}$ vs $\Delta$]
	{ \includegraphics[width=0.3\linewidth]{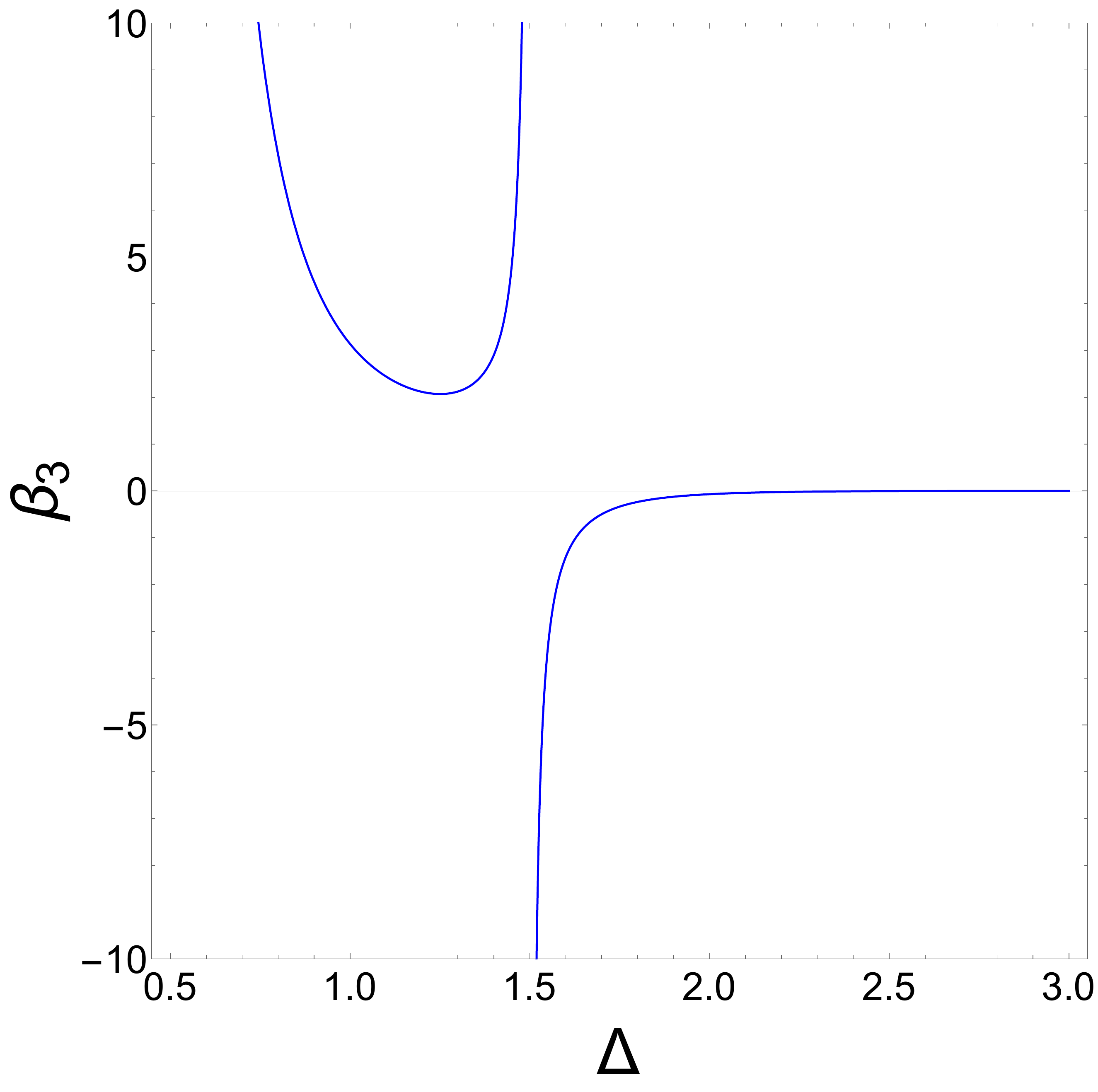}
	} 
	\subfigure[$G_{3}$ vs $\Delta$]
	{ \includegraphics[width=0.3\linewidth]{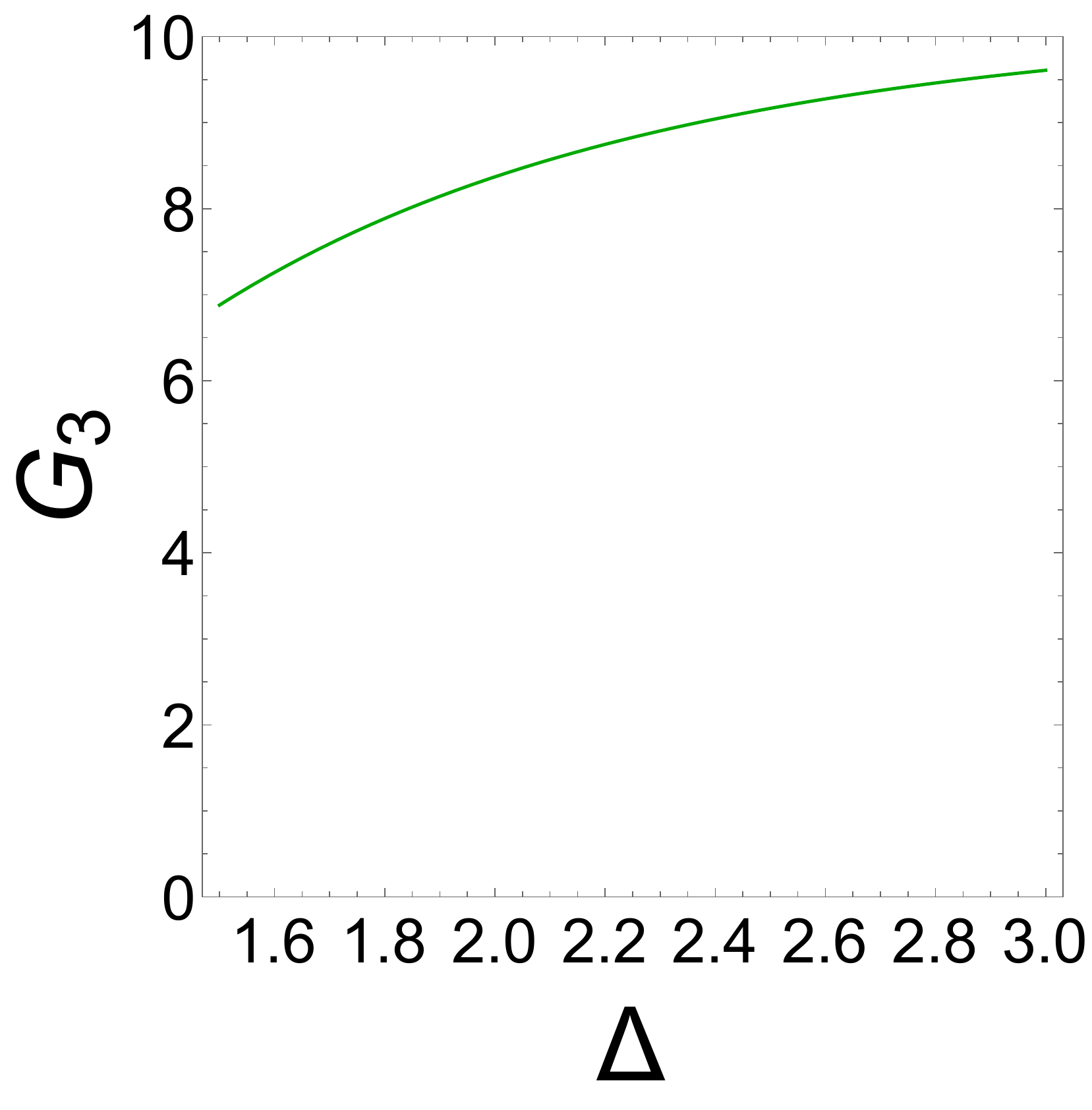}
	}   \caption{  Plots of  $\alpha_3$,  $\beta_3$ and  $G_3$ in eq.(\ref{si:671b}) over    $1/2<\Delta<3$.  }  \label{Gfunction}
\end{figure}

In ref. \cite{Horo2009}, it was numerically noticed that  $X$ is almost constant over the region $d/2 < \Delta < d$. To understand  this phenomena, we plot $\alpha_3$,  $\beta_3 $ and $G_{3}$ as function of $\Delta$    in the FIG. \ref{Gfunction}.   
In fact, 	one can show  that for $\Delta\gg \frac32$, 
\begin{eqnarray}   
	\alpha_{d} = \frac{\Delta }{4 (d-1) d}+\frac{1-3 \gamma_{E} -\psi\left( {1}/{2}\right)}{8 (d-1)} +\cdots , \label{jj:1a}
\end{eqnarray} 
so that for $d=3$, $\alpha_{3}=\frac{\Delta}{24}+0.077$. Notice that 
$\alpha_{d}$ is flat over the relevant regime  because the linear term grows with tiny slope. 
$\beta_{3}$, after vanishing at $\Delta=3$, saturate to 0 rapidly  like $ \sim -1/(4\Delta^{2})$.  
In addition,   $G_{3}$ moves  slowly in the FIG. \ref{Gfunction}.  All these collaborate  with the cancellation of the singularity at $\Delta=3/2$, to make the flatness of $X$ in $\Delta$ in the regime. 
Completely parallel reasoning  works  for $d=4$.  
It would be very   interesting to see if this is only for s-wave case or it continue to be so for $p$- and $d$-wave  case as well.  We will leave this as a future work.  
\begin{figure}[h]
	\centering
	\includegraphics[width=0.45\linewidth]{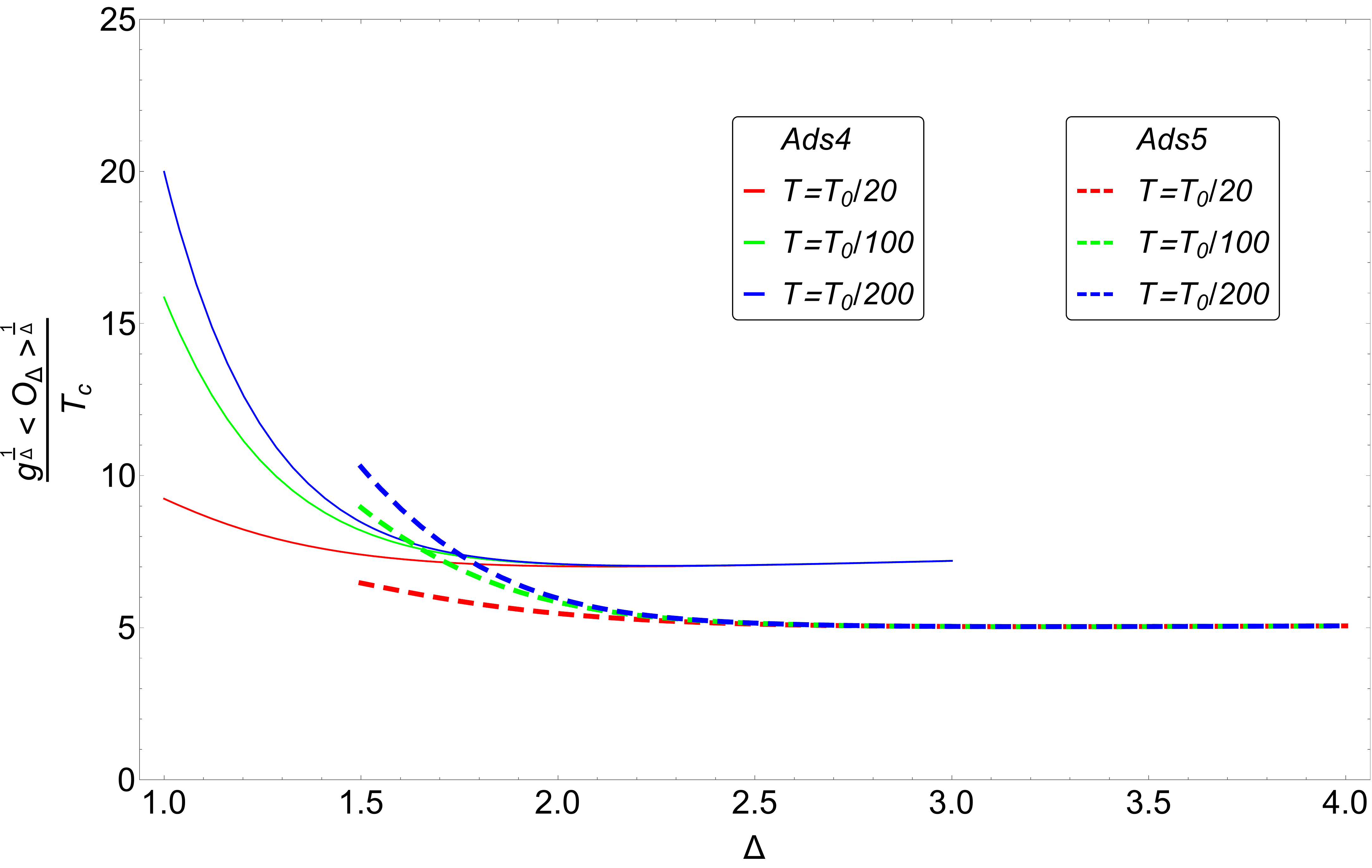}
	\caption{  $g^{\frac{1}{ \Delta}}\frac{\left< \mathcal{O}_{\Delta}\right>^{\frac{1}{\Delta}}}{T_c}$   vs $\Delta$ near $T=0$. Here, $T_{0}=(g \rho)^\frac{1}{d-1}$.  Notice the flatness  and the  $T$-independence over $\frac{d}2 < \Delta<d$.  }		\label{Odelta_zero_v1}
\end{figure} 
Fig.~\ref{Odelta_zero_v1}  is the plot of the results given in  Eq.(\ref{si:67}) for $d=3,4$.  
The solid lines are for  $d=3$,   and the dashed lines are for  $d=4$. 
$g^{\frac{1}{ \Delta}}\frac{\left< \mathcal{O}_{\Delta}\right>^{\frac{1}{\Delta}}}{T_c}$ is $\sim 7$ at $3/2 < \Delta <3$  for Ads$_4$, and  $\sim5$ at $2 < \Delta <4$  for AdS$_5$. These are in good agreement with  numerical results  of ref. \cite{Horo2009}.

A remark is in order to explain why 
analytic formulae  in $G_d$, $\alpha_{d}$ and $\beta_d$   were possible  in spite of the fact that the differential equations  in  the black hole background are  not of hypergeometric type, as we  can see from Eq.(\ref{eq:3}).     The simplification  happens   near $T=0$, where the higher order singularity  at  the horizon disappears as we  can see  from Eq.(\ref{eq:sim2}): there is only one regular singularity at $z=0$ and the order of the singularity is  independent of $d$ so that the differential equations reduces to  hypergeometric type. Details  are available in appendix \ref{small}, \ref{big}, \ref{smalla} and \ref{biga}.

We use $Y$ to   denote $g^{\frac{1}{\Delta}}\frac{\left< \mathcal{O}_{\Delta}\right>^{\frac{1}{\Delta}}}{T_0}$ which appears often. Here, $T_{0}=(g \rho)^\frac{1}{d-1}$. Then,  $Y$  satisfies 
\begin{equation}
	Y^{2(d-1)}
	= \widetilde{G_d}^{2(d-1)}   \left(\alpha_{d} +\beta_d \widetilde{\tau_{d}}^{d-2\Delta}  Y^{d-2\Delta} \right)
	\label{sss:67}
\end{equation} 
\begin{eqnarray}
	\hbox{where }\quad 	 \widetilde{G_d} &=& \Delta^{1/\Delta} \left(\frac{-2^{1+\nu}  }{\Gamma(-\nu)}\right)^{\frac{1}{d-1}}  \label{sss:671a}\\  
	\widetilde{\tau_{d}} &=&\frac{d}{4\pi \Delta^{1/\Delta}}\frac{T_0}{T} . \label{ssss:671b} 
\end{eqnarray}  
For derivation of this result, see the appendix \ref{small1} $\&$ \ref{smalla1}.
Fig.~\ref{Odelta_zero_v2}  is the plot of the results given in  Eq.(\ref{sss:67}) for $d=3,4$. 	 
We emphasize that although there is no $T_{c}$ in 
$\frac{d-2}{2}< \Delta < \frac{d-1}{2} $, there is    well defined condensation in this regime. 
\begin{figure}[h]
	\centering
	\includegraphics[width=0.5\linewidth]{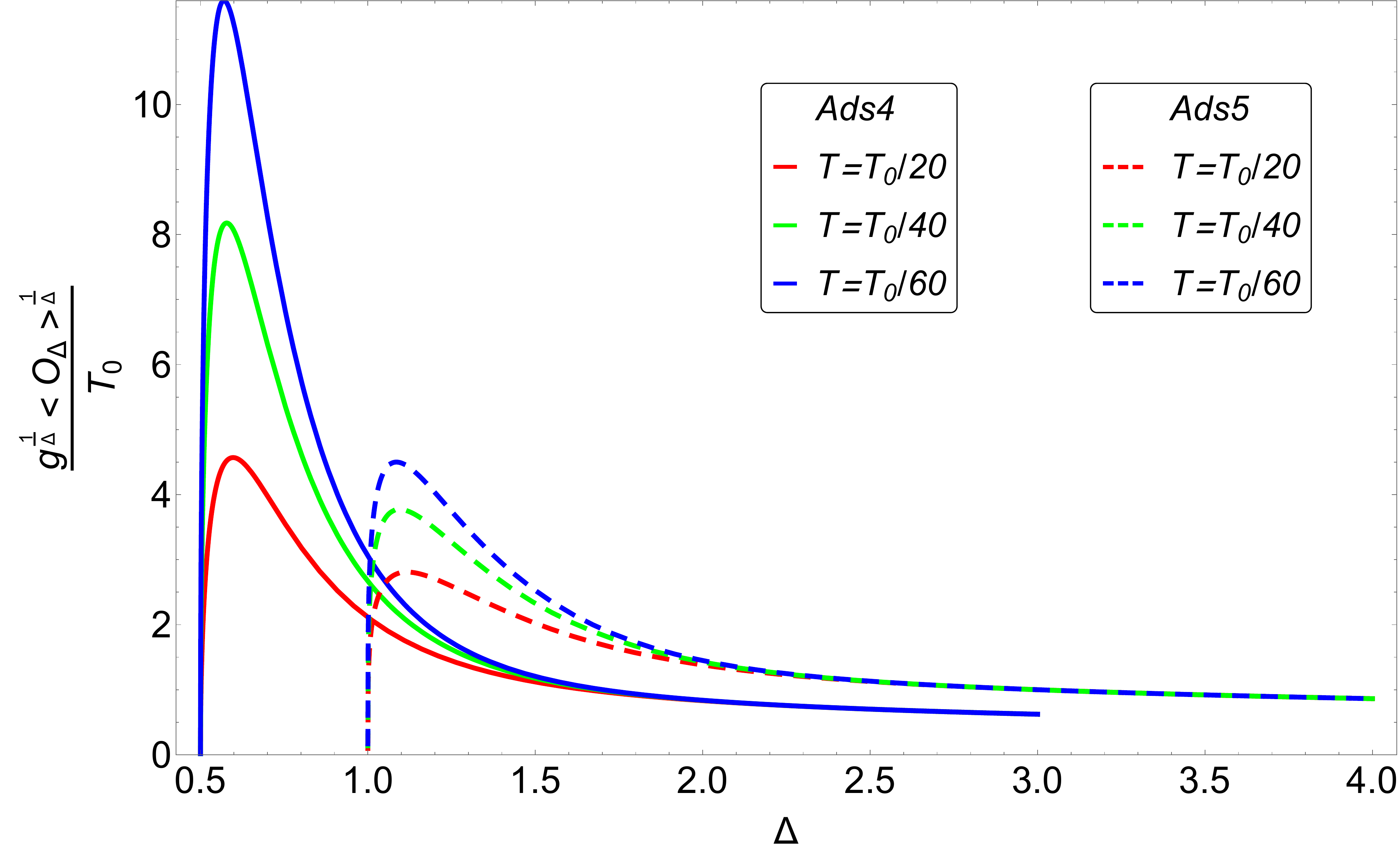}
	\caption{  $g^{\frac{1}{ \Delta}}\frac{\left< \mathcal{O}_{\Delta}\right>^{\frac{1}{\Delta}}}{T_0}$   vs $\Delta$ near $T=0$.  Notice again the flatness  and the  $T$-independence over $\frac{d}2 < \Delta<d$.  }		\label{Odelta_zero_v2}
\end{figure}

\subsubsection{Analytic calculation of $g^{\frac{1}{\Delta}}\frac{\left< \mathcal{O}_{\Delta}\right>^{\frac{1}{\Delta}}}{T_c}$  at $1\leq \Delta<3$}\label{small}
The Hawking temperature shows $r_h\rightarrow0$ as $T\rightarrow0$. We can say $z=r_h/r\rightarrow0$ at $r \gg r_h$ and the dominant
contribution comes from the neighborhood of the boundary $z=0$. So  near the $T=0$ we can simplify two coupled equations Eq.(\ref{eq:3}) and
Eq.(\ref{eq:3}) with Eq.(\ref{eq:11}) by letting $z\rightarrow0$:
\begin{subequations}
	\begin{equation}
		\frac{d^2 F }{d z^2} +  \frac{2(\Delta-1)}{z} \frac{d F}{d z} + \frac{g^2 \Phi^2}{r_h^2 }F =0
		\label{eq:48a}
	\end{equation}
	\begin{equation}
		\frac{d^2 \Phi}{d z^2} -\frac{ g^2\left< \mathcal{O}_{\Delta}\right>^2}{r_h^{2\Delta}}z^{2(\Delta-1)}F^2 \Phi =0.
		\label{eq:48b}
	\end{equation}
\end{subequations}
We use a boundary condition at the horizon, and
Eq.(\ref{eq:3}) with Eq.(\ref{eq:11}) is rewritten as
\begin{equation}
	-\frac{d^2 F}{d z^2} +\left( \frac{2+z^3}{z(1-z^3)}-\frac{2\Delta}{z}\right)\frac{d F}{d z}+\left(\frac{\Delta^2 z}{1-z^3}- \frac{g^2
		\Phi^2}{r_h^2(1-z^3)^2}\right)F=0
	\label{eq:49}
\end{equation}
and it provides us the following boundary condition at the horizon with Eq.(\ref{eq:3}),  
$\Phi(1)=0 $ 
{and}   $\Psi(1)<\infty$:
\begin{equation}
	3F^{'}(1)+\Delta^2 F(1)=0.
	\label{eq:50}
\end{equation}
By multiplying  $z$ to the eq. (\ref{eq:49}) and then taking the limit of $z\to 0$, we get $ F^ {'}(0)=0$.    
Note that $F(0)=1$ should be considered  as the normalization condition of $\vev{{\cal O}_{\Delta}}$ rather than as a boundary condition.  
Also for canonical system,  we regard the 
$\Phi'(0) =-\frac{\rho}{r_h} $ as BC 
and $\Phi(0)=\mu$ is not a BC but a value that  should be determined by $\rho$  from the horizon regularity condition  $\Phi(1)=0$. 
In Grand canonical system $\Phi(0)=\mu$ is the boundary condition and $\rho$ should be determined from it by the  $\Phi(1)=0$.  	Here we consider $\rho$ as the given parameter.

If we introduce $b$ by $\hbox{  for  } b^{\Delta}= \frac{ g\left< \mathcal{O}_{\Delta}\right>}{\Delta r_h^{\Delta}}
$, 	 the solution to Eq.(\ref{eq:48b}) for $\Phi$ with $F\approx 1$ is
\begin{eqnarray}
	\Phi(z)
	&=&\mathcal{A} r_h \sqrt{b z} K_{\frac{1}{2 \Delta }}\left(b^{\Delta } z^{\Delta }\right) \quad
	\label{eq:53b}
\end{eqnarray}
%
At the horizon $\Phi(1)\varpropto \exp(-b^{\Delta})\rightarrow 0$ because  $b\rightarrow\infty$ as $r_h \rightarrow 0$ ($T \rightarrow 0$), which takes care  the boundary condition
$\Phi(1)=0 $.
Substituting Eq.(\ref{eq:53b}) into Eq.(\ref{eq:48a}), $F$ becomes
\begin{equation}
	\frac{d^2 F }{d z^2} +  \frac{2(\Delta-1)}{z} \frac{d F}{d z} +g^2 b \mathcal{A}^2 z \left( K_{\frac{1}{2 \Delta }}\left(b^{\Delta } z^{\Delta
	}\right)\right)^2 F=0.
	\label{eq:54}
\end{equation}
$F(z)$ can be obtained iteratively starting from $F=1$. 
The result is 
\begin{subequations}    
	\begin{equation}
		F(z)= 1- g^2b\mathcal{A}^2 \int_{0}^{z} d\acute{z} \; \acute{z}^{2(1-\Delta)} \int_{0}^{\acute{z}} d\fH{z} \; \fH{z}^{2\Delta-1} \left( K_{\frac{1}{2
				\Delta }}\left(b^{\Delta } \fH{z}^{\Delta }\right)\right)^2
		\label{eq:55a}
	\end{equation}
	\begin{equation}
		F^{'}(z)=- g^2b\mathcal{A}^2 z^{2(1-\Delta)} \int_{0}^{z} d\fH{z}\;  \fH{z}^{2\Delta-1} \left( K_{\frac{1}{2 \Delta }}\left(b^{\Delta } \fH{z}^{\Delta
		}\right)\right)^2
		\label{eq:55b}
	\end{equation}
\end{subequations}
with the boundary condition  $ F^{'}(0)=0$ and normalized $F(0)=1$. Applying the boundary condition  Eq.(\ref{eq:50}) into Eq.(\ref{eq:55a}) and Eq.(\ref{eq:55b}),  we obtain
\begin{equation}
	g^2\mathcal{A}^2=\frac{\Delta^2 b^2}{3 F_{\Delta}^{'}(b)+ \Delta^2 F_{\Delta}(b)}
	\label{eq:56}
\end{equation}
where
\begin{subequations}
	\begin{equation}
		F_{\Delta}(b)=  \int_{0}^{b} dz \; z^{2-2\Delta} \int_{0}^{z} d\tilde{z} \; \tilde{z}^{2\Delta-1} \left( K_{\frac{1}{2 \Delta }}\left(
		\tilde{z}^{\Delta }\right)\right)^2
		\label{eq:57a}
	\end{equation}
	\begin{equation}
		F_{\Delta}^{'}(b)=  b^{3-2\Delta} \int_{0}^{b} dz\;  z^{2\Delta-1} \left( K_{\frac{1}{2 \Delta }}\left( z^{\Delta }\right)\right)^2.
		\label{eq:57b}
	\end{equation}
\end{subequations}
With $x=z^{\Delta}$, Eq.(\ref{eq:57b}) is simplified as
\begin{equation}
	F_{\Delta}^{'}(b)= \frac{ b^{3-2\Delta} }{\Delta} \int_{0}^{b^{\Delta}} dx\;  x \left( K_{\frac{1}{2 \Delta }}\left( x\right)\right)^2
	\approx  \frac{ b^{3-2\Delta} }{\Delta} \int_{0}^{\infty} dx\;  x \left( K_{\frac{1}{2 \Delta }}\left( x\right)\right)^2. 
	\label{eq:58}
\end{equation}
There is the integral formula \cite{GRAD1980}:
\begin{footnotesize}
	\begin{equation}
		\int_{0}^{\infty} dx\;  x^{-\lambda}  K_{\mu}\left( x\right) K_{\nu}\left( x\right)=\frac{2^{-2-\lambda}}{\Gamma(1-\lambda)} \Gamma\left(
		\frac{1-\lambda+\mu+\nu}{2}\right)\Gamma\left( \frac{1-\lambda-\mu+\nu}{2}\right)\Gamma\left( \frac{1-\lambda+\mu-\nu}{2}\right)\Gamma\left(
		\frac{1-\lambda-\mu-\nu}{2}\right)
		\label{eq:59}
	\end{equation}
\end{footnotesize}
where $Re \;\lambda < 1-|Re \;\mu|-|Re \;\nu|$. Using Eq.(\ref{eq:59}), Eq.(\ref{eq:58}) becomes
\begin{equation}
	F_{\Delta}^{'}(b)= \frac{\pi }{4 \Delta^2} \csc{\left( \frac{\pi}{2\Delta}\right)} b^{3-2\Delta} 
	\label{eq:60}
\end{equation}
Letting $x=\tilde{z}^{\Delta}$, Eq.(\ref{eq:57a}) is simplified as
\begin{equation}
	F_{\Delta}(b)= \frac{1}{\Delta} \int_{0}^{b} dz \; z^{2-2\Delta} \int_{0}^{z^{\Delta}} dx \; x \left( K_{\frac{1}{2 \Delta }}\left(  x\right)\right)^2
	\label{eq:61}
\end{equation}
We have the following integral formula:
\begin{equation}
	\int dx\;  x \left( K_{\nu}\left( x\right)\right)^2  = \frac{x^2}{2}\left\{ \left( K_{\nu}\left( x\right)\right)^2 - K_{\nu -1}\left( x\right) K_{\nu
		+1}\left( x\right) \right\}.
	\label{eq:62}
\end{equation}
And
\begin{equation}
	\lim_{x\rightarrow 0} K_{\nu}\left( x\right)=\frac{\Gamma(\nu)}{2}\left(\frac{x}{2}\right)^{-\nu}+\frac{\Gamma(-\nu)}{2}\left(\frac{x}{2}\right)^{\nu}
	\label{eq:63}
\end{equation}
As we apply Eq.(\ref{eq:59}), Eq.(\ref{eq:62}) and Eq.(\ref{eq:63}) into Eq.(\ref{eq:61}), we obtain
\begin{eqnarray}
	F_{\Delta}(b) &=&  \frac{\pi b^{3-2\Delta}}{4 \Delta^2 (3-2\Delta)} \csc{\left( \frac{\pi}{2\Delta}\right)} - \frac{\pi \epsilon^{3-2\Delta}}{4 \Delta^2 (3-2\Delta)} \csc{\left( \frac{\pi}{2\Delta}\right)}\nonumber\\ 
	&+&  \frac{1}{2\Delta^2}\lim_{\epsilon\rightarrow 0}\int_{\epsilon^{\Delta}}^{b^{\Delta}}dx x^{\frac{3-\Delta}{\Delta}}  K_{\frac{1}{2\Delta}}\left( x\right)^2
	-\frac{1}{2\Delta^2}\lim_{\epsilon\rightarrow 0}\int_{\epsilon^{\Delta}}^{b^{\Delta}}dx x^{\frac{3-\Delta}{\Delta}}  K_{\frac{1}{2\Delta}-1}\left( x\right) K_{\frac{1}{2\Delta}+1}\left( x\right)\nonumber\\  
	&\approx&  \frac{\pi b^{3-2\Delta}}{4 \Delta^2 (3-2\Delta)} \csc{\left( \frac{\pi}{2\Delta}\right)} - \frac{\pi \epsilon^{3-2\Delta}}{4 \Delta^2 (3-2\Delta)} \csc{\left( \frac{\pi}{2\Delta}\right)}\nonumber\\ 
	&+&  \frac{1}{2\Delta^2} \int_{0}^{\infty}dx x^{\frac{3-\Delta}{\Delta}}  K_{\frac{1}{2\Delta}}\left( x\right)^2
	-\frac{1}{2\Delta^2}\lim_{\epsilon\rightarrow 0}\int_{\epsilon^{\Delta}}^{b^{\Delta}}dx x^{\frac{3-\Delta}{\Delta}}  K_{\frac{1}{2\Delta}-1}\left( x\right) K_{\frac{1}{2\Delta}+1}\left( x\right)\nonumber\\  
	&=&  \frac{\pi b^{3-2\Delta}}{4 \Delta^2 (3-2\Delta)} \csc{\left( \frac{\pi}{2\Delta}\right)} - \frac{\pi \epsilon^{3-2\Delta}}{4 \Delta^2 (3-2\Delta)} \csc{\left( \frac{\pi}{2\Delta}\right)}\nonumber\\ 
	&+&   \frac{\sqrt{\pi}\Gamma\left(\frac{1}{\Delta} \right)\Gamma\left(\frac{3}{2\Delta} \right)\Gamma\left( \frac{2}{\Delta}\right)}{8\Delta^2\Gamma\left( \frac{3+2\Delta}{2\Delta}\right)} 
	-\frac{1}{2\Delta^2}\lim_{\epsilon\rightarrow 0}\int_{\epsilon^{\Delta}}^{b^{\Delta}}dx x^{\frac{3-\Delta}{\Delta}}  K_{\frac{1}{2\Delta}-1}\left( x\right) K_{\frac{1}{2\Delta}+1}\left( x\right)
	\label{eq:64}
\end{eqnarray}
here, we introduce small $\epsilon$, and take zero at the end of calculations.  

There are two different formulas:
\begin{footnotesize}
	\begin{eqnarray}
		&&\int dx\; x^{\lambda} K_{\nu}\left( x\right)K_{ \mu}\left( x\right) = \pi ^2 2^{-\mu -\nu -3} \csc (\pi  \mu ) \csc (\pi  \nu ) x^{\lambda -\mu -\nu +1} \left\{4^{\mu +\nu } \Gamma (-\mu -\nu +1) \Gamma \left(\frac{1}{2} (\lambda -\mu -\nu +1)\right)\right.\nonumber\\
		&&\times		\pFq[4]{3}{4}{\frac{1}{2} (-\mu -\nu +1),\frac{1}{2} (-\mu -\nu +2),\frac{1}{2} (\lambda -\mu -\nu +1)}{1-\mu ,1-\nu ,-\mu -\nu +1,\frac{1}{2} (\lambda -\mu -\nu +3)}{x^2}  \nonumber\\
		&&- 4^{\nu } x^{2 \mu } \Gamma (\mu -\nu +1) \Gamma \left(\frac{1}{2} (\lambda +\mu -\nu +1)\right)
		\pFq[4]{3}{4}{\frac{1}{2} (\mu -\nu +1),\frac{1}{2} (\mu -\nu +2),\frac{1}{2} (\lambda +\mu -\nu +1) }{\mu +1,1-\nu ,\mu -\nu +1,\frac{1}{2} (\lambda +\mu -\nu +3) }{x^2} \nonumber\\
		&&-4^{\mu } x^{2 \nu } \Gamma (-\mu +\nu +1) \Gamma \left(\frac{1}{2} (\lambda -\mu +\nu +1)\right)   
		\pFq[4]{3}{4}{ \frac{1}{2} (-\mu +\nu +1),\frac{1}{2} (-\mu +\nu +2),\frac{1}{2} (\lambda -\mu +\nu +1)}{1-\mu ,\frac{1}{2} (\lambda -\mu +\nu +3),\nu +1,-\mu +\nu +1 }{x^2}   \nonumber\\
		&&+\left.  \Gamma (\mu +\nu +1) x^{2 \mu +2 \nu } \Gamma \left(\frac{1}{2} (\lambda +\mu +\nu +1)\right)  \pFq[4]{3}{4}{\frac{1}{2} (\mu +\nu +1),\frac{1}{2} (\mu +\nu +2),\frac{1}{2} (\lambda +\mu +\nu +1) }{\mu +1,\frac{1}{2} (\lambda +\mu +\nu +3),\nu +1,\mu +\nu +1 }{x^2}  \right\} \hspace{1cm}\label{eq:75}
	\end{eqnarray}
\end{footnotesize}
\begin{footnotesize}
	\begin{equation}
		\pFq[4]{p}{q}{a,a_2,\cdots,a_p}{a-1,b_2,\cdots,b_q}{z}= \pFq[4]{p-1}{q-1}{a_2,\cdots,a_p}{b_2,\cdots,b_q}{z} + \frac{z a_2\cdots a_p}{(a-1)b_2 \cdots b_q}
		\pFq[4]{p-1}{q-1}{a_2+1,\cdots,a_p+1}{b_2+1,\cdots,b_q+1}{z} \label{eq:77}
	\end{equation}
\end{footnotesize}
And the asymptotic formula for the $_2F_3$ hypergeometric function as $|z|\rightarrow\infty$  is written by\cite{Olve2010}:
\begin{footnotesize}
	\begin{eqnarray}
		\pFq[4]{2}{3}{a_1, a_2}{b_1, b_2, b_3}{z} &=& \frac{\Gamma(b_1)\Gamma(b_2)\Gamma(b_3)}{2\sqrt{\pi}\Gamma(a_1)\Gamma(a_2)} (-z)^{\chi} \left(
		\exp(-i(\pi \chi +2\sqrt{-z}))+\exp(i(\pi \chi +2\sqrt{-z}))+ \mathcal{O}\left(\frac{1}{\sqrt{-z}} \right)\right) \nonumber\\
		&&+ \frac{\Gamma(b_1)\Gamma(b_2)\Gamma(b_3)\Gamma(a_2-a_1)}{\Gamma(b_1-a_1)\Gamma(b_2-a_1)\Gamma(b_3-a_1)\Gamma(a_2)}(-z)^{-a_1}\left(1+
		\mathcal{O}\left(\frac{1}{z}  \right)   \right) \nonumber\\
		&&+ \frac{\Gamma(b_1)\Gamma(b_2)\Gamma(b_3)\Gamma(a_1-a_2)}{\Gamma(b_1-a_2)\Gamma(b_2-a_2)\Gamma(b_3-a_2)\Gamma(a_1)}(-z)^{-a_2}\left(1+
		\mathcal{O}\left(\frac{1}{z}  \right) \right)
		\label{eq:79}
	\end{eqnarray}
\end{footnotesize}
at $\chi=\frac{1}{2}\left(a_1+a_2-b_1-b_2-b_3+\frac{1}{2}\right)$ and wherein the case of simple poles (i.e. $a_1- a_2 \not\in \mathbb{Z} $).

After some long but simple ccalculations using the properties Eq.(\ref{eq:75}), Eq.(\ref{eq:77}) and Eq.(\ref{eq:79}), an integral in Eq.(\ref{eq:64}) is shows
\begin{equation} 
	\int_{\epsilon}^{b^{\Delta}}dx x^{\frac{3-\Delta}{\Delta}}  K_{\frac{1}{2\Delta}-1}\left( x\right) K_{\frac{1}{2\Delta}+1}\left( x\right) = -\frac{3 \sqrt{\pi } \Gamma \left(1+\frac{1}{\Delta }\right) \Gamma \left(\frac{3}{2 \Delta }-1\right) \Gamma \left(\frac{2}{\Delta }\right)}{8 \Gamma \left(\frac{\Delta +3}{2 \Delta }\right)} 	+\frac{\pi \epsilon^{3-2\Delta}}{2  (3-2\Delta)} \csc{\left( \frac{\pi}{2\Delta}\right)}	
	\label{ii:77}
\end{equation} 
with $b\rightarrow \infty$. Substitute Eq.(\ref{ii:77}) into Eq.(\ref{eq:64}), and we have
\begin{equation}
	F_{\Delta}(b) =\frac{\pi b^{3-2\Delta}}{4 \Delta^2 (3-2\Delta)} \csc{\left( \frac{\pi}{2\Delta}\right)} -\frac{\sqrt{\pi } \Gamma \left(\frac{3}{2 \Delta }-1\right) \Gamma \left(\frac{1}{\Delta }\right) \Gamma \left(\frac{2}{\Delta }\right)}{8 \Delta ^2 \Gamma \left(\frac{\Delta +3}{2 \Delta }\right)} \label{ii:78}
\end{equation} 
Putting  Eq.(\ref{eq:60}) and Eq.(\ref{ii:78}) into Eq.(\ref{eq:56}), we have
\begin{equation}
	g^2\mathcal{A}^2=\frac{b^2}{-\frac{\sqrt{\pi } \Gamma \left(\frac{3}{2 \Delta }-1\right) \Gamma \left(\frac{1}{\Delta }\right) \Gamma \left(\frac{2}{\Delta }\right)}{8 \Delta ^2 \Gamma \left(\frac{\Delta +3}{2 \Delta }\right)}+ \frac{\pi  (3-\Delta )^2 \csc \left(\frac{\pi }{2 \Delta }\right)}{4 \Delta ^4 (3-2 \Delta )}b^{3-2\Delta}}
	\label{eq:65}
\end{equation}      
Apply  Eq.(\ref{eq:63}) into Eq.(\ref{eq:53b}) using Eq.(\ref{eq:4}), we deduce
\begin{equation}
	\frac{\rho}{r_h^2}=-\frac{\Gamma \left(\frac{-1}{2\Delta }\right)}{2^{1+\frac{1}{2\Delta}}}\mathcal{A} b
	\label{eq:66}
\end{equation}
As we combine $
T_c =\frac{3}{4\pi}r_c=\frac{3}{4\pi}\sqrt{\frac{\rho}{\lambda_3}}
$, Eq.(\ref{eq:36}), Eq.(\ref{eq:65}) and Eq.(\ref{eq:66}) with $ b= \left(\frac{ g\left< \mathcal{O}_{\Delta}\right>}{\Delta r_h^{\Delta}}\right)^{\frac{1}{\Delta}}$   in the form of $X$; here, $X:=$	$\frac{g^{1/\Delta}\left< \mathcal{O}_{\Delta}\right>^{1/\Delta}}{T_c}$ for simple notation, we obtain  the condensate at $T\approx 0$:
\begin{equation}
	X^{4}
	= G_3^{4} \left(\alpha_{3} +\beta_3 \tau_{3}^{3-2\Delta} X^{3-2\Delta} \right)
	\label{si:67}
\end{equation} 
\begin{eqnarray}
	\hbox{where }\quad 		G_3 &=& \frac{4\pi \Delta^{1/\Delta}}{3}\left(\frac{-2^{1+\nu} \lambda_{g,3}}{\Gamma(-\nu)}\right)^{\frac{1}{2}}  \nonumber\\  
	\alpha_{3} &=&-\frac{\sqrt{\pi}\Gamma\left(\frac{3-2\Delta}{2\Delta}\right)\Gamma\left(\frac{1}{\Delta}\right)\Gamma\left(\frac{2}{\Delta}\right)}{8\Delta^2 \Gamma\left(\frac{3+\Delta}{2\Delta}\right)} \label{si:671b}\\  
	\beta_3 &=& \frac{\nu \pi  (3-\Delta)^2 \csc(\nu \pi)}{2\Delta^3 \left(3-2\Delta\right)}. 
	\nonumber
\end{eqnarray}  	
with    $\nu=\frac{1}{2\Delta}$ and $\tau_{3}=\frac{3}{4\pi \Delta^{1/\Delta}}\frac{T_c}{T} $.  

\begin{figure}[h]
	\centering
	\includegraphics[scale=.6]{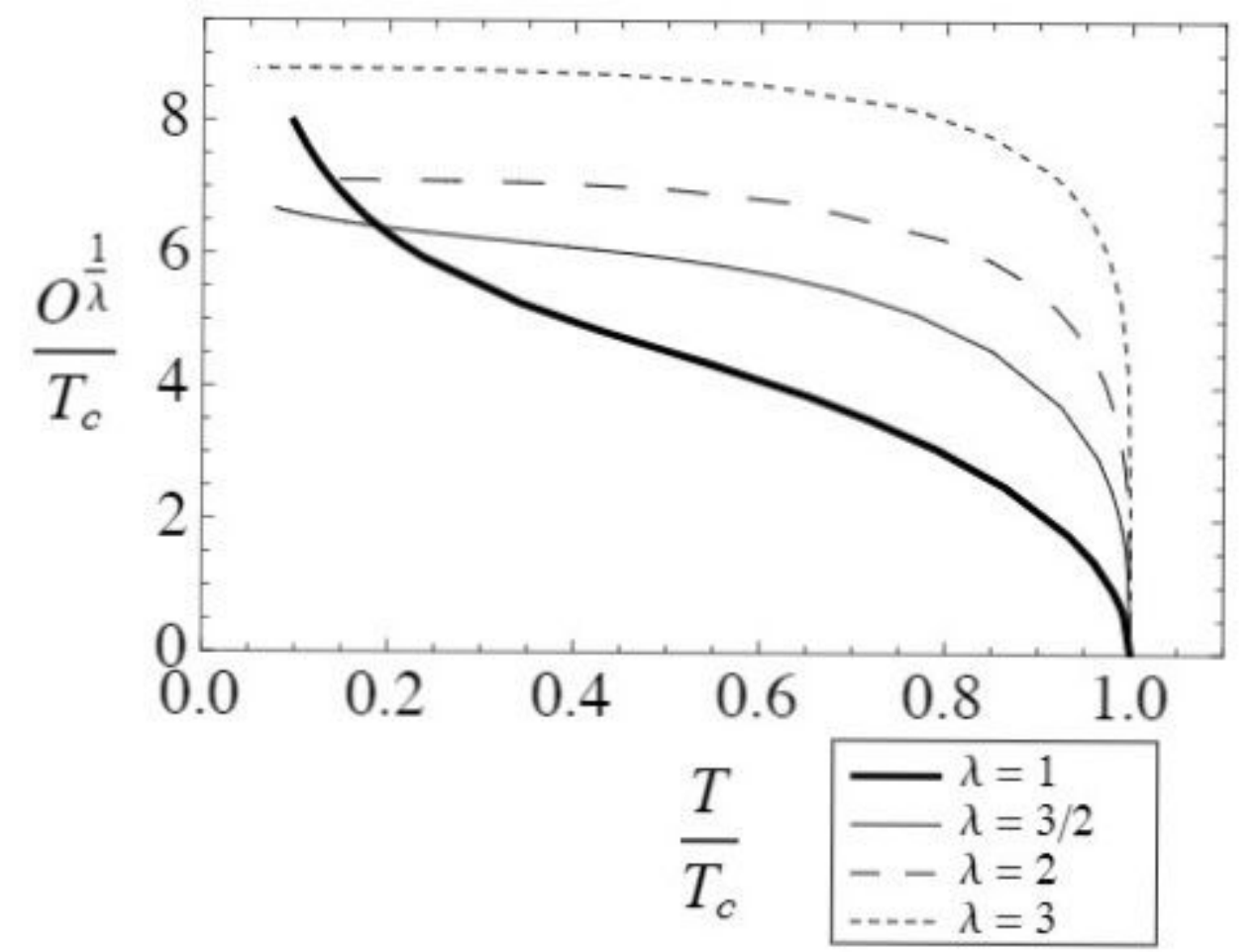}
	\caption{  Result of  ref. \cite{Horo2009}: The condensate as a function of temperature. Here, $\lambda =\Delta$.
	}
	\label{Horowi}
\end{figure}

The authors of ref. \cite{Siop2010} argued that $X$ approaches to zero as $\Delta \rightarrow 3$, while   Horowitz et.al \cite{Horo2009}'s numerical calculation got a finite value  $X= 8.8$ at $\Delta=3$ (see FIG. \ref{Horowi}). On the other hand,  our  calculation show  $X= 7.2$ at $\Delta=3$. Our result is good agreement with the one of ref \cite{Horo2009}.

\subsubsection{Analytic calculation of $g^{\frac{1}{\Delta}}\frac{\left< \mathcal{O}_{\Delta}\right>^{\frac{1}{\Delta}}}{T_c}$   at $\Delta=3/2$}\label{big}
\begin{figure}[ht!]
	\centering
	\subfigure[ $X$  vs $T/T_c$ at $\Delta=3/2$]
	{\includegraphics[width=0.45\linewidth]{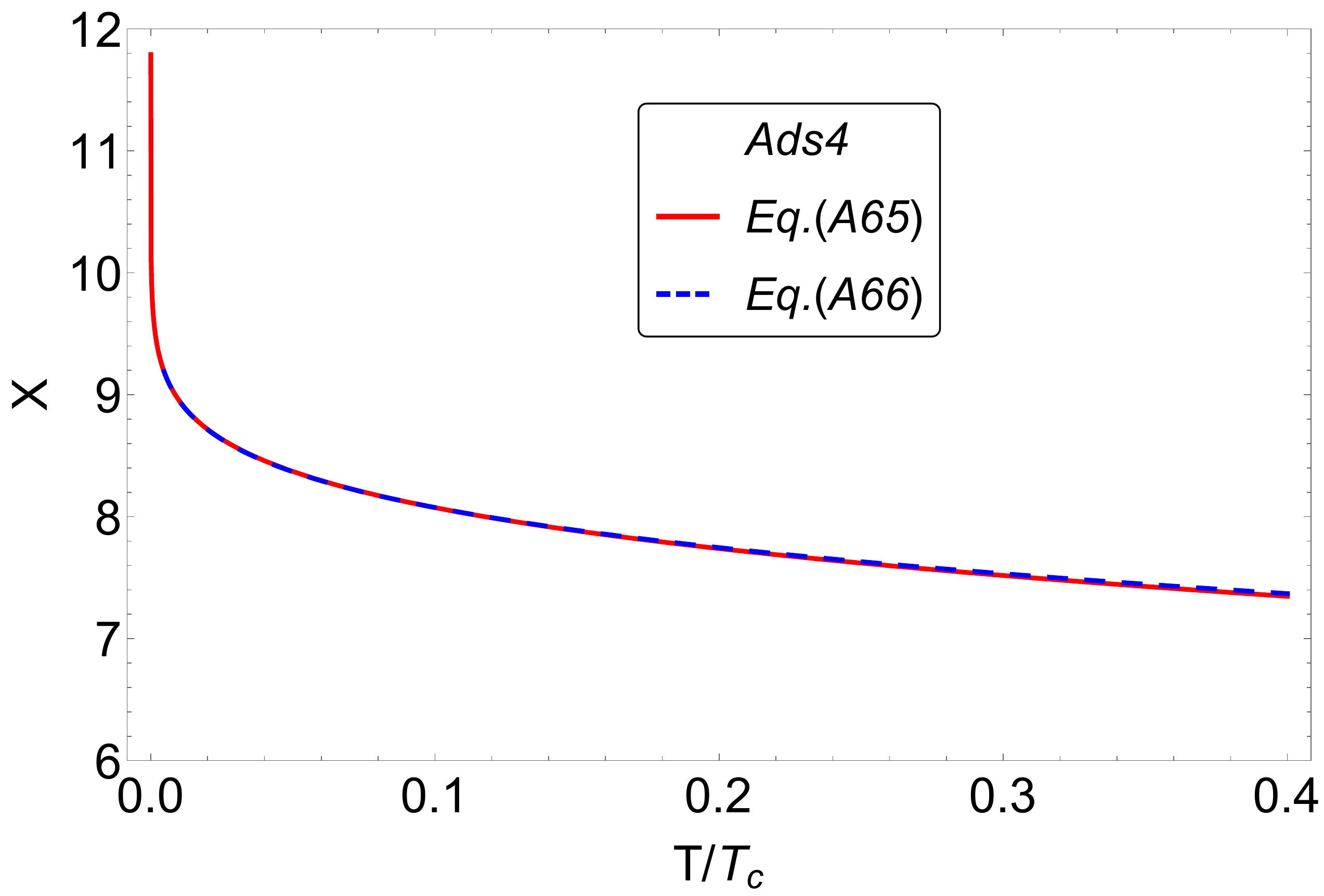}}
	\subfigure[$X^{4}$-$\log(T/T_c)$ graph at $\Delta=3/2$. ]
	{\includegraphics[width=0.45\linewidth]{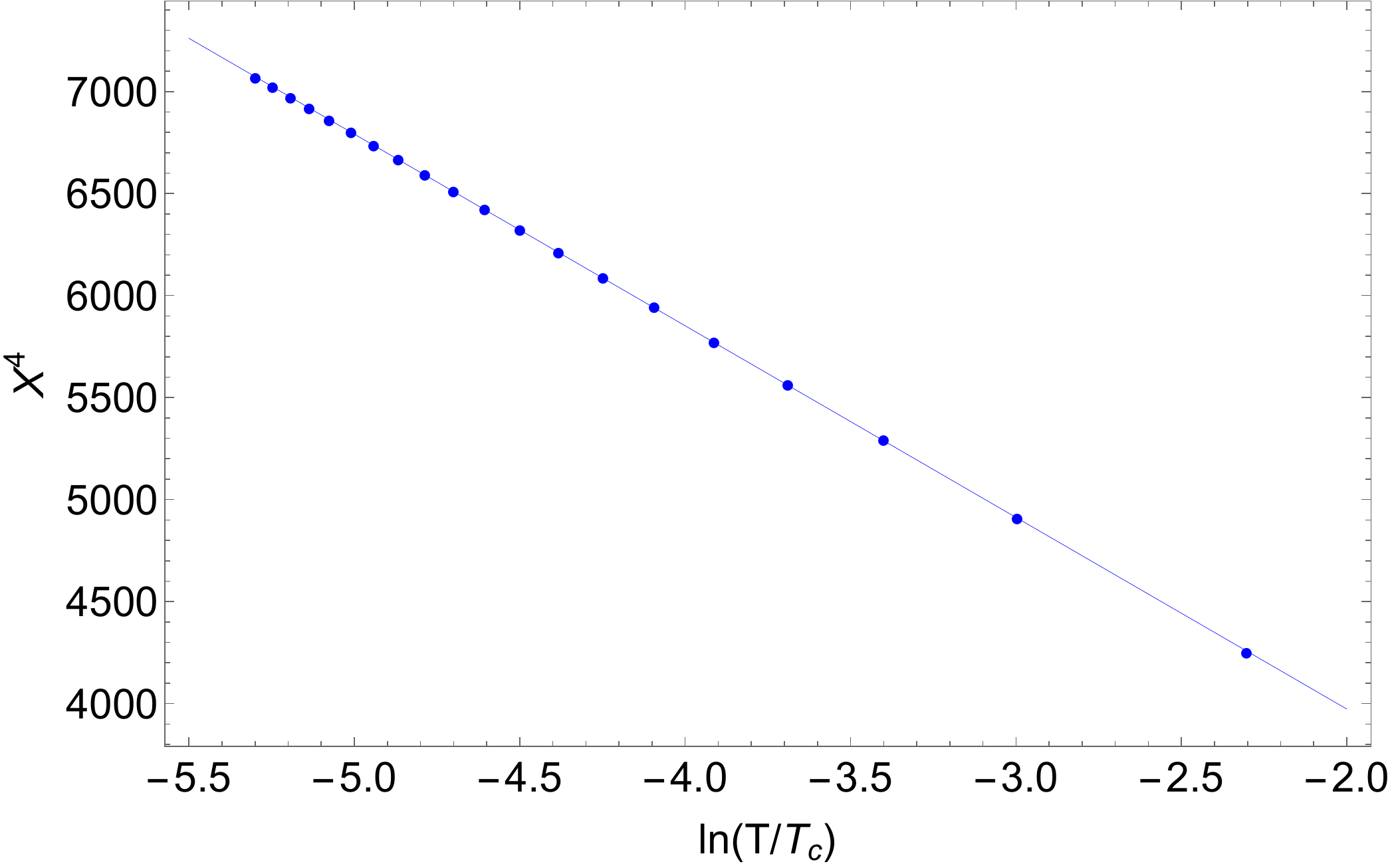}}
	\caption{\small
		(a) $X$  vs $T/T_c$: red  colored curves for $X$ is a plot of Eq.(\ref{ppp:4}). And blue dashed curves is a plot of  Eq.(\ref{oo:62}). These two curves are almost indentical for low temperature.  \\
		(b)$X^{4}$-$\log(T/T_c)$ graph at $\Delta=3/2$: The slope of  blue  dotted line  for $X^{4}$ is $-939$. }
	\label{logplot}
\end{figure}

$\alpha_{3}$ and $\beta_3 $	in Eq.(\ref{si:671b}) have series expansions at $\Delta=3/2$:
\begin{scriptsize}   
	\begin{eqnarray}   
		\alpha_{3} &=&\frac{\pi \csc \left(\frac{2 \pi }{3}\right)}{18 \left(\Delta -\frac{3}{2}\right)}+\frac{\pi   \csc \left(\frac{2 \pi }{3}\right) \left(3+3 (-\log (4))-4\psi\left(2-\frac{2}{3}\right)-2 \psi\left(\frac{2}{3}\right)\right)}{3^4}+\mathcal{O} \left(\Delta -\frac{3}{2}\right) \label{ppp:1}\\  
		\beta_3 &=& -\frac{\pi \csc \left(\frac{2 \pi }{3}\right)}{18 \left(\Delta -\frac{3}{2}\right)}+\frac{\pi  \left(18+ \pi  \cot \left(\frac{2 \pi }{3}\right) \right) \csc \left(\frac{2 \pi }{3}\right)}{3^4}+\mathcal{O}\left(\Delta -\frac{3}{2}\right). 
		\label{ppp:2}
	\end{eqnarray} 
\end{scriptsize}    
As Eq.(\ref{ppp:1}) and Eq.(\ref{ppp:2}) are substituted into Eq.(\ref{si:67}) with taking the limit $\Delta \rightarrow 3/2$, we obtain
\begin{equation}      
	X^{4}
	= G_3^{4} \left(\rho_{3} +\frac{\sigma_3}{2} \left(\frac{1-\tau_{3}^{3-2\Delta} X^{3-2\Delta}}{\Delta-\frac{3}{2}} \right)\right)
	\label{ppp:3}
\end{equation} 	
where
\begin{eqnarray}   
	\sigma_3 &=&\frac{\pi \csc \left(\frac{2 \pi }{3}\right)}{9} \nonumber\\  
	\rho_3 &=& \frac{\sigma_3}{3}\left(21-3\ln(4)+\pi \cot\left(\frac{2\pi}{3} \right)-4\psi(4/3)-2\psi(2/3) \right) \nonumber\\
	&=& \frac{\sigma_3}{3}\left(3-\pi\cot\left(\frac{2\pi}{3} \right)-\ln(4)-2\psi(2/3)\right)  
	\nonumber
\end{eqnarray}  	
here, $\psi(z)$ is the digamma function. By using L'Hopital's rule,  Eq.(\ref{ppp:3}) becomes
\begin{eqnarray}   
	X^{4} &=& G_3^{4} \left(\rho_{3} -\frac{\sigma_3}{2} \frac{\partial }{\partial \Delta}\left(\tau_{3}^{3-2\Delta} X^{3-2\Delta} \right)\right) \nonumber\\
	&=&  G_3^{4}  \left(\rho_{3} +\sigma_3 \ln\left(\tau_{3}X \right)\right) \label{ppp:4} 
\end{eqnarray} 	 
Fig.~\ref{logplot} (b) tells us that $X\sim \ln(T_c/T)^{1/4}$ for low temperature; 
Numerical result tells us that $X^{4}$-$\log(T/T_c)$ plot  demonstrates our arguements with high precision. 

And $X$ is numerically 
\begin{small}
	\bea
	X &&\approx 6.76 \left( 1+0.45 \ln\left(\frac{T_c}{T}\right)\right)^{1/4} 
	\label{oo:62}
	\eea    
\end{small}  
\subsubsection{Analytic calculation of $g^{\frac{1}{\Delta}}\frac{\left< \mathcal{O}_{\Delta}\right>^{\frac{1}{\Delta}}} {\sqrt{g \rho}}$   at $1/2<\Delta<3$}\label{small1}
Apply  Eq.(\ref{eq:65}) into Eq.(\ref{eq:66}) with $T  =\frac{3}{4\pi}r_h$ with $ b= \left(\frac{ g\left< \mathcal{O}_{\Delta}\right>}{\Delta r_h^{\Delta}}\right)^{\frac{1}{\Delta}}$   in the form of $Y$; here, $Y:=$	$\frac{g^{1/\Delta}\left< \mathcal{O}_{\Delta}\right>^{1/\Delta}}{\sqrt{g \rho}}$ for simple notation, we obtain  the condensate at $T\approx 0$: 		
\begin{equation}
	Y^{4}
	= \widetilde{G_3}^{4}   \left(\alpha_{3} +\beta_3 \widetilde{\tau_{3}}^{3-2\Delta}  Y^{3-2\Delta} \right)
	\label{s4:1}
\end{equation} 
\begin{eqnarray}
	\hbox{where }\quad 	 \widetilde{G_3} &=& \Delta^{1/\Delta} \left(\frac{-2^{1+\nu}  }{\Gamma(-\nu)}\right)^{\frac{1}{2}}  \label{s4:2}\\  
	\widetilde{\tau_{3}} &=&\frac{d}{4\pi \Delta^{1/\Delta}}\frac{\sqrt{g \rho}}{T}   \label{s4:3} 
\end{eqnarray} 
with    $\nu=\frac{1}{2\Delta}$. Here, $\alpha_{3}$ and $\beta_3$ are in Eq.(\ref{si:671b}).   
\subsubsection{Analytic calculation of $g^{\frac{1}{\Delta}}\frac{\left< \mathcal{O}_{\Delta}\right>^{\frac{1}{\Delta}}} {\sqrt{g \rho}}$   at $\Delta=3/2$}\label{big1}
\begin{figure}[ht!]
	\centering
	\subfigure[ $Y$  vs $T/\sqrt{g \rho}$ at $\Delta=3/2$]
	{\includegraphics[width=0.45\linewidth]{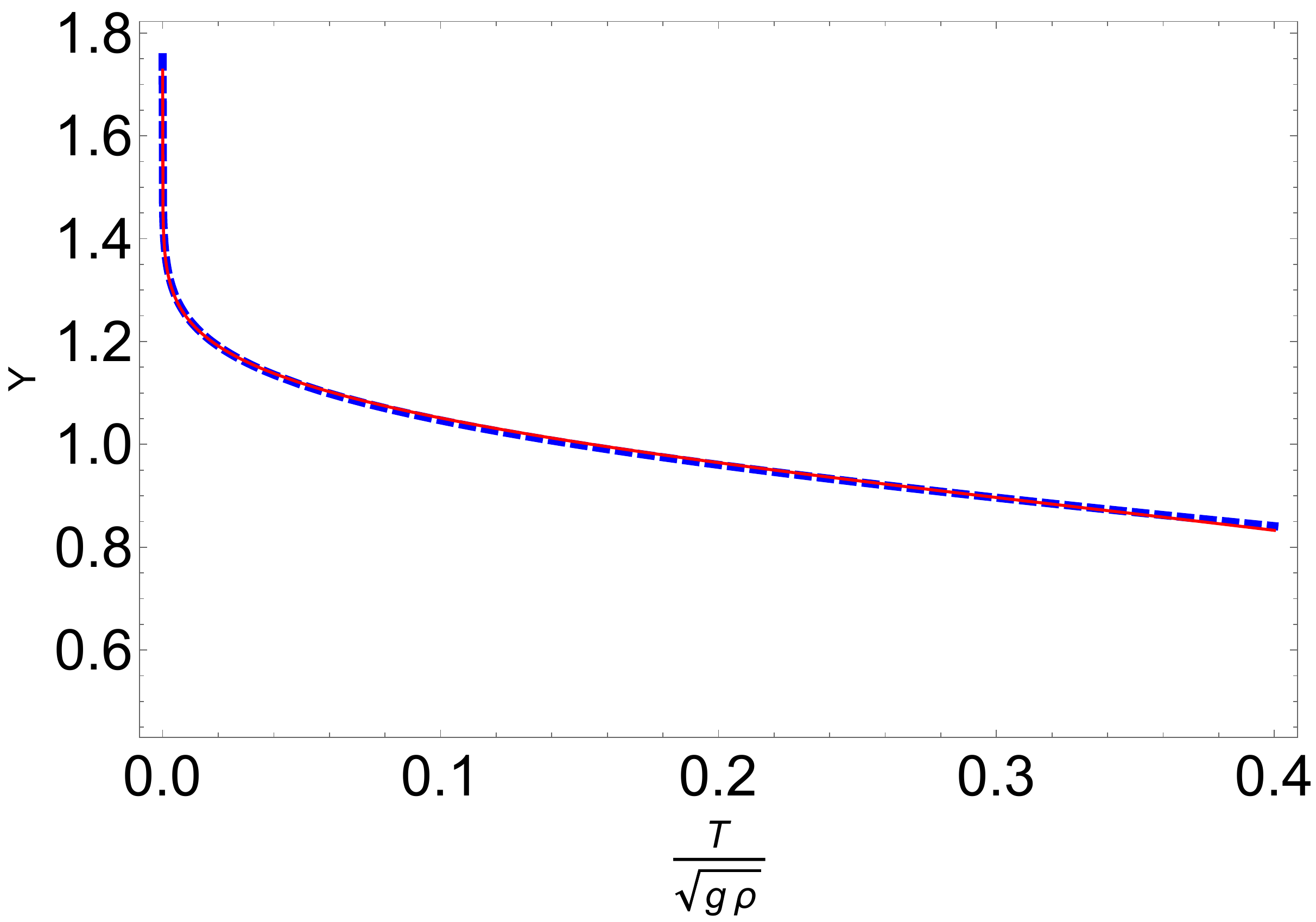}}
	\subfigure[$Y^{4}$-$\log(T/\sqrt{g \rho})$ graph at $\Delta=3/2$. ]
	{\includegraphics[width=0.45\linewidth]{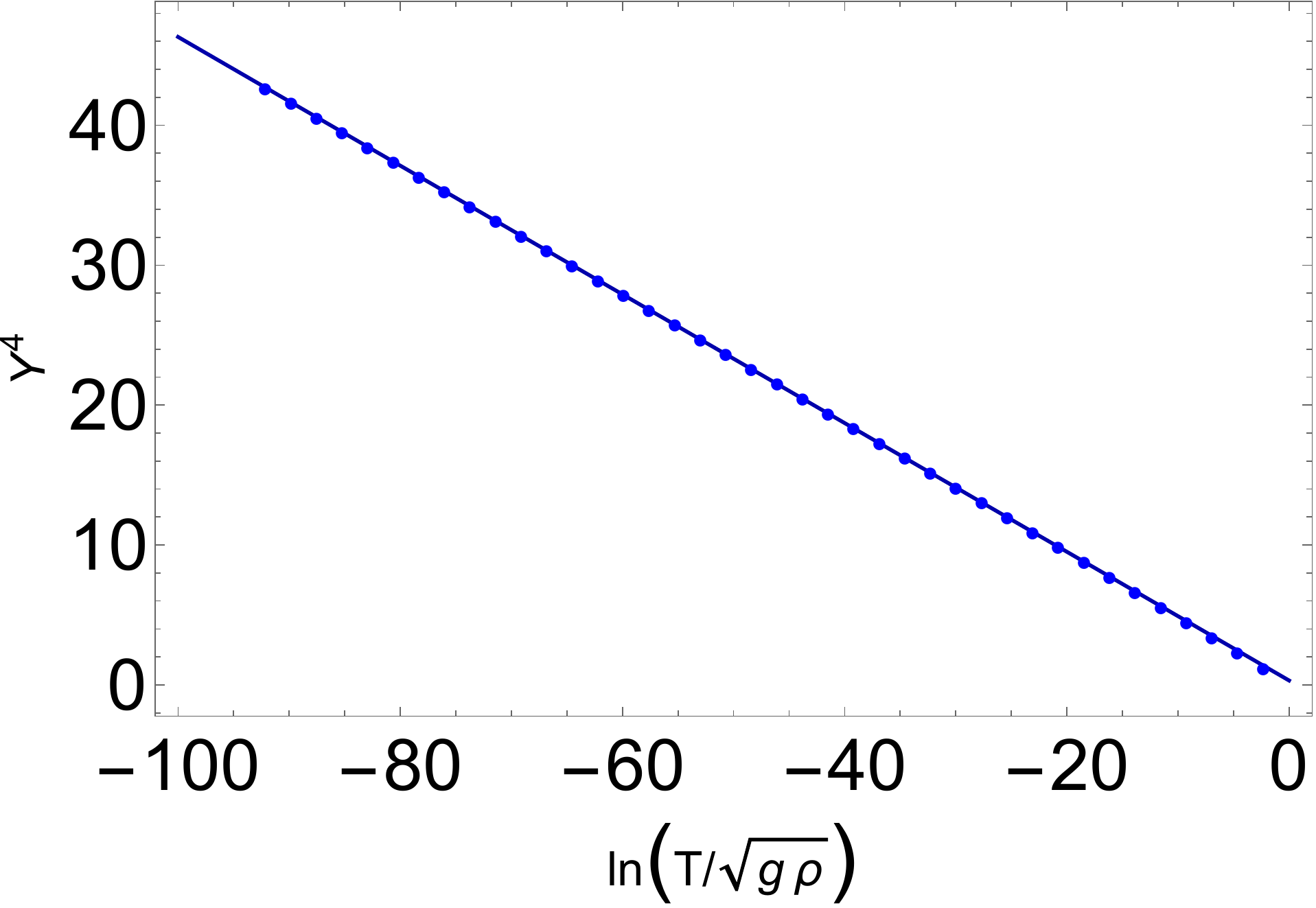}}
	\caption{\small
		(a) $Y$  vs $T/\sqrt{g \rho}$: red  colored curves for $Y$ is a plot of Eq.(\ref{s4:4}). And blue dashed curves is a plot of  Eq.(\ref{s4:5}). These two curves are almost indentical for low temperature.  \\
		(b)$Y^{4}$-$\log(T/\sqrt{g \rho})$ graph at $\Delta=3/2$: The slope of  blue  dotted line  for $Y^{4}$ is $-0.46$. }
	\label{logplot10}
\end{figure}

As Eq.(\ref{ppp:1}) and Eq.(\ref{ppp:2}) are substituted into Eq.(\ref{s4:1}) with taking the limit $\Delta \rightarrow 3/2$, we obtain
\begin{equation}      
	Y^{4}
	= \widetilde{G_3}^{4}  \left(\rho_{3} +\frac{\sigma_3}{2} \left(\frac{1-\widetilde{\tau_{3}}^{3-2\Delta} X^{3-2\Delta}}{\Delta-\frac{3}{2}} \right)\right).
	\label{s4:3}
\end{equation} 		
By using L'Hopital's rule,  Eq.(\ref{s4:3}) becomes
\begin{eqnarray}   
	Y^{4} &=& \widetilde{G_3}^{4} \left(\rho_{3} -\frac{\sigma_3}{2} \frac{\partial }{\partial \Delta}\left(\widetilde{\tau_{3}}^{3-2\Delta} Y^{3-2\Delta} \right)\right) \nonumber\\
	&=&  \widetilde{G_3}^{4}  \left(\rho_{3} +\sigma_3 \ln\left(\widetilde{\tau_{3}}Y \right)\right) \label{s4:4} 
\end{eqnarray} 	 
Fig.~\ref{logplot10} (b) tells us that $Y\sim \ln(\sqrt{g \rho}/T)^{1/4}$ for low temperature; 
Numerical result tells us that $Y^{4}$-$\log(T/\sqrt{g \rho})$ plot  demonstrates our arguements with high precision. 

And $Y$ is numerically 
\begin{small}
	\bea
	Y &&\approx 0.44 \left( 1+ 13.47 \ln\left(\frac{\sqrt{g \rho}}{T}\right)\right)^{1/4} 
	\label{s4:5}
	\eea    
\end{small}  

\subsection{ The Conductivity Gap }\label{Gap}
\begin{figure}[h]
	\subfigure[ $z_0$  vs $\Delta$. ]
	{\includegraphics[width=0.52\linewidth]{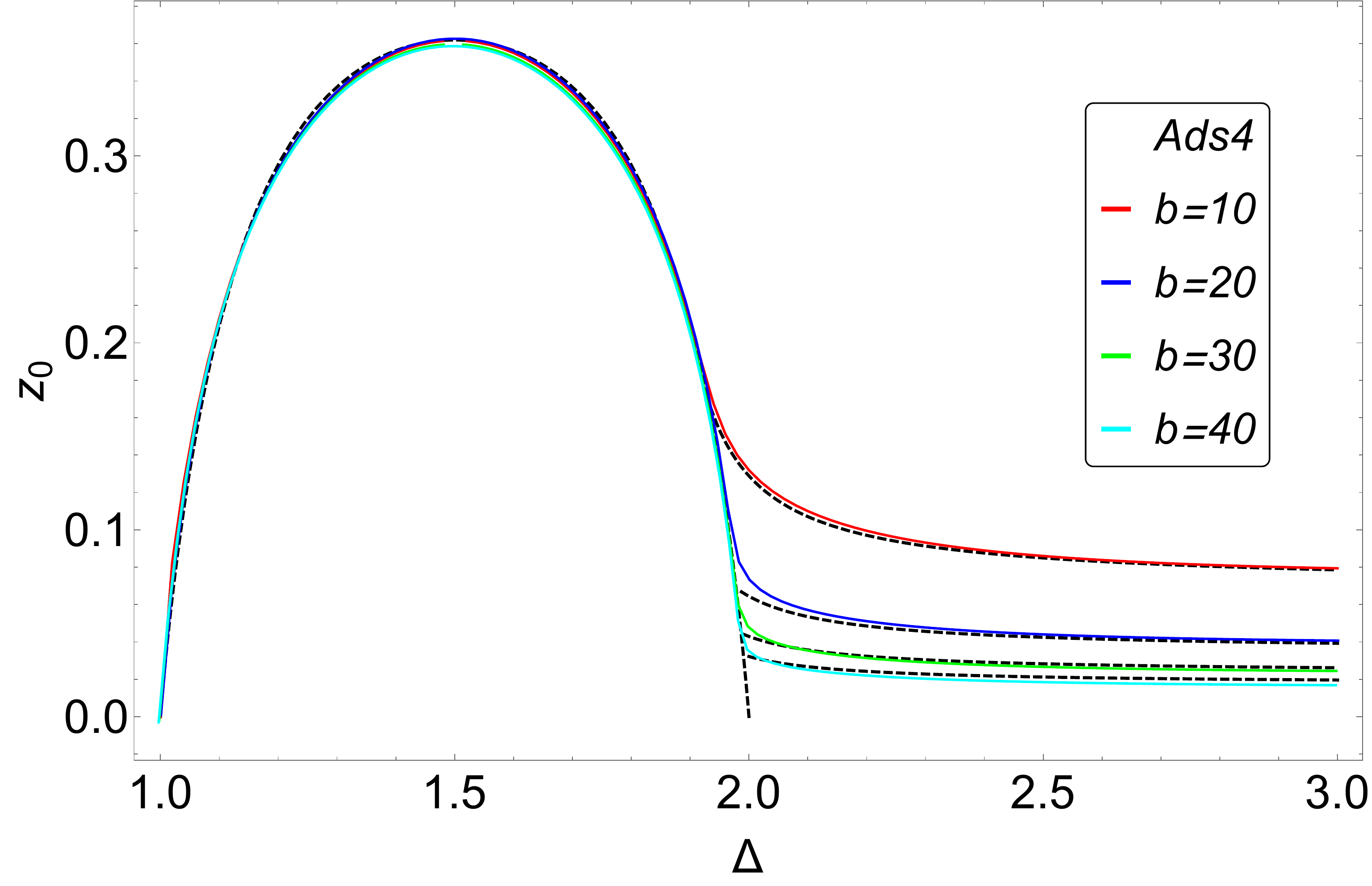}}
	\hfill
	\subfigure[ $V(z)$  vs $z$. ]
	{\includegraphics[width=0.52\linewidth]{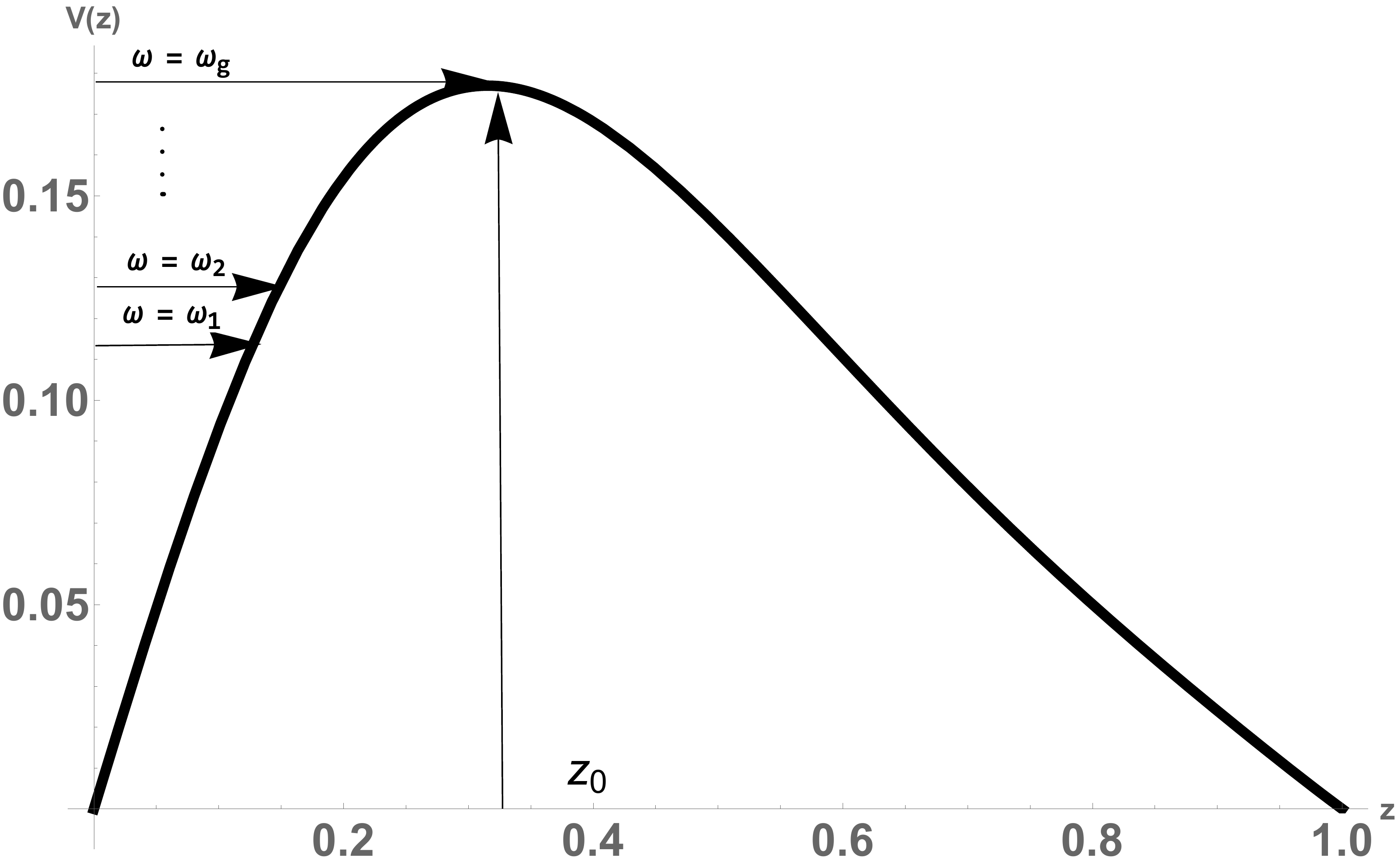}}
	\caption{\small
		(a)   $z_0$ for  various $b$'s. Solid colored curves are numerical expression Eq.(\ref{cod:29}) 
		and dashed curves are analytic expression  Eqs.(\ref{codd:31}, \ref{codd:32}).   The local maximum $z_0 \approx 0.362$  is at $\Delta=3/2$.    
		(b)  a rough picture $V(z)$ in terms of a $z$ coordinate. 
		$\omega= \omega_g$ at  $z=z_0$. $\omega_n$ is the $n^{th}$ pole.   $\omega_1< \omega_2< \cdots $ are resonant frequencies, but  $\omega_g$ is the approximate value of the gap in AC conductivities. }
	\label{z.zero}
\end{figure}

Now we begin to discuss the resonant frequencies. 
The Eq.(\ref{cod:1}) takes the form of 	a  Schr\"odinger equation with energy $\omega$: 
\begin{equation}
	-\frac{d^2 A_x }{d r_{\star}^2} + V(r_{\star}) A_x=\omega^2 A_x , 
	\label{cod:27}
\end{equation}	 
where,  $ V(r_{\star})$ is re-expression of $V(z)=   \frac{g^2 \left< \mathcal{O}_{\Delta}\right>^2}{r_{+}^{2\Delta-2}}(1-z^3) z^{2\Delta-2} F(z)^2$ in terms of the tortoise coordinate $r_{\star}$, 
\begin{equation}
	r_{\star}= \int \frac{dr}{f(r)}=\frac{1}{6 r_{+}}\left[\ln\frac{(1-z)^3}{1-z^3}-2\sqrt{3}\tan^{-1}\frac{\sqrt{3}\;z}{2+z}\right], 
	\label{cod:28}
\end{equation}	
where  the integration constant is chosen such that boundary is at $r_{\star} =0$.	
We follow \cite{Horo2009} to define  the size of the gap in AC conductivity $\omega_{g}$  by     
\be
\omega_g =  \sqrt{V_{\mbox{max}}}. 
\ee
%
Here, there is no solution for $\omega_g$ at $1/2 <\Delta <1$. Because $\lim_{z\rightarrow 0}V(z)\rightarrow \infty$.jh  
Then, we can construct an analytic expression of $\omega_{g} $. 
First introduce  $z_0$ at which $V$ is maximum:  
\begin{equation}
	\frac{d V}{d z}\Bigg|_{z=z_0}= \frac{d}{d z} (1-z^3) z^{2\Delta-2} F(z)^2\Bigg|_{z=z_0}   =0 .
	\label{cod:29}
\end{equation}		 
Then it can be numerically calculated as a function  of $\Delta$ and $b$, and the result can be fit by following expressions. 
\bea
z_0 &&\approx 0.41  \sum_{k=1}^{\infty}\frac{\sin\left(\pi(\Delta-1)(2k-1)\right)}{k^{2.64}},  \;\;   
\hbox{ for }  1\leq\Delta <2
\label{codd:31}, 
\\ 
z_0 &&\approx \left(\frac{0.10}{\Delta-1.83}+0.70\right) \frac{1}{b}, \;\;  
\hbox{ for }  2<\Delta \leq 3.
\label{codd:32}		
\eea
Notice that from the first expression, we can see that  there is no $b$ dependence. 
This result  is    plotted   in the Figure \ref{z.zero}(a). Notice that the numerical data is fit very well by our formula.   
Using these data,  $\omega_{g}$ is given by
\begin{equation}
	\omega_{g} = \sqrt{V_{\mbox{max}}} = \frac{g \left< \mathcal{O}_{\Delta}\right> }{r_{+}^{ \Delta-1}}\sqrt{1-z_0^3} z_0^{ \Delta-1} F(z_0) \approx \frac{g \left< \mathcal{O}_{\Delta}\right> }{r_{+}^{ \Delta-1}}  z_0^{ \Delta-1} F(z_0) .
	\label{codd:33}
\end{equation} 
The expression for 	$F(z_0)$ is cumbersome and it is given in the 
appendix \ref{appendix4}. 
The solution of Eq.(\ref{codd:33}) according to the regimes of $\Delta$ is given in Table.~\ref{omegag1} ealier in the introduction and summary section.
For derivation of these results, see the appendix \ref{appendix4}.

\begin{figure}[h]
	\centering
	\includegraphics[width=0.6\linewidth]{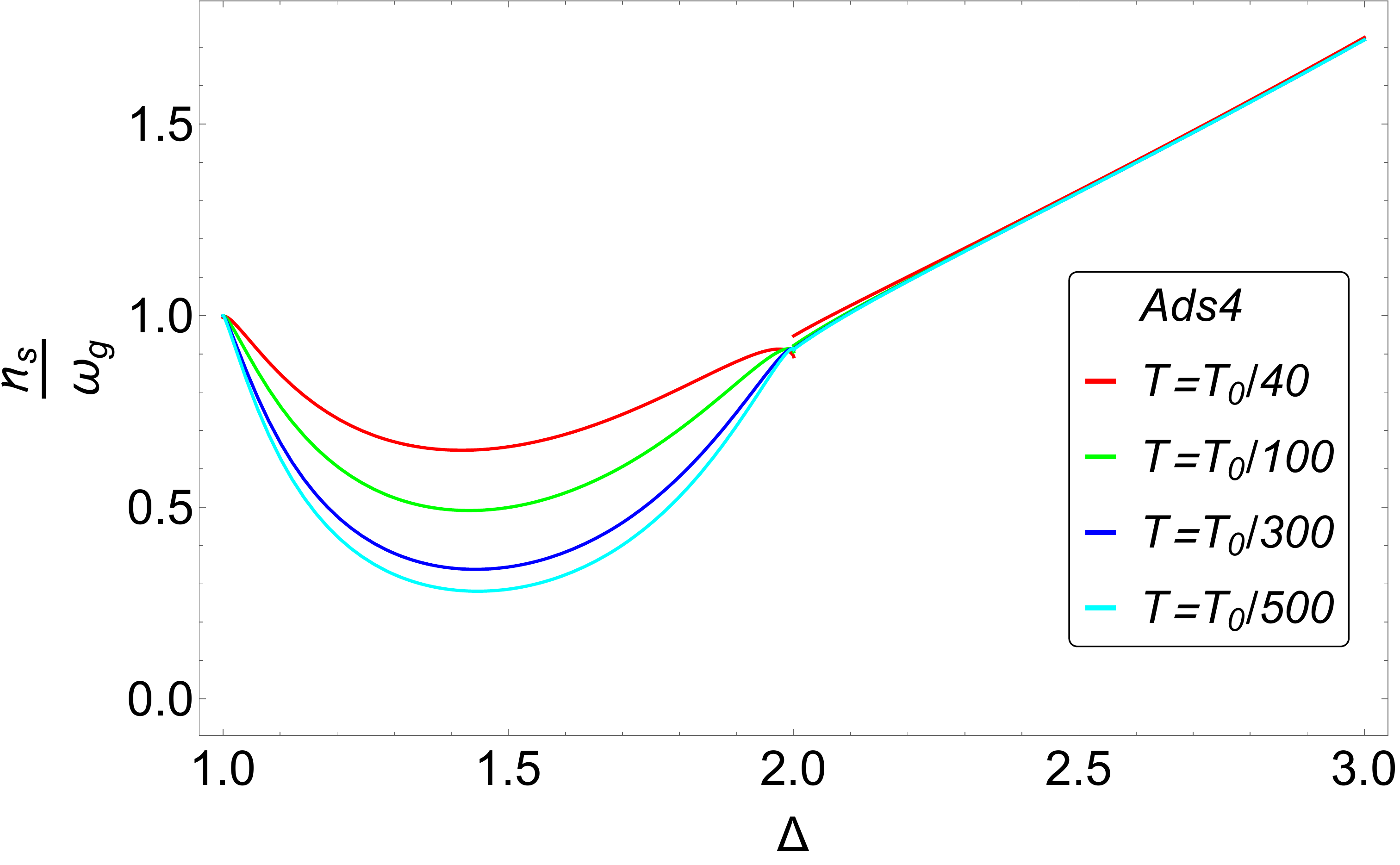}
	\caption{ $ n_s/\omega_g$  vs    $\Delta$ for  $d=3$. Here, $T_{0}=(g \rho)^\frac{1}{2}$. In the regime $2\leq \Delta \leq3$,   it is   independent of   $T$  but only   $\Delta$ dependence.} 
	\label{ns.omegag}
\end{figure}

Using the result of the Cooper pair density $n_{s}$ given in Eq.(\ref{codd:26}) 
and the expression of $\omega_{g}$,  
we can calculate the ratio $n_{s}/\omega_{g}$. 
FIG. \ref{ns.omegag}  is the plot of this result.  
Interestingly, in the regime $2\leq \Delta \leq3$, we have linearity 
between ${n_s}$ and ${\omega_g} $. 
\begin{equation}
	\frac{n_s}{\omega_g}   \approx 0.8\Delta -0.7 ,
	\nonumber
\end{equation}
Notice   that   in this regime of $\Delta$,  there is no $b$ dependence in the ratio due to the cancellation of $b$-dependent pieces of $n_{s}$ and $\omega_{g}$.    

\subsubsection{Maxwell perturbations and the conductivity at near the zero temperature}\label{appendix3}
The Maxwell equation at zero spatial momentum and with a time dependence of the form  $e^{-i\omega t}$ gives	
\begin{equation} 
	r_{+}^2 (1-z^3)^2 \frac{d^2 A_x }{d z^2} -3 r_{+}^2 z^2 (1-z^3)\frac{d A_x }{d z}+\left( \omega^2-V(z) \right) A_x=0
	\nonumber
\end{equation}	
where $A_x$ is any component of the perturbing electromagnetic potential along the boundary and 
$$V(z)= \frac{g^2 \left< \mathcal{O}_{\Delta}\right>^2}{r_{+}^{2\Delta-2}}(1-z^3) z^{2\Delta-2} F(z)^2$$
with $F$ defined before.
We introduce $A_x(z)=(1-z^3)^{-\frac{i}{3} \hat{\omega}} G(z)$ where $\hat{\omega}=\omega/r_{+}$.  Because we require $A_x\propto (1-z^3)^{-\frac{i}{3} \hat{\omega}}$  near $z=1$ corresponding to ingoing wave boundary conditions at the horizon. Then, the wave equation  reads
\begin{equation} 
	(1-z^3)  \frac{d^2 G }{d z^2} -3 \left( 1-\frac{2i \hat{\omega}}{3} \right) z^2 \frac{d G }{d z}+\left( \frac{\hat{\omega}^2(1+z)(1+z^2)}{1+z+z^2} +2i\hat{\omega}z - \frac{g^2\left< \mathcal{O}_{\Delta}\right>^2}{r_{+}^{2\Delta}} z^{2\Delta-2} F^2(z)  \right)G=0
	\nonumber
\end{equation}  
We have the following limiting form:
\begin{equation} 
	\lim_{z\rightarrow 0}  K_{\nu}\left( z\right)\sim \frac{\Gamma(\nu)}{2}\left(\frac{z}{2}\right)^{-\nu} \sum_{k=0}^{1}\frac{\left(\frac{z}{2}\right)^{2k}}{(1-\nu)_k k!}+\frac{\Gamma(-\nu)}{2}\left(\frac{z}{2}\right)^{\nu} \sum_{k=0}^{1}\frac{\left(\frac{z}{2}\right)^{2k}}{(1+\nu)_k k!}
	\label{cod:3}
\end{equation}
at $\nu \ne \mathbb{Z} $. 

We obtain the analytic expressions of $F(z)$ in the following way:
\begin{small}
	\bea
	F(z) &&\approx  1- \frac{(bz)^4 \mathbb{J}_1+ (bz)^3 \mathbb{J}_2+(bz)^2 \mathbb{J}_3}{-\frac{\sqrt{\pi } \Gamma \left(\frac{3}{2 \Delta }-1\right) \Gamma \left(\frac{1}{\Delta }\right) \Gamma \left(\frac{2}{\Delta }\right)}{8 \Delta ^2 \Gamma \left(\frac{\Delta +3}{2 \Delta }\right)}+ \frac{\pi  (3-\Delta )^2 \csc \left(\frac{\pi }{2 \Delta }\right)}{4 \Delta ^4 (3-2 \Delta )}b^{3-2\Delta}},  \;\;   
	\hbox{ for }  z\leq 1/b
	\label{cod:4},
	\\
	F(z) &&\approx 1- \frac{-\frac{\sqrt{\pi } \Gamma \left(\frac{3}{2 \Delta }-1\right) \Gamma \left(\frac{1}{\Delta }\right) \Gamma \left(\frac{2}{\Delta }\right)}{8 \Delta ^2 \Gamma \left(\frac{\Delta +3}{2 \Delta }\right)}+ \frac{\pi  \csc \left(\frac{\pi }{2 \Delta }\right)}{4 \Delta ^2 (3-2 \Delta )}(bz)^{3-2\Delta}}{-\frac{\sqrt{\pi } \Gamma \left(\frac{3}{2 \Delta }-1\right) \Gamma \left(\frac{1}{\Delta }\right) \Gamma \left(\frac{2}{\Delta }\right)}{8 \Delta ^2 \Gamma \left(\frac{\Delta +3}{2 \Delta }\right)}+ \frac{\pi  (3-\Delta )^2 \csc \left(\frac{\pi }{2 \Delta }\right)}{4 \Delta ^4 (3-2 \Delta )}b^{3-2\Delta}} , \;\;  
	\hbox{ for } z> 1/b.
	\label{cod:7}
	\eea
\end{small}
where
\begin{eqnarray}
	\mathbb{J}_1 &=&  2^{-\frac{1}{\Delta }-8}   \Gamma \left(-1-\frac{1}{2 \Delta }\right)^2 \left(\frac{8 \Delta +4}{\Delta ^2}+\frac{8 (2 \Delta +1) (bz)^{2 \Delta }}{\Delta  \left(4 \Delta ^2+9 \Delta +2\right)}+\frac{(bz)^{4 \Delta }}{6 \Delta ^2+7 \Delta +1}\right)  \nonumber\\  
	\mathbb{J}_2 &=& - \frac{1}{24} \pi   \csc \left(\frac{\pi }{2 \Delta }\right)\left(\frac{\Delta ^2 \left(48 \Delta +(2 \Delta +3) (bz)^{2 \Delta }+36\right)(bz)^{2 \Delta }}{(2 \Delta +3) (4 \Delta +3) \left(4 \Delta ^2-1\right)}+4 \right)\nonumber\\  
	\mathbb{J}_3 &=& 2^{\frac{1}{\Delta }-7}   \Gamma \left(\frac{1}{2 \Delta }-1\right)^2\left(\frac{8 \Delta -4}{\Delta ^2}+\frac{4 (2 \Delta -1) (bz)^{2 \Delta }}{\Delta  \left(4 \Delta ^2+3 \Delta -1\right)}+\frac{(bz)^{4 \Delta }}{12 \Delta ^2+4 \Delta -1}\right)   
	\label{cod:5} 
\end{eqnarray}   
As we apply  Eq.(\ref{cod:3}) and Eq.(\ref{eq:65}) into Eq.(\ref{eq:55a}), we obtain   Eq.(\ref{cod:4}). 

Replacing $b$ by $bz$ in  Eq.(\ref{eq:57a}) and Eq.(\ref{ii:78}), we have 	
\begin{eqnarray}
	F_{\Delta}(bz) &=&  \int_{0}^{bz} dz \; z^{2-2\Delta} \int_{0}^{z} d\tilde{z} \; \tilde{z}^{2\Delta-1} \left( K_{\frac{1}{2 \Delta }}\left(
	\tilde{z}^{\Delta }\right)\right)^2 \nonumber\\
	&=&	b^3\int_{0}^{z} dz \; z^{2-2\Delta} \int_{0}^{z} d\tilde{z} \; \tilde{z}^{2\Delta-1} \left( K_{\frac{1}{2 \Delta }}\left(
	b^{\Delta }\tilde{z}^{\Delta }\right)\right)^2  \nonumber\\
	&=&	\frac{\pi \csc{\left( \frac{\pi}{2\Delta}\right)}  }{4 \Delta^2 (3-2\Delta)} (bz)^{3-2\Delta} -\frac{\sqrt{\pi } \Gamma \left(\frac{3}{2 \Delta }-1\right) \Gamma \left(\frac{1}{\Delta }\right) \Gamma \left(\frac{2}{\Delta }\right)}{8 \Delta ^2 \Gamma \left(\frac{\Delta +3}{2 \Delta }\right)}
	\label{cod:6}
\end{eqnarray}
Substitute Eq.(\ref{cod:6}) and Eq.(\ref{eq:65}) into   Eq.(\ref{eq:55a}). We obtain  Eq.(\ref{cod:7}). 
\begin{figure}[h]
	\centering
	\includegraphics[scale=.35]{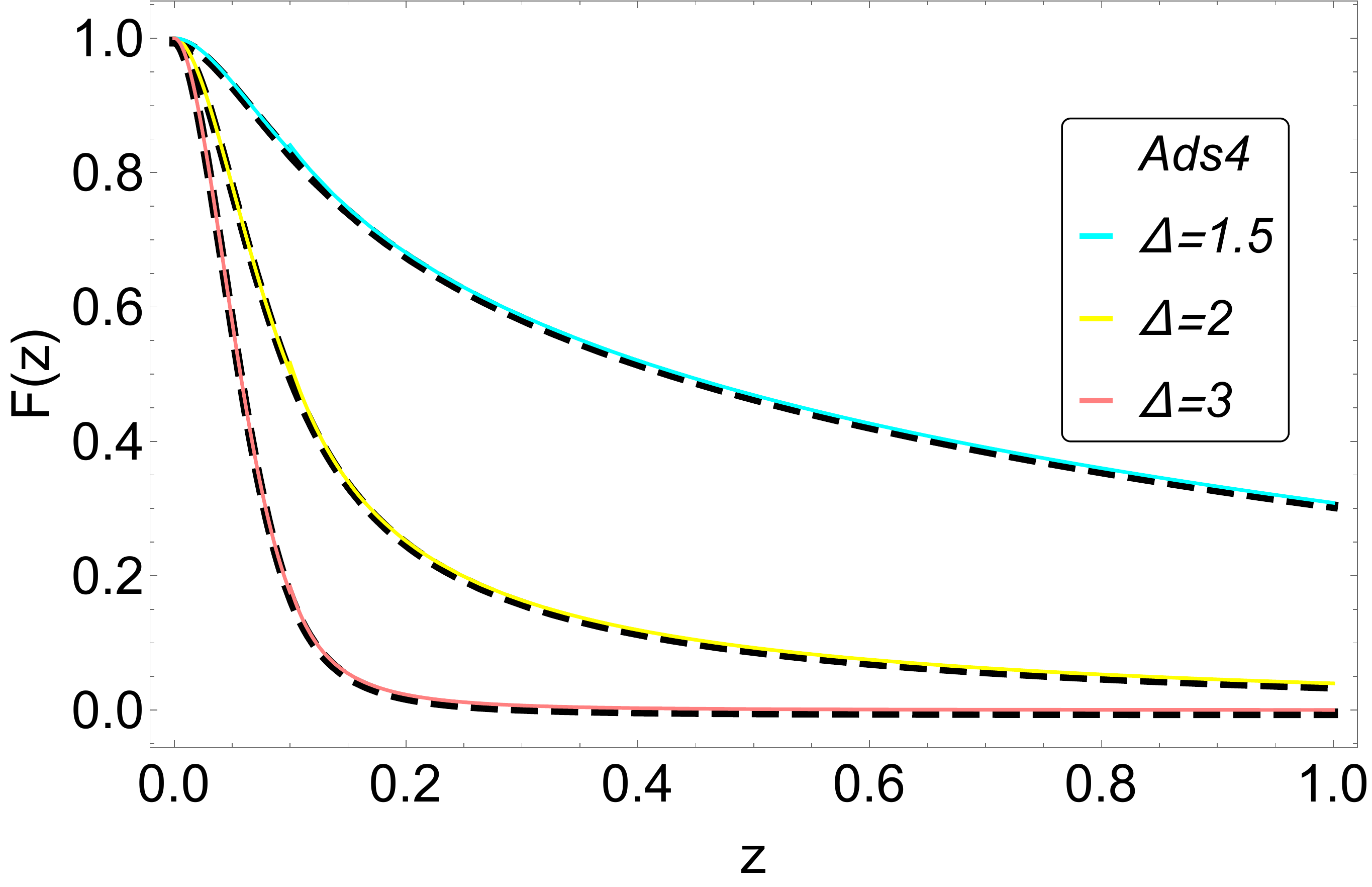}
	\caption{  The field $F$ for $\Delta =1.5$ (light blue), $\Delta =2$ (yellow), $\Delta =3$ (pink) at $b=10$. Solid colored curves are analytic expression Eq.(\ref{cod:4}) and Eq.(\ref{cod:7}) and dashed curves are exact numerical results Eq.(\ref{eq:55a}) with Eq.(\ref{eq:65}) (almost indistinguishable).
	}
	\label{Numer.delta}
\end{figure}  

Fig. \ref{Numer.delta} shows how the data fits by above formulas Eq.(\ref{cod:4}) and Eq.(\ref{cod:7}).	   

%
$F(z)\approx 1$ at $\Delta=1$ in  Eq.(\ref{cod:4}) and Eq.(\ref{cod:7}). And the solution of  Eq.(\ref{cod:10}) is  
\begin{equation} 
	G(z)=   \exp \left( i z\left(-\frac{\omega}{3}+\sqrt{ \omega^2+\frac{i \omega}{3}-\frac{g^2\left< \mathcal{O}_{1}\right>^2}{r_{+}^{2}}}\right)\right)  +R\;  \exp \left( i z\left(-\frac{\omega}{3}-\sqrt{ \omega^2+\frac{i \omega}{3}-\frac{g^2\left< \mathcal{O}_{1}\right>^2}{r_{+}^{2}}}\right)\right) 
	\label{cod:11}
\end{equation}
Here, $R$ is a reflection coefficient.  $T \rightarrow 0$  is equivalent to sending the horizon to  infinity. Then infalling boundary condition corresponds to $R=0$.   Then  it gives  the   conductivities to be  
\begin{equation}
	\sigma(\omega) =\frac{g\left< \mathcal{O}_{1}\right>}{\omega} \sqrt{\left(1+\frac{i r_{+}}{3 \omega} \right)\left(\frac{\omega}{g\left< \mathcal{O}_{1}\right> } \right)^2-1}  
	\label{cod:12}
\end{equation}
via Eq.(\ref{cod:9}).   

For $\Delta=2$ in Eq.(\ref{cod:10}), We may substitute the trial function
\begin{equation}
	F(z) =\frac{\tanh(1.5bz)}{1.5bz}
	\label{cod:13}
\end{equation}
which is satisfied with Eq.(\ref{cod:4}) and Eq.(\ref{cod:7}) numerically. Also, this trial function obey the correct boundary conditions ($F(0)=1$, $F^{\prime}(0)=1$ and $\lim_{z\rightarrow \infty}F(z) \propto (b z)^{3-2\Delta}$). Here, $b= \frac{\sqrt{g\left< \mathcal{O}_{2}\right>}}{\sqrt{2}r_{+}}$.  
Then at low temperature  Eq.(\ref{cod:10}) reads
\begin{equation}
	\frac{d^2 G }{d z^2} +\frac{2i \hat{\omega} }{3}\frac{d G }{d z} +\left(\frac{8}{9} \hat{\omega}^2 +\frac{i \hat{\omega}}{3} -\frac{16 b^2}{9}\tanh^2(1.5bz) \right)G=0 
	\label{cod:14}
\end{equation}
whose general solution is given in terms of Legendre functions,
\begin{equation} 
	G(z)=  \exp \left(-\frac{i \hat{\omega}z}{3}\right) P_{\nu}^{\mu}\left( \tanh(1.5 bz) \right)+ R\;  \exp \left(-\frac{i \hat{\omega}z}{3}\right) P_{\nu}^{-\mu}\left( \tanh(1.5 bz) \right)
	\label{cod:15}
\end{equation}
where $\nu=\frac{-9+\sqrt{337}}{18}$ and $\mu=\frac{2i\sqrt{\hat{\omega}^2+\frac{i\hat{\omega}}{3}-\left( \frac{4}{3}b\right)^2}}{3b}$. 
Similar to $\Delta=1$, we choose   infalling boundary condition corresponds to $R=0$. 
This exact result then produces the nonzero conductivities
\begin{small}
	\begin{eqnarray}
		\sigma(\omega) &=&\frac{3i\sqrt{g\left< \mathcal{O}_{2}\right>}}{\sqrt{2}  \omega } \frac{\Gamma\left( \frac{1}{36}\left(27-\sqrt{337}-16i\sqrt{P(\omega)} \right)\right)\Gamma\left( \frac{1}{36}\left(27+\sqrt{337}-16i\sqrt{P(\omega)} \right)\right)}{\Gamma\left( \frac{1}{36}\left(9-\sqrt{337}-16i\sqrt{P(\omega)} \right)\right)\Gamma\left( \frac{1}{36}\left(9+\sqrt{337}-16i\sqrt{P(\omega)} \right)\right)}  \nonumber\\
		&=&\frac{3i\sqrt{g\left< \mathcal{O}_{2}\right>}}{  \sqrt{2}\omega } \frac{\Gamma\left(   0.24-\frac{4i}{9}\sqrt{P(\omega)} \right) }{\Gamma\left(  -0.26-\frac{4i}{9}\sqrt{P(\omega)} \right) }  \frac{\Gamma\left(   1.26-\frac{4i}{9}\sqrt{P(\omega)} \right)}{\Gamma\left(   0.76-\frac{4i}{9}\sqrt{P(\omega)} \right)} 
		\label{cod:16}
	\end{eqnarray}
\end{small}
via Eq.(\ref{cod:9}) where
\begin{equation} 
	P(\omega) =     \frac{9}{8}\left(1+\frac{i r_{+}}{3 \omega} \right)\left(\frac{\omega}{\sqrt{g\left< \mathcal{O}_{2}\right>}}\right)^2  -1.
	\nonumber
\end{equation} 
Here we apply the following limiting form:
\begin{equation} 
	\lim_{z\rightarrow 0}  P_{\nu}^{\mu}\left( z \right)= \frac{2^{\mu}\sqrt{2}}{\Gamma\left( 1+\frac{\nu}{2}-\frac{\mu}{2}\right)\Gamma\left( \frac{1}{2}-\frac{\nu}{2}-\frac{\mu}{2}\right)}
	\label{cod:17}
\end{equation}
The solution of  Eq.(\ref{cod:21}) is	
\begin{equation} 
	H_0(z)=   \sqrt{bz} K_{\frac{1}{2\Delta}}\left( b^{\Delta}z^{\Delta}\right)  +R  \sqrt{bz} I_{\frac{1}{2\Delta}}\left( b^{\Delta}z^{\Delta}\right)  
	\label{cod:23}
\end{equation} 
Here, we take $R=0$: The other solution $\sqrt{bz} I_{\frac{1}{2\Delta}}\left( b^{\Delta}z^{\Delta}\right) $ is rejected because it is  is monotonically increasing as $z$ increases for large $b$. By substituting Eq.(\ref{cod:23}), the solution to the field equation Eq.(\ref{cod:22}) for $H_1$ is
\begin{footnotesize} 
	\begin{eqnarray} 
		H_1(z) &=&   \sqrt{bz} K_{\frac{1}{2\Delta}}\left( b^{\Delta}z^{\Delta}\right) \nonumber\\
		&&+\frac{i\sqrt{b}}{6} z^{7/2} \left\{ 2\pi \csc\left(\frac{\pi}{2\Delta}\right)\left(I_{-\frac{1}{2\Delta}}\left( b^{\Delta}z^{\Delta}\right)+I_{\frac{1}{2\Delta}}\left( b^{\Delta}z^{\Delta}\right)    \right) \pFq[4]{2}{3}{\frac{1}{2},\frac{3}{2\Delta}}{1-\frac{1}{2\Delta},1+\frac{1}{2\Delta},1+\frac{3}{2\Delta} }{ b^{2\Delta}z^{2\Delta}} \right. \nonumber\\
		&&- 6\Delta \frac{2^{1/\Delta}}{\sqrt{bz}} \left(\Gamma\left(1+\frac{1}{2\Delta}\right)\right)^2 I_{\frac{1}{2\Delta}}\left( b^{\Delta}z^{\Delta}\right) \pFq[4]{2}{3}{\frac{1}{2}-\frac{1}{2\Delta},\frac{1}{\Delta}}{1-\frac{1}{\Delta},1-\frac{1}{2\Delta},1+\frac{1}{\Delta} }{ b^{2\Delta}z^{2\Delta}}   \nonumber\\
		&&\left.-3\Delta \frac{\sqrt{bz} }{2^{1/\Delta}}  \left(\Gamma\left(1-\frac{1}{2\Delta}\right)\right)^2 I_{-\frac{1}{2\Delta}}\left( b^{\Delta}z^{\Delta}\right) \pFq[4]{2}{3}{\frac{1}{2}+\frac{1}{2\Delta},\frac{2}{\Delta}}{1+\frac{1}{\Delta},1+\frac{1}{2\Delta},1+\frac{2}{\Delta} }{ b^{2\Delta}z^{2\Delta}}  \right\}
		\label{cod:24}
	\end{eqnarray} 
\end{footnotesize} 	
Eq.(\ref{cod:23}) and Eq.(\ref{cod:24}) give us the nonzero conductivities
\begin{equation}
	\lim_{\omega \rightarrow 0}	\sigma(\omega) = \frac{2\pi \Delta \csc\left( \frac{\pi}{2\Delta} \right)}{(2\Delta)^{1/\Delta} \left(\Gamma\left(\frac{1}{2\Delta}\right)\right)^2}\frac{g^{1/\Delta}\left< \mathcal{O}_{\Delta}\right>^{1/\Delta}}{\omega}i
	\label{cod:25}
\end{equation}	
And we obtain	
\begin{equation}
	\frac{n_s}{T_c} = \frac{2\pi \Delta \csc\left( \frac{\pi}{2\Delta} \right)}{(2\Delta)^{1/\Delta} \left(\Gamma\left(\frac{1}{2\Delta}\right)\right)^2}\frac{g^{1/\Delta}\left< \mathcal{O}_{\Delta}\right>^{1/\Delta}}{T_c} 
	\label{cod:26}
\end{equation}	  
here, $n_s$ is also the coefficient of the pole in the imaginary part $\Im{\sigma(\omega)}\sim n_s/\omega $ as
$\omega \rightarrow 0$.

\subsection{ The Resonant Frequencies}

There is a maximum of $z_0$ at $\Delta =3/2$ and the resonance, by which $\sigma(\omega)$ diverges, occurs only  in the vicinity of  $\Delta =3/2$.   This can be understood using standard WKB matching formula.  The resonance occurs when there exists $\omega$ satisfying \cite{Horo2011} 
\begin{equation}
	\int_{r_{*0}}^{0}\sqrt{\omega^2- V(r_{\star})} dr_{\star} +\frac{\pi}{4} =n\pi ,
	\nonumber
\end{equation}
for an integer $n$ and  $r_{*0}<0$ is the position  at which $V$ has the maximum:
$	\frac{d V}{d r_{\star}}( r_{*0}) =0$. 
The above   equation can be converted to $z$ coordinate  to give the following expression:
\begin{equation}
	\frac{1}{r_{+}}\int_{0}^{z_0   }\frac{\sqrt{\omega^2-V(z)}}{1-z^3} dz  =\left(n-\frac{1}{4}\right)\pi .
	\nonumber
\end{equation} 
At $\Delta=3/2$,    we have   
\begin{eqnarray} 
	\int_{1/b}^{z_0} \sqrt{\left(\frac{\omega}{T_c}\right)^2- \left(\frac{2\pi}{3}\frac{T}{T_c}\right)^2 b^3 z \left(\frac{ 4-3\ln z}{2+\ln b}\right)^2}   dz   
	=\frac{4\pi^2}{3}\left(n-\frac{1}{4}\right)\frac{T}{T_c} ,
	\label{co3d:1}
\end{eqnarray}	
where $z_0 = 0.362$ from Eq.(\ref{codd:31}), and 
\begin{eqnarray}
	b &=& 1.23\left(1+0.45\ln\left(\frac{T_c}{T}\right)\right)^{1/4} \frac{T_c}{T}, 
	\nonumber\\
	\frac{\omega_{g}}{T_c} &=& \frac{7}{10}\frac{ {X^{3/2} \left(\frac{T_c}{T}\right)^{1/2}}}{\ln \left(X \frac{T_c}{T}\right)}, \quad 
	\hbox{ with } X=\frac{g^{1/\Delta}\left< \mathcal{O}_{\Delta}\right>^{1/\Delta}}{T_c}.\nonumber 
\end{eqnarray}	
Resonant $\omega_i$'s  exist only when $z_0$ is large enough. We can see  that $z_0$ is maximum at $\Delta=3/2$ from the Fig \ref{z.zero}(b). It turns out that only near the $\Delta=3/2$ because for other values which is much bigger or smaller than $z_{0}$,  the barrier is too thick for the resonance to happen. 	   
For $\frac{T_c}{T}=0.1$,  we have  $\frac{\omega_1}{T_c}=10.44$  which is in good agreement with the $ {\omega_1}/{T_c}=10.4$ \cite{Horo2009} if we set $g=1$. 
In general, as $ {T}/{T_c}  $ decreases, the number of poles increases. 	
These results are summarized in the Table \ref{poles}.  For derivation of these results, see the appendix \ref{appendix4}.   

\begin{table}[!htb]
	\begin{tabular}{|l||l|l|l|l|l|}
		\hline
		\textbf{ Poles of $\sigma(\omega)$} & $\frac{\omega_1}{T_c}$     &  $\frac{\omega_2}{T_c}$     &  $\frac{\omega_3}{T_c}$     &  $\frac{\omega_4}{T_c}$    &  $\frac{\omega_5}{T_c}$       \\ \hline\hline
		$\frac {T}{T_c}=0.1 $ with $\omega_{g}/T_c=10.9$      & 10.44 &       &       &       &          \\ \hline
		$\frac {T}{T_c}=0.05 $ with $\omega_{g}/T_c=13.4$       & 11.85 & 13.11 &       &       &               \\ \hline
		$\frac {T}{T_c}=0.04 $  with $\omega_{g}/T_c=14.62$     & 12.19 & 13.65 & 14.4  &       &              \\ \hline
		$\frac {T}{T_c}=0.03 $   with $\omega_{g}/T_c=16.34$     & 12.6  & 14.26 & 15.21 & 15.84 & 16.25      \\ \hline
	\end{tabular}
	\caption{  $\frac{\omega_n}{T_c}$ and $\frac {T}{T_c}$  at  lower $T $'s. where $n=1,2,3,\cdots$. A position of the pole is obtained from  Eq.(\ref{co3d:1}) with given $\frac  {T}{T_c}$ by applying Mathematica program. } \label{poles}
\end{table}

\subsubsection{Expression for the schr\"odinger wave equation of the conductivity at near the zero temperature}\label{appendix4}	
Eq.(\ref{cod:1}) takes the form of 	a  schr\"odinger equation with energy $\omega$: 
\begin{equation}
	-\frac{d^2 A_x }{d r_{\star}^2} + V(r_{\star}) A_x=\omega^2 A_x.
	\nonumber
\end{equation}	 
Here,  $ V(r_{\star})$ is re-expression of $V(z)=   \frac{g^2 \left< \mathcal{O}_{\Delta}\right>^2}{r_{+}^{2\Delta-2}}(1-z^3) z^{2\Delta-2} F(z)^2$ in terms of the tortoise coordinate $r_{\star}$, 
\begin{equation}
	r_{\star}= \int \frac{dr}{f(r)}=\frac{1}{6 r_{+}}\left[\ln\frac{(1-z)^3}{1-z^3}-2\sqrt{3}\tan^{-1}\frac{\sqrt{3}\;z}{2+z}\right], 
	\nonumber
\end{equation}	
where  the integration constant is chosen such that boundary is at $r_{\star} =0$.	
%
%
\begin{figure}[h]
	\centering
	\includegraphics[width=0.5\linewidth]{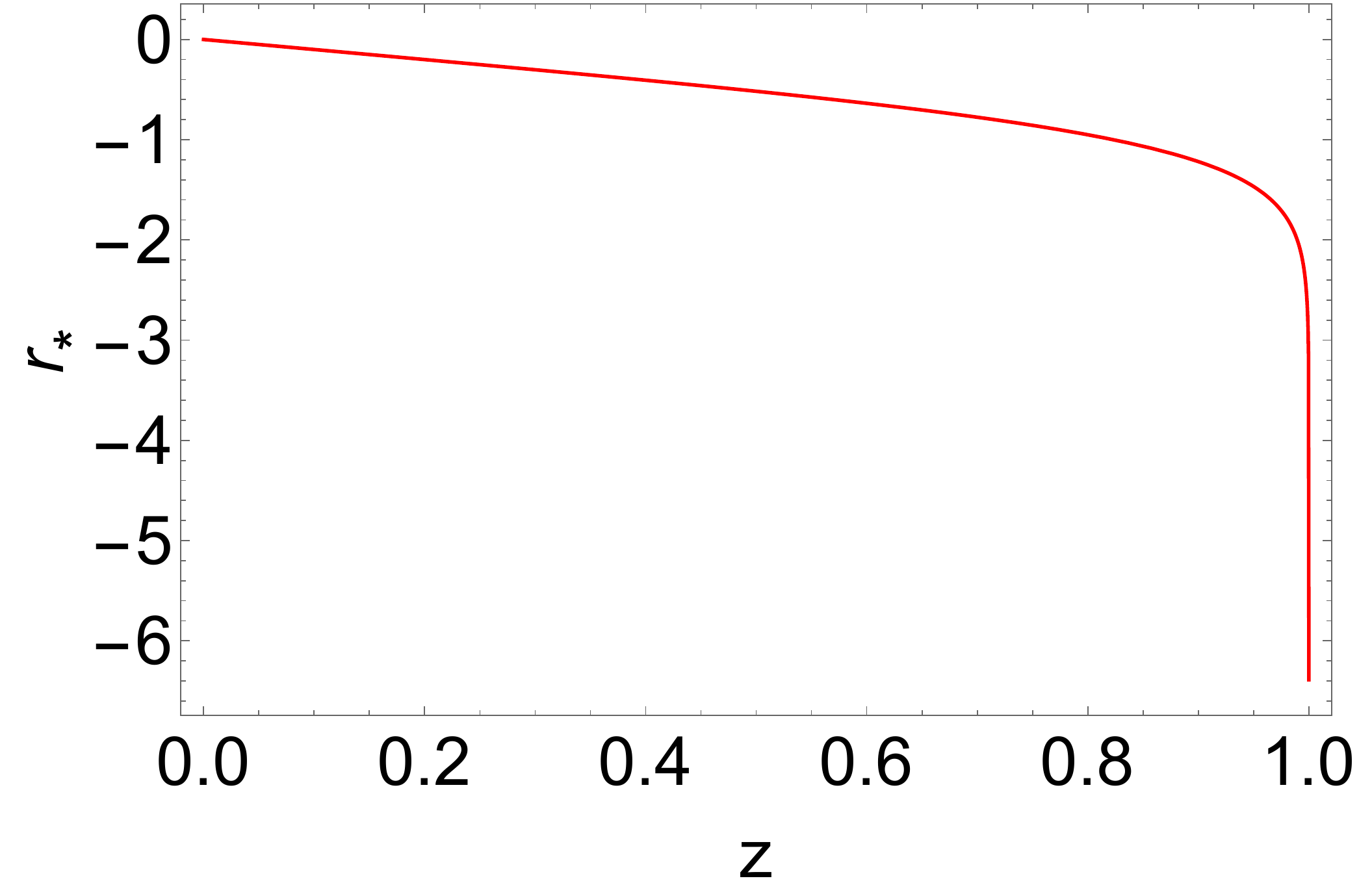}
	\caption{  $r_{\star}$  vs $z$ Here, $r_{+}=1$ for simplicity.   }		\label{r.star}
\end{figure}  

\begin{figure}[ht!]
	\centering
	\subfigure[ $V(r_{\star})$  vs $r_{\star}$]
	{\includegraphics[width=0.48\linewidth]{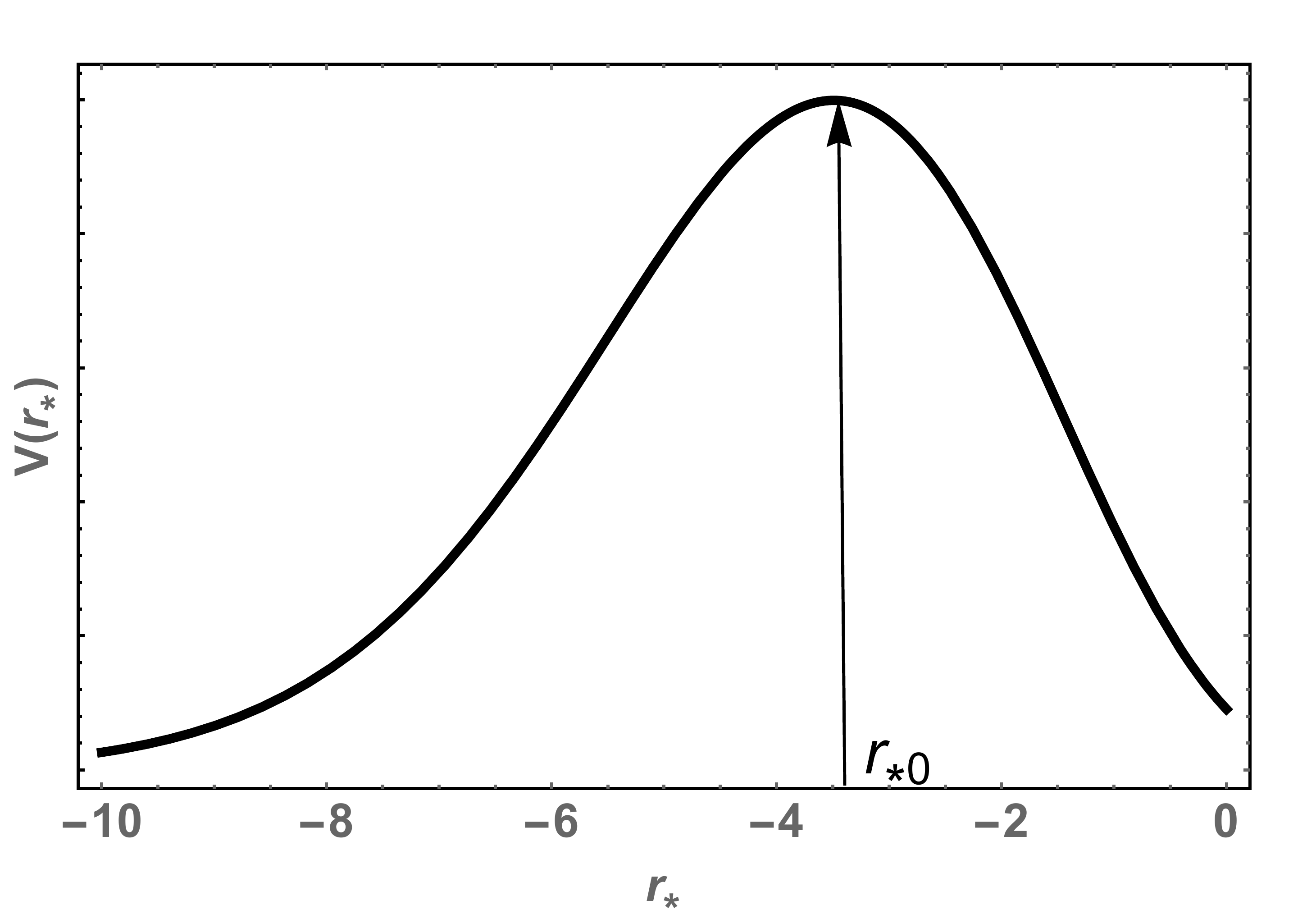}}
	\subfigure[$V(z)$  vs $z$. ]
	{\includegraphics[width=0.48\linewidth]{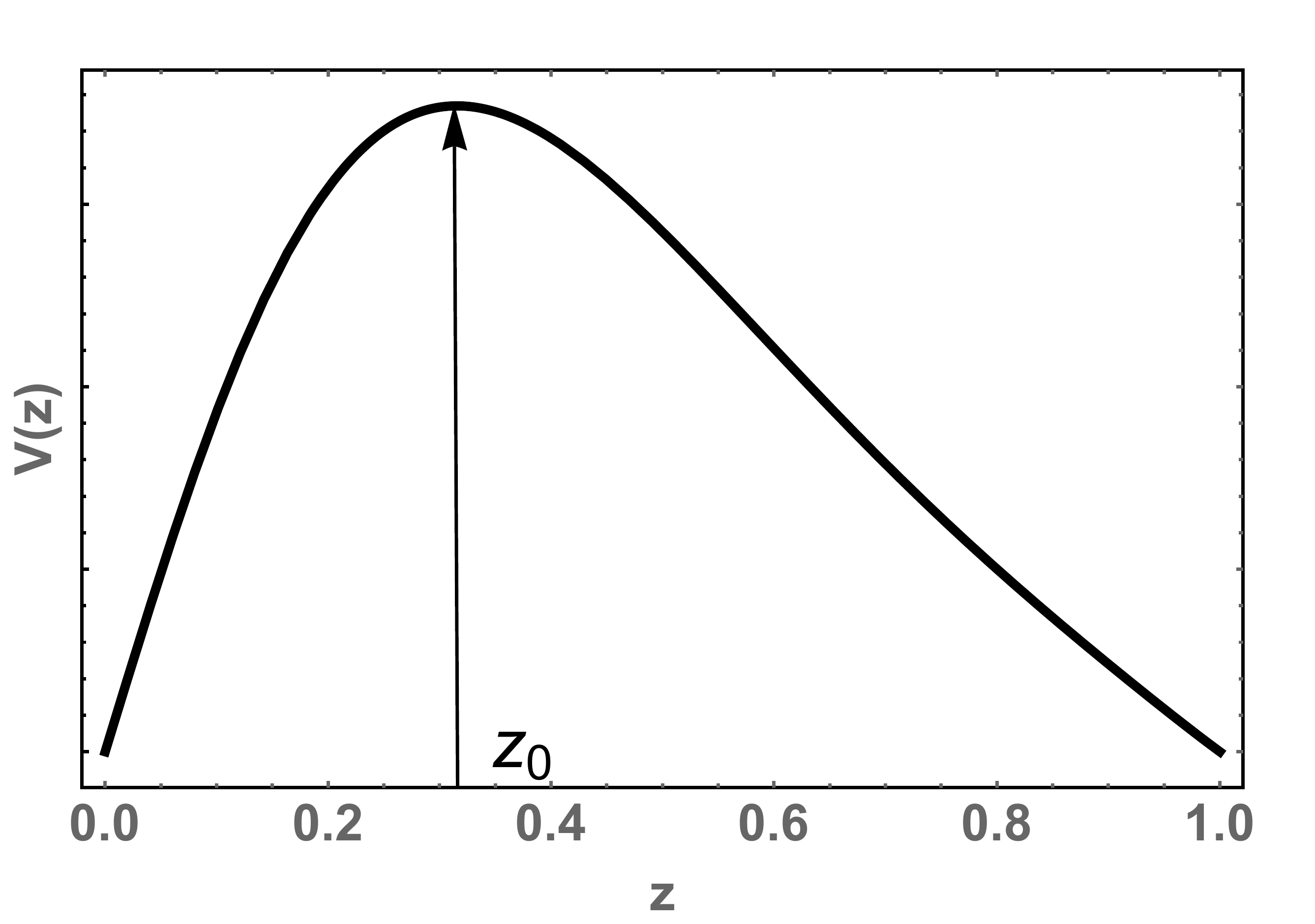}}
	\caption{\small
		(a)  a rough picture $V(r_{\star})$ in terms of a $r_{\star}$ coordinate,
		(b) a rough picture $V(z)$ in terms of a $z$ coordinate.}
	\label{V.rz}
\end{figure}

FIG. \ref{r.star}. shows that the horizon corresponds to $r_{\star}=-\infty$.
We can easily show	that $V(r_{\star}= 0) = 0 $ if $\Delta>1$, $V(r_{\star}  = 0) $ is a nonzero constant if  $\Delta=1$, and $V(r_{\star})$   diverges as $r_{\star}  \rightarrow 0$ if $1/2 < \Delta < 1$. FIG. \ref{V.rz}. can  show that $V(z)$ always vanishes at the horizon (or $V(r_{\star})$   vanishes at $r_{\star} \rightarrow -\infty$).
The maximum value of $V(r_{\star})$ (or $V(z)$ ) always exists at $r_{\star}=r_{*0}$ (or $z=z_0$) if $  \Delta \geq 1$.   	
As we substitute  Eq.(\ref{cod:7}) into Eq.(\ref{cod:29}), we obtain a polynomial equation such as	  
\begin{equation}
	z_0^{2\Delta} (3-\Delta)^2(2+z_0^3+2\Delta(z_0^3-1)) +\Delta^2 z_0^3(4-2\Delta+(2\Delta-7)z_0^3)  =0.
	\label{cod:30}
\end{equation}		 
And its numerical solution is 
\begin{equation}
	z_0 \approx 0.41249 \sum_{k=1}^{\infty}\frac{\sin\left(\pi(\Delta-1)(2k-1)\right)}{k^{2.6376}}
	\label{cod:31}
\end{equation}
where $1\leq\Delta <2$. A dashed curve at $1\leq\Delta <2$ in FIG. \ref{z.zero} (a). indicates  Eq.(\ref{cod:31}), and we see that there are no $b $ (or $T$) dependence.  

As we substitute  Eq.(\ref{cod:4}) into Eq.(\ref{cod:29}), we obtain a polynomial equation, and we see $z_0 \propto 1/b$. Its numerical solution is	 
\begin{equation}
	z_0 \approx \left(\frac{0.1}{\Delta-1.83}+0.7\right) \frac{1}{b}
	\label{cod:32}
\end{equation}	   
where  $2\leq\Delta \leq 3$.	  
A dashed curve at $2\leq\Delta \leq 3$ in FIG. \ref{z.zero}. indicates  Eq.(\ref{cod:32}), and we see that there are  $b $  dependence.  

And $\omega_g$ is given by
\begin{equation}
	\omega_g = \sqrt{V_{\mbox{max}}} = \frac{g \left< \mathcal{O}_{\Delta}\right> }{r_{+}^{ \Delta-1}}\sqrt{1-z_0^3} z_0^{ \Delta-1} F(z_0) \approx \frac{g \left< \mathcal{O}_{\Delta}\right> }{r_{+}^{ \Delta-1}}  z_0^{ \Delta-1} F(z_0).
	\label{cod:33}
\end{equation} 
We obtain the analytic expressions of $	\frac{\omega_g}{T_c}$ in the following way: 
\bea
\frac{\omega_g}{T_c} &&=   \left(\frac{3z_0}{4\pi}\frac{T_c}{T}\right)^{\Delta-1}\left(\frac{g^{1/\Delta} \left< \mathcal{O}_{\Delta}\right>^{1/\Delta} }{T_c}\right)^{\Delta} F_{<}(z_0),  \;\;   
\hbox{ for } 1\leq\Delta <2
\label{cod:34},
\\
\frac{\omega_g}{T_c} &&=  \left(\Delta^{1/\Delta }\rho_{\Delta}\right)^{\Delta -1}\frac{g^{1/\Delta} \left< \mathcal{O}_{\Delta}\right>^{1/\Delta} }{T_c}  F_{>}(\rho_{\Delta}) , \;\;  
\hbox{ for } 2\leq\Delta \leq3,
\label{cod:36}
\eea 
where
\begin{small}
	\bea
	F_{<}(z_0)  &&=  1- \frac{-\frac{\sqrt{\pi } \Gamma \left(\frac{3}{2 \Delta }-1\right) \Gamma \left(\frac{1}{\Delta }\right) \Gamma \left(\frac{2}{\Delta }\right)}{8 \Delta ^2 \Gamma \left(\frac{\Delta +3}{2 \Delta }\right)}+ \frac{\pi  \csc \left(\frac{\pi }{2 \Delta }\right)}{4 \Delta ^2 (3-2 \Delta )}(b z_0)^{3-2\Delta}}{-\frac{\sqrt{\pi } \Gamma \left(\frac{3}{2 \Delta }-1\right) \Gamma \left(\frac{1}{\Delta }\right) \Gamma \left(\frac{2}{\Delta }\right)}{8 \Delta ^2 \Gamma \left(\frac{\Delta +3}{2 \Delta }\right)}+ \frac{\pi  (3-\Delta )^2 \csc \left(\frac{\pi }{2 \Delta }\right)}{4 \Delta ^4 (3-2 \Delta )}b^{3-2\Delta}}, 
	\label{cod:35}	\\
	F_{>}(\rho_{\Delta})  &&= 1+\frac{\Delta^2 \rho_{\Delta}^2 \Gamma\left(\frac{3+\Delta}{2\Delta}\right) \left[\rho_{\Delta}\mathbb{M}_1+\rho_{\Delta}^2	\mathbb{M}_2 +  \mathbb{M}_3\right] }{6\sqrt{\pi}\Gamma\left(\frac{1}{\Delta}\right)\Gamma\left(\frac{2}{\Delta}\right)\Gamma\left(-1+\frac{3}{2\Delta}\right)},
	\label{cod:37}
	\eea
\end{small}	 
with
\begin{eqnarray}
	\mathbb{M}_1 &=& 8\pi   \csc\left(\frac{\pi}{2\Delta}\right)\left( -1+\frac{3 \Delta^2 \rho_{\Delta}^{2\Delta} }{3-2\Delta(4\Delta^2+6\Delta-1)} \right),\nonumber\\  
	\mathbb{M}_2 &=& \frac{3\left(\Gamma\left(\frac{-1}{2\Delta}\right)\right)^2}{2^{1/\Delta}}\left( \frac{2+\Delta(9+4\Delta+2\rho_{\Delta}^{2\Delta})}{(2+\Delta)(1+2\Delta)(1+4\Delta)}\right),\nonumber\\  
	\mathbb{M}_3 &=& 6 \left(\Gamma\left(\frac{1}{2\Delta}\right)\right)^2 2^{1/\Delta}\left( \frac{4\Delta^2-1+\Delta(3+ \rho_{\Delta}^{2\Delta})}{8\Delta^3+2\Delta^2-5\Delta +1}\right),\nonumber\\  
	\rho_{\Delta} &=& \left(\frac{0.1}{\Delta-1.83}+0.7\right).  	\label{cod:38} 
\end{eqnarray}   

\begin{figure}[h]
	\centering
	\includegraphics[width=0.5\linewidth]{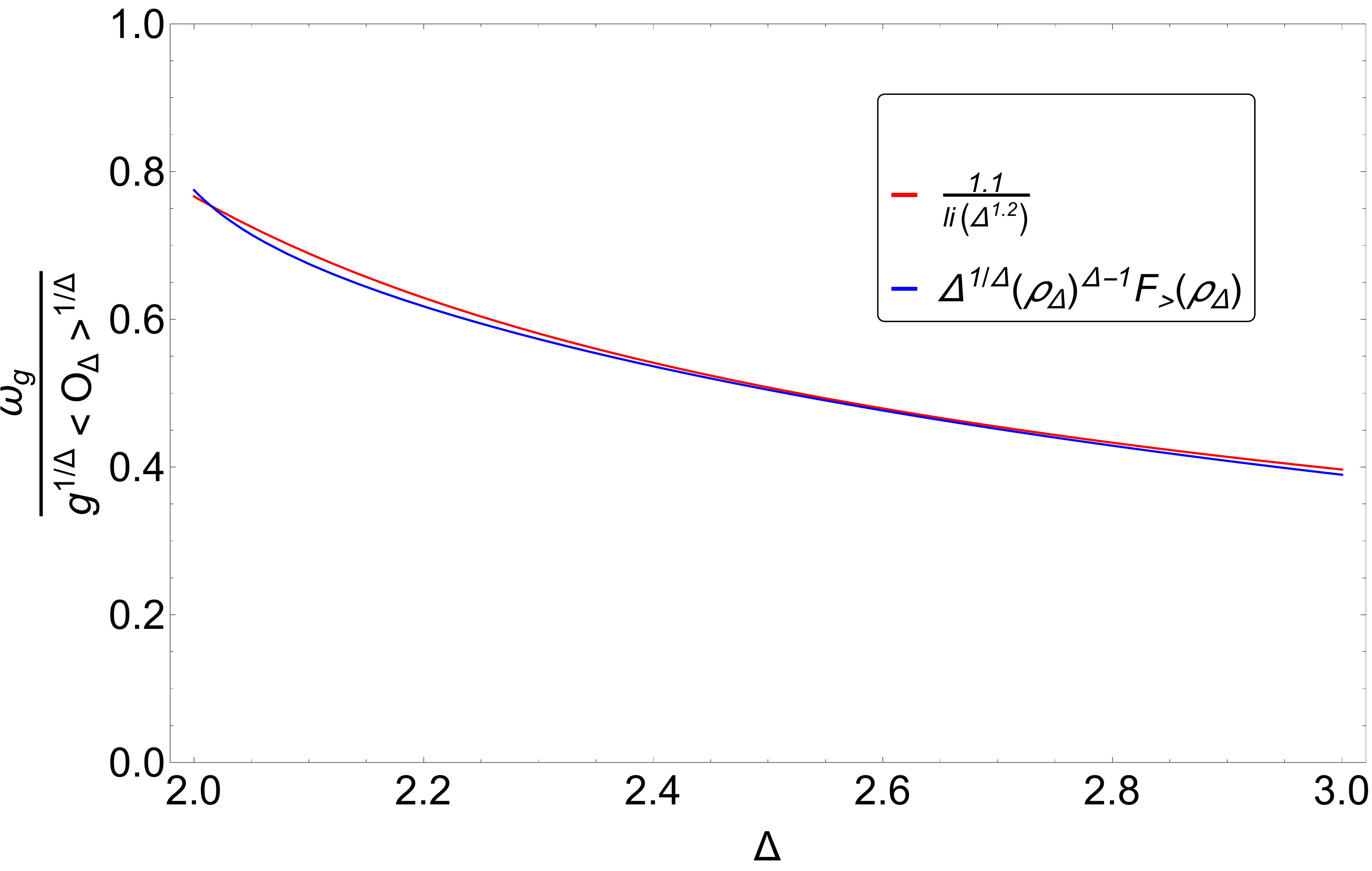}
	\caption{ Solid red curve is a plot of $\frac{1.1}{\mbox{li}(\Delta^{1.2})} $  and blue one is  a plot of $\left(\Delta^{1/\Delta }\rho_{\Delta}\right)^{\Delta -1}F_{>}(\rho_{\Delta})$ with Eq.(\ref{cod:37}) in Eq.(\ref{cod:36}) (almost indistinguishable).  }		
	\label{li.function}
\end{figure}

Substitute Eq.(\ref{cod:7}) with Eq.(\ref{cod:31}) into Eq.(\ref{cod:33}) and we obtain Eq.(\ref{cod:34}). 
Also, substitute Eq.(\ref{cod:4}) with Eq.(\ref{cod:32}) into Eq.(\ref{cod:33}) and we obtain Eq.(\ref{cod:36}). 
A numerical result tells us that Eq.(\ref{cod:36}) approximately is
\begin{equation}
	\frac{\omega_g}{T_c} \approx  \frac{1.1}{\mbox{li}(\Delta^{1.2})}\frac{g^{1/\Delta} \left< \mathcal{O}_{\Delta}\right>^{1/\Delta} }{T_c},   
	\label{cod:39}
\end{equation}
here, $\mbox{li} (x)$ is an logarithmic integral function. See Fig.\ref{li.function}.

And we can classify Eq.(\ref{cod:34}) into the following way:
\begin{enumerate}
	\item
	As $1\leq \Delta \ll 3/2$,
	\begin{equation}
		\frac{\omega_g}{T_c} = \left(\frac{3z_0}{4\pi}\frac{T_c}{T}\right)^{\Delta-1} \left( 1-\left(\frac{\Delta}{3-\Delta} z_0^{3/2-\Delta}\right)^2\right) X^{\Delta}
		\label{cod:40}
	\end{equation}
	\item
	As $  \Delta = 3/2$,
	\begin{equation}
		\frac{\omega_g}{T_c} = \frac{7}{10}  
		\frac{ {X^{3/2} \left(\frac{T_c}{T}\right)^{1/2}}}{\ln \left(X \frac{T_c}{T}\right)}
		\label{cod:41}
	\end{equation}
	\item
	As $3/2\ll \Delta <2$,
	\begin{footnotesize} 
		\begin{equation}
			\frac{\omega_g}{T_c} = \left(\frac{3z_0}{4\pi}\frac{T_c}{T}\right)^{2-\Delta} \frac{\sqrt{\pi} \Gamma\left(\frac{3+\Delta}{2+\Delta}\right)\csc\left(\frac{\pi}{2\Delta}\right)}{\Delta^{3/\Delta}\Gamma\left(\frac{2}{\Delta}\right)\Gamma\left(\frac{3}{2\Delta}\right)\Gamma\left(1+\frac{1}{\Delta}\right)}\left( 1-\left(\frac{\Delta}{3-\Delta} z_0^{\Delta-3/2}\right)^2\right) X^{3-\Delta}
			\label{cod:42}
		\end{equation}
	\end{footnotesize} 
\end{enumerate}
Here, $X=\frac{g^{1/\Delta} \left< \mathcal{O}_{\Delta}\right>^{1/\Delta} }{T_c}$.

\begin{figure}[h]
	\centering
	\includegraphics[width=0.5\linewidth]{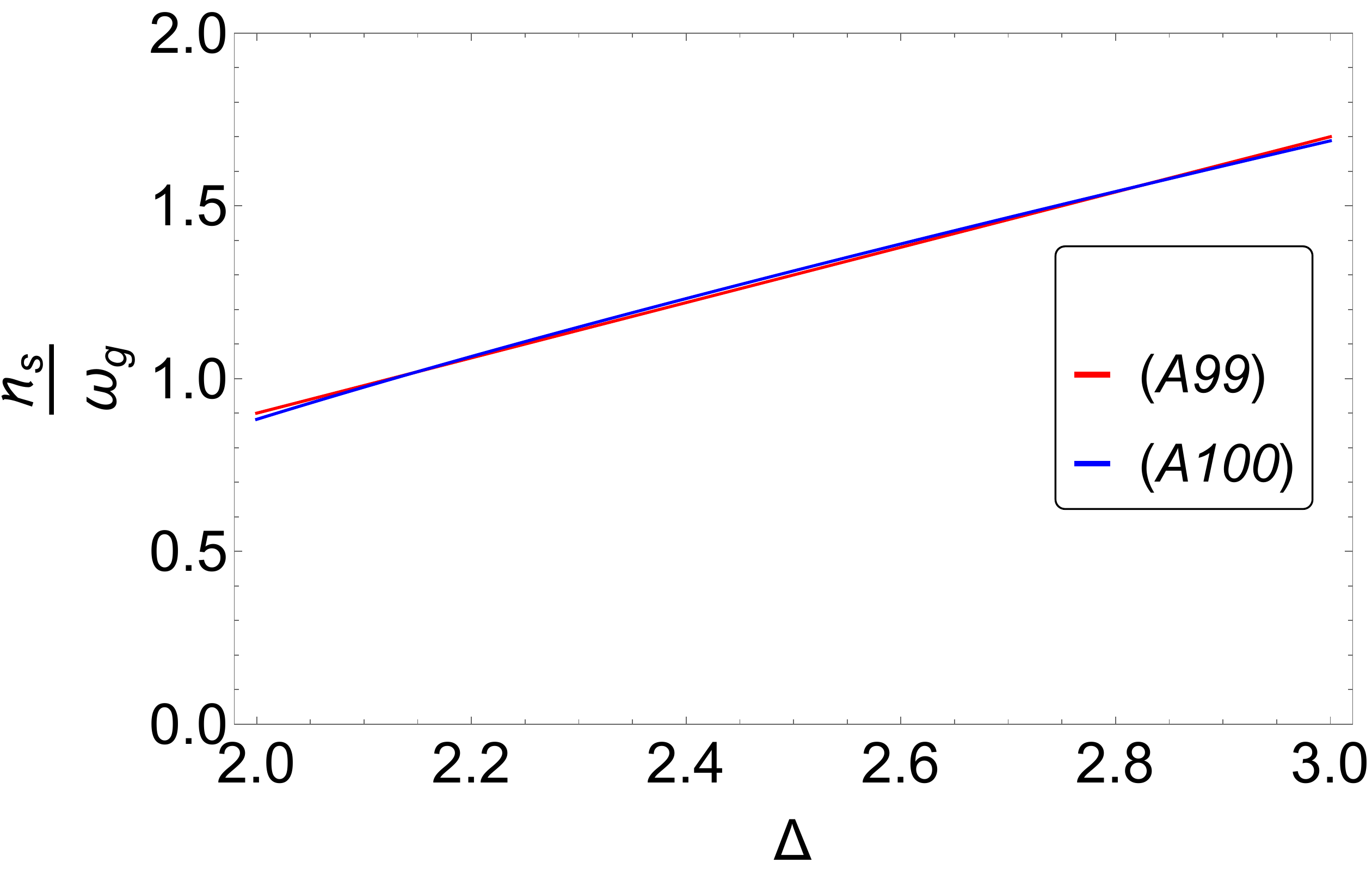}
	\caption{ Solid red curve is a plot of Eq.(\ref{cod:46})  and blue one is  a plot of Eq.(\ref{cod:47}) (almost indistinguishable).  }		
	\label{ns.wg}
\end{figure}

From Eq.(\ref{cod:26}), Eq.(\ref{cod:40}), Eq.(\ref{cod:41}) and Eq.(\ref{cod:42}) , we find a relation between $n_s$ and the gap frequency $\omega_g$:
\begin{enumerate}
	\item
	As $1\leq \Delta \ll 3/2$,
	\begin{equation}
		\frac{n_s}{\omega_g} =  \frac{2\pi \Delta \csc\left(\frac{\pi}{2\Delta}\right)}{(2\Delta)^{1/\Delta}\left(\Gamma\left(\frac{1}{2\Delta}\right)\right)^2}\frac{1}{ 1-\left(\frac{\Delta}{3-\Delta} z_0^{3/2-\Delta}\right)^2} \left(\frac{3z_0}{4\pi} X\right)^{1-\Delta} 
		\label{cod:43}
	\end{equation}
	\item
	As $  \Delta = 3/2$,
	\begin{equation}
		\frac{n_s}{\omega_g}= \frac{\ln \left(X \frac{T_c}{T}\right)}{\sqrt{X \frac{T_c}{T}}} 
		\label{cod:44}
	\end{equation}
	\item
	As $3/2\ll \Delta <2$,
	\begin{footnotesize} 
		\begin{equation}
			\frac{n_s}{\omega_g} = \frac{2\sqrt{\pi} \Delta^{1+3/\Delta}  }{(2\Delta)^{1/\Delta}\left(\Gamma\left(\frac{1}{2\Delta}\right)\right)^2}\frac{\Gamma\left( \frac{2}{\Delta}\right)\Gamma\left( \frac{3}{2\Delta}\right)\Gamma\left( 1+\frac{1}{\Delta}\right)}{\Gamma\left( \frac{3+\Delta}{2\Delta}\right)}\frac{1}{ 1-\left(\frac{\Delta}{3-\Delta} z_0^{\Delta-3/2}\right)^2} \left(\frac{3z_0}{4\pi} X\right)^{ \Delta-2} 
			\label{cod:45}
		\end{equation}
	\end{footnotesize} 
	\item
	As $2\leq \Delta \leq3$,
	\begin{equation}
		\frac{n_s}{\omega_g} = \frac{2\pi \Delta \csc\left(\frac{\pi}{2\Delta} \right)}{1.1 (2\Delta)^{1/\Delta}\left( \Gamma\left(\frac{1}{2\Delta}\right)\right)^2}  \mbox{li}(\Delta^{1.2})  
		\label{cod:46}
	\end{equation}
\end{enumerate} 
A numerical result	 tells us that Eq.(\ref{cod:46}) is approximately
\begin{equation}
	\frac{n_s}{\omega_g}   \approx 0.8\Delta -0.7,
	\label{cod:47}
\end{equation}
here, $z_0$ is 	 Eq.(\ref{cod:31}). See Fig.\ref{ns.wg}.

As we see Fig 2. \cite{Horo2009}, $\sigma(\omega)$ has a spike at $\frac{\omega}{T_c}\approx 10.4$ and $\Delta= 3/2$ for $A\lowercase{d}S_{4}$. In FIG. \ref{V.rz} (b), resonance occurs well when the distance between $z=0$ and $z=z_0$ is the maximum. 
FIG. \ref{z.zero} (a) shows that there is a maximum of $z_0$ at $\Delta =3/2$. So, the resonance, by which $\sigma(\omega)$ diverges, occurs only  in the vicinity of  $\Delta =3/2$. This can be understood using standard WKB matching formula: The resonance occurs when there exists $\omega$ satisfying \cite{Horo2011} 
\begin{equation}
	\int_{r_{*0}}^{0}\sqrt{\omega^2-V(r_{\star})} dr_{\star} +\frac{\pi}{4}=n\pi,
	\label{cod:48}
\end{equation}
for an integer $n$ and  $r_{*0}<0$ is the position  at which $V$ has the maximum:
$	\frac{d V}{d r_{\star}}( r_{*0}) =0$. 
The above   equation can be converted to $z$ coordinate  to give the following expression: 
\begin{equation}
	\frac{1}{r_{+}}\int_{0}^{z_0}\frac{\sqrt{\omega^2-V(z)}}{1-z^3} dz  =\left(n-\frac{1}{4}\right)\pi.
	\label{cod:49}
\end{equation}	
By applying Eq.(\ref{cod:7}) into Eq.(\ref{cod:49}), we obtain	 
\begin{eqnarray}
	&&	\frac{1}{r_{+}}\int_{0}^{z_0}\frac{\sqrt{\omega^2-V(z)}}{1-z^3} dz  \approx  \frac{1}{r_{+}}\int_{1/b}^{z_0}\frac{\sqrt{\omega^2- \left(\frac{g \left< \mathcal{O}_{\Delta}\right>}{r_{+}^{\Delta-1}}\right)^2 z^{2\Delta-2}(1-z^3)\left(\frac{(\beta-\beta_0 z^{3-2\Delta})b^{3-2\Delta}}{\alpha+\beta b^{3-2\Delta}}\right)^2}}{1-z^3} dz\nonumber\\
	& \approx& \frac{1}{r_{+}}\int_{1/b}^{z_0} \sqrt{\omega^2- \left(\frac{g \left< \mathcal{O}_{\Delta}\right>}{r_{+}^{\Delta-1}}\right)^2 z^{2\Delta-2} \left(\frac{(\beta-\beta_0 z^{3-2\Delta})b^{3-2\Delta}}{\alpha+\beta b^{3-2\Delta}}\right)^2}   dz=\left(n-\frac{1}{4}\right)\pi,
	\label{cod:50}
\end{eqnarray}	
where	
\begin{eqnarray}
	\alpha &=& -\frac{\sqrt{\pi } \Gamma \left(\frac{3}{2 \Delta }-1\right) \Gamma \left(\frac{1}{\Delta }\right) \Gamma \left(\frac{2}{\Delta }\right)}{8 \Delta ^2 \Gamma \left(\frac{\Delta +3}{2 \Delta }\right)},  \nonumber\\  
	\beta &=&  \frac{\pi  (3-\Delta )^2 \csc \left(\frac{\pi }{2 \Delta }\right)}{4 \Delta ^4 (3-2 \Delta )},\nonumber\\  
	\beta_{0} &=&  \frac{\pi   \csc \left(\frac{\pi }{2 \Delta }\right)}{4 \Delta ^2 (3-2 \Delta )}.
	\label{cod:51} 
\end{eqnarray}  	
At $\Delta=3/2$,   Eq.(\ref{cod:50}) becomes
\begin{equation}
	\int_{1/b}^{z_0} \sqrt{\left(\frac{\omega}{T_c}\right)^2- \left(\frac{2\pi}{3}\frac{T}{T_c}\right)^2 b^3 z \left(\frac{ 4-3\ln z}{2+\ln b}\right)^2}   dz=\frac{4\pi^2}{3}\left(n-\frac{1}{4}\right)\frac{T}{T_c}, 
	\label{cod:52}
\end{equation}	
where $z_0 = 0.362$ from Eq.(\ref{codd:31}), and 
\begin{eqnarray}
	b &=& 1.23\left(1+0.45\ln\left(\frac{T_c}{T}\right)\right)^{1/4} \frac{T_c}{T},
	\nonumber\\
	\frac{\omega_{g}}{T_c} &=& \frac{7}{10}\frac{ {X^{3/2} \left(\frac{T_c}{T}\right)^{1/2}}}{\ln \left(X \frac{T_c}{T}\right)}, \quad 
	\hbox{ with } X=\frac{g^{1/\Delta}\left< \mathcal{O}_{\Delta}\right>^{1/\Delta}}{T_c}.	\label{cod:53}
\end{eqnarray}
Here, 	 Eq.(\ref{cod:53}) is derived from  Eq.(\ref{oo:62}). 
\begin{table}[!htb]
	\scalebox{0.9}{	\begin{tabular}{|l||l|l|l|l|l|l|l|l|l|}
			\hline
			\textbf{ Poles of $\sigma(\omega)$} & $\frac{\omega_1}{T_c}$     &  $\frac{\omega_2}{T_c}$     &  $\frac{\omega_3}{T_c}$     &  $\frac{\omega_4}{T_c}$    &  $\frac{\omega_5}{T_c}$     &  $\frac{\omega_6}{T_c}$   &  $\frac{\omega_7}{T_c}$    &  $\frac{\omega_8}{T_c}$     &  $\frac{\omega_9}{T_c}$     \\ \hline\hline
			$\frac{T}{T_c}=0.1 $ with $\omega_g/T_c=10.9$      & 10.44 &       &       &       &       &      &      &       &       \\ \hline
			$\frac{T}{T_c}=0.05 $ with $\omega_g/T_c=13.4$       & 11.85 & 13.11 &       &       &       &      &      &       &       \\ \hline
			$\frac{T}{T_c}=0.04 $  with $\omega_g/T_c=14.62$     & 12.19 & 13.65 & 14.4  &       &       &      &      &       &       \\ \hline
			$\frac{T}{T_c}=0.03 $   with $\omega_g/T_c=16.34$     & 12.6  & 14.26 & 15.21 & 15.84 & 16.25 &      &      &       &       \\ \hline
			$\frac{T}{T_c}=0.02 $  with $\omega_g/T_c=19.18$      & 13.11 & 15.0  & 16.15 & 16.97 & 17.6  & 18.1 & 18.5 & 18.82 & 19.07 \\ \hline
	\end{tabular}}
	\caption{ $\frac{\omega_n}{T_c}$ and $\frac {T}{T_c}$  at  lower $T $'s. where $n=1,2,3,\cdots$. A position of the pole is obtained from  Eq.(\ref{cod:52}) with given $\frac  {T}{T_c}$ by applying Mathematica program. } \label{pole}
\end{table}  

\section{   H\lowercase{olographic superconductors with} A\lowercase{d}S\lowercase{$_{5}$}}
\subsection{Near the critical temperature}\label{Tc.v1}
\subsubsection{ Computation of $T_c$ by applying matrix algorithm and Pincherle’s Theorem  } \label{hahaha}
At the critical temperature $T_c$, $\Psi =0$, so  Eq.(\ref{eq:3}) tells us $\Phi^{''}=0$. Then, we can set
\begin{equation}
	\Phi(z)= \lambda_4 r_c (1-x) \hspace{1cm}\mbox{where}\;\;\lambda_4 =\frac{\rho}{r_c^3}
	\label{qq:9}
\end{equation}
where $x=z^2$. As $T\rightarrow T_c$, the field equation $\Psi$ approaches to
\begin{equation}
	-\frac{d^2 \Psi }{d x^2} +\frac{1+x^2}{x(1-x^2)}\frac{d \Psi }{d x}+  \frac{m^2}{4 x^2(1-x^2)} \Psi =\frac{\lambda_{g,4}^2}{4x(1+x)^2}\Psi
	\label{qq:10}
\end{equation}
where $\lambda_{g,4}= g\lambda_4$. Factoring out the behavior near the boundary $z=0$ and the horizon, we define
\begin{equation}
	\Psi(x)= \frac{\left< \mathcal{O}_{\Delta}\right>}{\sqrt{2}r_h^{\Delta}}x^{\frac{\Delta}{2}}F(x)  \hspace{1cm}\mbox{where}\;\;F(x)=
	(1+x)^{-\lambda_{g,4}/2}y(x)
	\label{qq:11}
\end{equation}
Then,  $F$ is normalized as $F(0)=1$ and  we obtain
\begin{equation}
	\frac{d^2 y }{d x^2} + \left( \frac{\Delta-1}{x}+\frac{1}{x-1}+\frac{1-\lambda_{g,4}}{x+1}\right) \frac{d y}{d x}  +\frac{\frac{(\Delta-\lambda_{g,4})^2}{4}x-\frac{\lambda_{g,4}}{2}\left( \frac{\lambda_{g,4}}{2}-\Delta+1\right)}{x(x-1)(x+1)} y=0.
	\label{qq:12}
\end{equation}
Eq.(\ref{qq:12}) is the Heun differential equation that has four regular singular points at $z=0,1,-1,\infty$ \cite{Ronv1995}.
Substituting $y(x)= \sum_{n=0}^{\infty } d_n x^{n}$ into (\ref{qq:12}), we obtain the following three term  recurrence relation:
\begin{equation}
	\alpha_n\; d_{n+1}+ \beta_n \;d_n + \gamma_n \;d_{n-1}=0  \quad
	\hbox{  for  } n \geq 1, \label{qq:13}
\end{equation}
with
\begin{equation}
	\begin{cases} \alpha_n= (n+1)(n+ \Delta-1) \cr
		\beta_n= -\frac{\lambda_{g,4}}{2}\left(2n+\Delta-1-\frac{\lambda_{g,4}}{2}\right) \cr
		\gamma_n=-\left( n -1+\frac{\Delta}{2}-\frac{\lambda_{g,4}}{2}\right)^2
	\end{cases}
	\label{qq:14}
\end{equation}
The first three $d_{n}$'s are given by   $\alpha_0 d_1+ \beta_0 d_0=0$ and  $d_{-1}=0$.
Eq.(\ref{qq:11}), Eq.(\ref{qq:13}) and Eq.(\ref{qq:14}) tells us the following boundary condition
\begin{equation}
	F^{'}(0)=0.
	\label{qq:15}
\end{equation}
We rewrite Eq.(\ref{qq:13}) as
\begin{equation}
	d_{n+1}+ A_n\;d_n + B_n\;d_{n-1}=0,
	\label{qq:26}
\end{equation}
where $A_n$ and $B_n$ have asymptotic expansions of the form
\begin{equation}
	\begin{cases} A_n= \frac{\beta_n}{\alpha_n} \sim \sum_{j=0}^{\infty}\frac{a_j}{n^j}   \cr
		B_n= \frac{\gamma_n}{\alpha_n} \sim  \sum_{j=0}^{\infty}\frac{b_j}{n^j}
	\end{cases}
	\label{qq:27}
\end{equation}
with
\begin{equation}
	\begin{cases} a_0= 0, \quad a_1= -\lambda_{g,4},\quad a_2= \frac{\lambda_{g,4}}{2}\left( \Delta+1+\frac{\lambda_{g,4}}{2}\right) \cr
		b_0= -1, \quad b_1 =2+\lambda_{g,4},\quad b_2 = \frac{1}{4}\left( \lambda_{g,4}+\Delta \right)^2+2+\lambda_{g,4}
	\end{cases}
	\label{qq:28}
\end{equation}

The radius of convergence, $\rho$, satisfies characteristic equation associated with Eq.(\ref{qq:26})   \cite{Miln1933,Perr1921,Poin1885} :
\be 
\rho^2+ a_0\rho +b_0 =\rho^2 -1= 0,
\ee
whose roots  are given by 
\begin{equation}
	\rho_1 = 1\hspace{1cm} \rho_2 = -1 \label{qq:17}
\end{equation}
So   for a three--term recurrence relation in Eq.(\ref{qq:13}), the radius of convergence is 1 for all two cases.   Since the solutions should converge at the horizon, $y(z)$ should be convergent at
$|z|\leq1$. According to Pincherle’s Theorem  \cite{Jone1980},
we have a convergent solution of $y(z)$ at $|z|=1$ if only if the three term recurrence relation Eq.(\ref{qq:13}) has a minimal solution. 
Since we have two different roots $\rho_i$'s, so Eq.(\ref{qq:26}) has two linearly independent solutions $d_1(n)$, $d_2(n)$. One can show that  \cite{Jone1980} for large $n$, 
\begin{equation}
	d_i(n)\sim \rho_i^n n^{\alpha_i}\sum_{r=0}^{\infty} \frac{\tau_i(r)}{n^r},\quad i=1,2,3
	\label{qq:29}
\end{equation}
with
\begin{equation}
	\alpha_i=\frac{a_1\rho_i +b_1 }{a_0\rho_i +2b_0 },\quad i=1,2
	\label{qq:30}
\end{equation}
and $\tau_i(0)=1$.
In particular, we obtain
\begin{equation}
	\tau_i(1)= \frac{-2\rho_i^2 \alpha_i(\alpha_i-1)-\rho_i(a_2+\rho_i a_1+\alpha_i(\alpha_i-1)a_0/2)-b_2}{2\rho_i^2(\alpha_i-1)+\rho_i(a_1+(\rho_i-1)a_0)+b_1},\quad i=1,2
	\label{qq:31}
\end{equation}
Substituting Eq.(\ref{qq:17}) and Eq.(\ref{qq:28}) into Eq.(\ref{qq:29})--Eq.(\ref{qq:31}), we obtain
\begin{equation}
	\begin{cases}  d_1(n)\sim n^{-1}\left( 1+ \frac{\Delta^2+4\Delta\lambda_{g,4}+2(\lambda_{g,4}^2+\lambda_{g,4}+12)}{8n} \right)  \cr
		d_2(n)\sim   (-1)^n  n^{-1-\lambda_{g,4}}\left( 1+ \frac{\lambda_{g,4}^2+\frac{11}{4}\lambda_{g,4}+\frac{\Delta^2}{8}+3}{n} \right)
	\end{cases}
	\label{qq:32}
\end{equation}
Since $\lambda_{g,4}>0$,
\begin{equation}
	\lim_{n\rightarrow\infty}\frac{ d_2(n)}{ d_1(n)}=0
	\label{qq:34}.
\end{equation} 
So $d_2(n)$ is a minimal solution. Also,
\begin{equation}
	\begin{cases} \sum|d_1(n)|\sim \sum \frac{1}{n}\rightarrow\infty  \cr
		\sum|d_2(n)|\sim \sum n^{-1-\lambda_{g,4}}<\infty
	\end{cases}
	\label{qq:35}
\end{equation}
Therefore, $y(z)=\sum_{n=0}^{\infty}d_n x^n$ is convergent at $x=1$ if only if $d_n$ is a minimal solution.
Eq.(\ref{eq:24}) with $\delta_2=\delta_3=\cdots=\delta_N=0$ becomes a polynomial of degree $N$ with respect to $\lambda_{g,4}$.
Put Eq.(\ref{qq:14}) into Eq.(\ref{eq:24}) where $\delta_i =0$ at $i\in \{2,3,\cdots,N\}$  and we  choose $N=15$ 

For algorithm to find $\lambda_{g,4}$ for a given $\Delta$, 
\begin{enumerate}
	\item  
	Choose an $N$. 
	\item
	Put Eq.(\ref{qq:14}) into Eq.(\ref{eq:24}).  
	\item
	Define a function returning the determinant of system Eq.(\ref{eq:24}).  
	\item
	Find the roots of interest of this function.
	\item
	Increase $N$ until those roots become constant to within the desired precision \cite{Leav1990}.
\end{enumerate}
\subsubsection{ Unphysical region of $\Delta$}\label{unphysical2}
\begin{figure}[h]
	\subfigure[  $\lambda_{g,4}$  vs $\Delta$. ]
	{\includegraphics[width=0.52\linewidth]{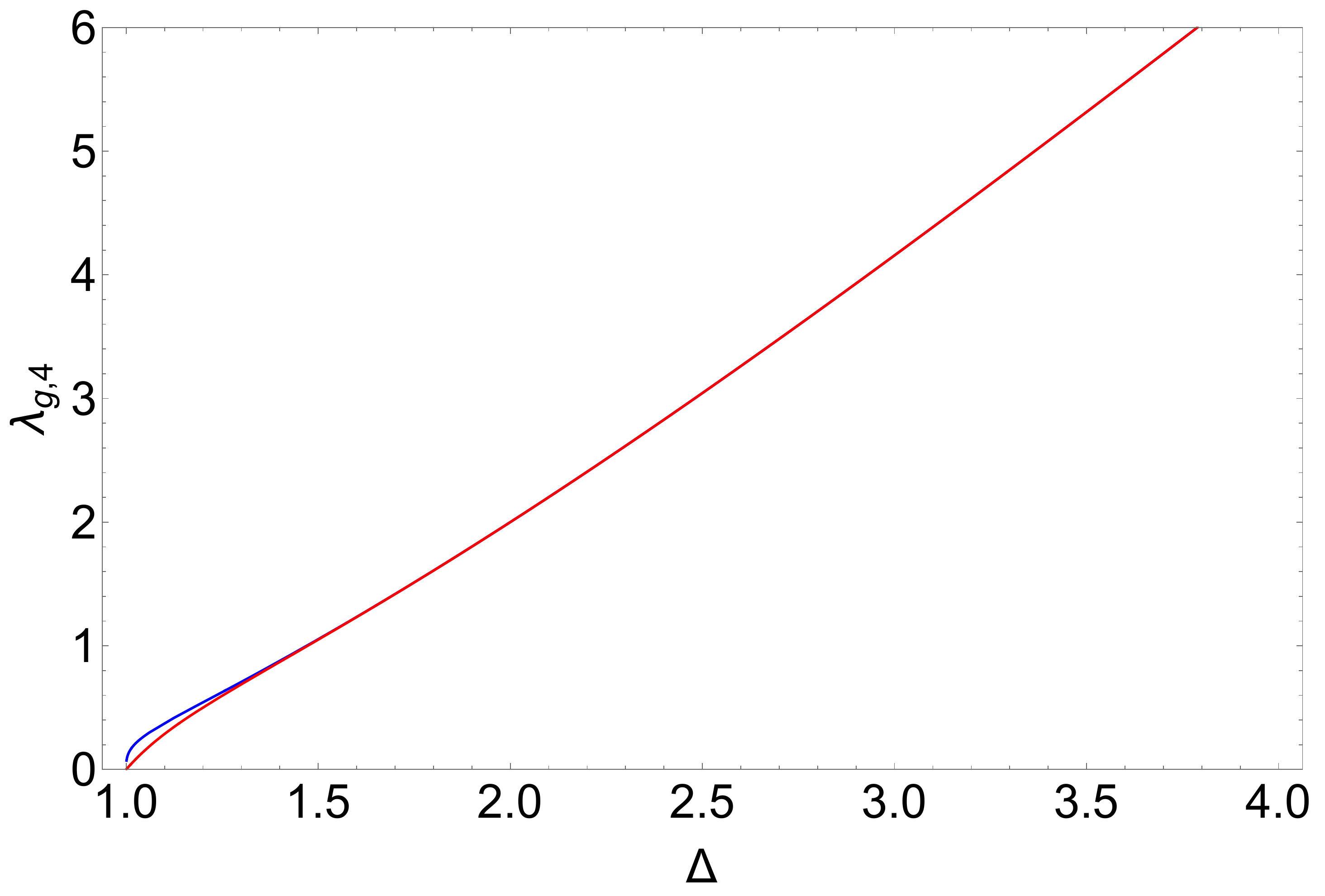}}
	\hfill
	\subfigure[  $\lambda_{g,4}$  vs $\Delta$. ]
	{\includegraphics[width=0.52\linewidth]{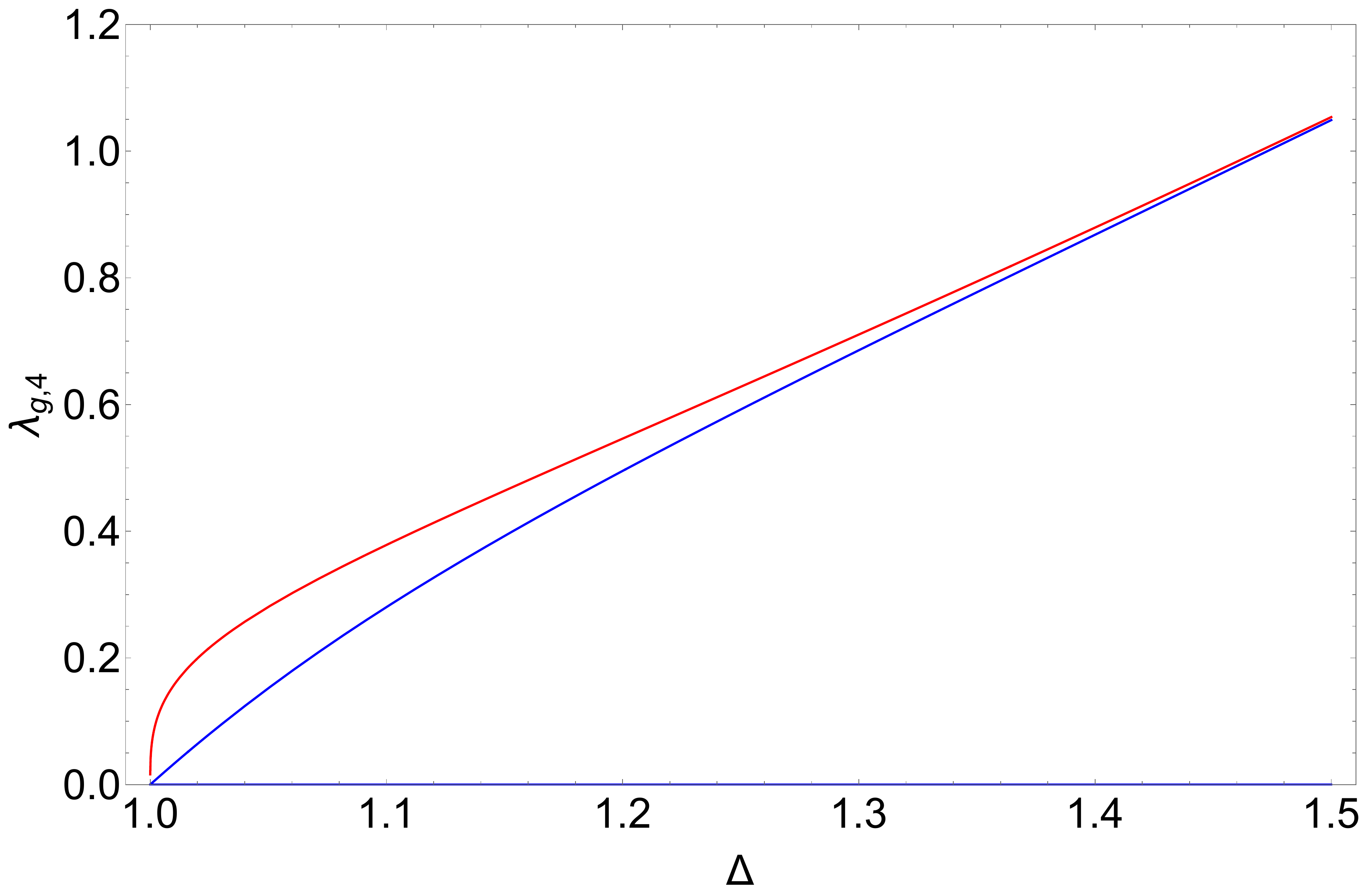}}
	\caption{\small
		(a) $\lambda_{g,4}$  vs $\Delta$: Blue and  red  colored curves for $\lambda_{g,4}$ are obtained by letting $d_N =0$, with $N=29, 30 $ respectively. There are two branches in  $1<\Delta<3/2$. And such  brances merge for $\Delta\geq3/2$. 
		(b) Zoom in on the left plot.     }
	\label{Lambda_big}
\end{figure}

We use Mathematica to compute the determinants to locate their roots. We are only interested in smallest positive real roots of $\lambda_{g,4}$. For computation of roots, we choose $N=30$.
For the smallest value of $\lambda_{g,4}$, we can find an approximate fitting function that is given by
\be
\lambda_{g,4} \approx 1.18 \Delta^{4/3} - 0.97 \;\;\mbox{at}\;\; 3/2\leq \Delta\leq 4\label{qq:36}
\ee
Fig.~\ref{Lambda_big}	shows us that there is no convergent solution for $1< \Delta<3/2$. Because there are two branches, and these branches does not converge to single value  $\lambda_{g,4}$ no matter how we increase $N$.  
Fig.~\ref{Lambda_big} shows that these two branches merge to the single value $\lambda_{g,4} \approx 1$ as $N$ increases.	
We are interested why two branches occur near $\Delta =1$ regardless of the size of $N$. 
 
Eq.(\ref{eq:24}) can be simplified using the formula for 
the determinant of a block matirix,
\begin{equation}
	\det \begin{pmatrix}
		A & B \\
		C & D 
	\end{pmatrix}= \det(A)\det(D-C A^{-1} B), \quad {\rm with }
\end{equation}
%
$ \begin{footnotesize}
	A= \begin{pmatrix}
		\beta_0 & \alpha_0 \\
		\gamma_1 & \beta_1 
	\end{pmatrix}  , 
	\quad B=\begin{pmatrix}
		0 & 0 &     \cdots & 0 \\
		\alpha_1 &     0 & \cdots & 0 
	\end{pmatrix}  , 
	\quad C=\begin{pmatrix}
		0 & \gamma_2 \\
		0 & 0 \\
		0 & 0 \\
		0 & 0 \\
		\vdots & \vdots \\
		0 & 0 
	\end{pmatrix} ,
	\quad D=\begin{pmatrix}
		\beta_2 & \alpha_2 &  &  &  &  \\
		\gamma_3 & \beta_3 & \alpha_3 &  &  &  \\
		 & \gamma_4 & \beta_4 & \alpha_4 &  &  \\
		&  & \ddots & \ddots & \ddots &  \\
		&  &  & \gamma_{N-1} & \beta_{N-1} & \alpha_{N-1} \\
		&  &  &  & \gamma_{N} & \beta_{N} 
	\end{pmatrix}. 
\end{footnotesize} $
By explicit computation,  we can see  the factor $\det(A)=\frac{\lambda^3}{16}(\lambda -4)$ at $\Delta =1$ so that the minimal real root is  $\lambda_{g,4}=0$.   
Near $\Delta =1$, we can expand the determinant as a series in $ \varepsilon=\Delta- 1 \ll 1$ and  $0< \lambda_{g,4}\ll 1$.  After some calculations, we found that    
$d_N =0$  gives     following results:  
\begin{enumerate}
	\item For $N=2m$ where $m=1,2,3,\cdots$,  
	\be
	\lambda_{g,4}^2 \sum_{n=0}^{2N}  \alpha_{0,n} \lambda_{g,4}^n +\varepsilon 	\lambda_{g,4} \sum_{n=0}^{2N}  \beta_{0,n} \lambda_{g,4}^n + \mathcal{O}(\varepsilon^2) =0 \nonumber
	\ee
		This leads us   $\lambda_{g,4} \sim \varepsilon  \sim(\Delta-1)$ as far as $\alpha_{0,0}\beta_{0,0}\neq 0$, which can be confirmed by explicit computation. 	
	\item For $N=2m+1$,  
	\be
	\lambda_{g,4}^3 \sum_{n=0}^{2N-1}  \alpha_{1,n} \lambda_{g,4}^n +\varepsilon    \sum_{n=0}^{2N+1}  \beta_{1,n} \lambda_{g,4}^n + \mathcal{O}(\varepsilon^2) =0 \nonumber
	\ee
	leading to   $\lambda_{g,4} \sim (\Delta-1)^{1/3}$. 
\end{enumerate}
This proof tells us that two branches should be occured near $\Delta=1$. 

We calculated  121 different values of $\lambda_{g,4}$'s  at various $\Delta$ and the result is the blue colored curves in Fig.~\ref{Tcsingular}.
These data fits well by above formula. 

The authors of ref\cite{Siop2012}  got   $\lambda_{g,4}$'s by using variational method using the fact that the eigenvalue $\lambda_{g,4}$
minimizes the expression
\be
\lambda_{g,4}^2 = \frac{\int_{0}^{1}dz\; z^{2\Delta-3}\left( (1-z^4)\left[F^{'}(z)\right]^2+\Delta^2 z^2 \left[F(z)\right]^2\right)}{\int_{0}^{1}dz\;
	z^{2\Delta-3}\frac{1-z^2}{1+z^2}\left[F(z)\right]^2}
\label{qq:37}
\ee
for $\Delta>1$. This integral does not converge at $\Delta=1$ because of $\ln(z)$.  The trial function  used is $F(z)=1-\alpha z^3$  where $\alpha$ is the variational parameter. 
Their result is given by the  green colored dots in Fig.~\ref{Tcsingular} and ours by the blue   curves. The differences    are  appreciable  at $\Delta >1.8$. Our results are consistently lower.  The variational method show us that there are  numerical values of $\lambda_{g,4}$ for $1<\Delta<3/2$. But our method tells us that the region $1<\Delta<3/2$ is not valid for analytic solutions because of non-convergence $\lambda_{g,4}$.
The critical temperature which is given by 
$
T_c =\frac{1}{\pi}r_c=\frac{1}{\pi}\left(\frac{\rho}{\lambda_4}\right)^{\frac{1}{3}}
$
which  can be calculated by Eq.(\ref{qq:36}) and the Fig.~\ref{Tcsingular}(b) demonstrate the result.
Notice that figure 1 in ref. \cite{siopsis2011holographic} show us that  $T_c$ is divergent at $\Delta =1$ and it is a monotonically decreasing function of $\Delta$. 


\subsubsection{  The analytic solution of  $g\frac{\left< \mathcal{O}_{\Delta}\right>}{T_c^{\Delta}}$}
Substituting Eq.(\ref{qq:11}) into  Eq.(\ref{eq:3}), the field equation $\Phi$ becomes
\be
\frac{d^2 \Phi}{d x^2} =   \frac{g^2\left< \mathcal{O}_{\Delta}\right>^2}{ 4 r_h^{2\Delta}}\frac{x^{ \Delta-2} F^{2}(x)}{1-x^2 }  \Phi
\label{qq:39}
\ee
where $\frac{g^2\left< \mathcal{O}_{\Delta}\right>^2}{ 4 r_h^{2\Delta}}$ is small because of $T\approx T_c$. The above equation has the   expansion around Eq.(\ref{qq:9}) with small correction:
\be
\frac{ \Phi}{r_h} =   \lambda_4(1-x)+ \frac{g^2\left< \mathcal{O}_{\Delta}\right>^2}{4 r_h^{2\Delta}}\chi_1(z) 
\label{qq:40}
\ee
We have $\chi_1(1)=\chi_1^{'}(1)=0$ due to  the boundary conditions 
$\Phi(1)=0 $ .
Taking derivative of  Eq.(\ref{qq:40}) twice with respect to $x$ and using the result in Eq.(\ref{qq:39}),
\be
\chi_1^{''}=\frac{x^{ \Delta-2}F^2(z)}{1-x^2}\left\{ \lambda_4(1-x)+ \frac{g^2\left< \mathcal{O}_{\Delta}\right>^2}{4 r_h^{2\Delta}}\chi_1 \right\}\approx \frac{\lambda_4 x^{ \Delta-2}F^2(z)}{1+x}.
\label{qq:41}
\ee
Integrating Eq.(\ref{qq:41}) gives us
\be
\chi_1^{'}(0) = -\lambda_4 \mathcal{C}_4 \quad
\hbox{  for  }\mathcal{C}_4 = \int_{0}^{1}dx\;\frac{  x^{ \Delta-2}F^2(x)}{1+x}
\label{qq:42}
\ee
Eq.(\ref{qq:11}) with Eq.(\ref{qq:14}) shows
\begin{equation}
	F(z)= (1+x)^{-\lambda_{g,4}/2}y(x)\approx (1+x)^{-\lambda_{g,4}/2} \sum_{n=0}^{10}d_n x^n
	\label{qq:43}
\end{equation}
Here, we ignore $d_n x^n$ terms if $n\geq11$ because   $0<|d_n|\ll1$ numerically and $y(x)$ converges at $0\leq x\leq1$.
We can calculate the numerical value of $\mathcal{C}_4$ by putting Eq.(\ref{qq:36}) and Eq.(\ref{qq:14}) into Eq.(\ref{qq:42}). We calculated  121 different values of $\sqrt{1/\mathcal{C}_4}$'s  at various $\Delta$, which is drawn as dots  in Fig.~\ref{M.eigen}. Then we tried to find an approximate fitting function. The result is given as follows, 
\be
\sqrt{ \frac{1}{\mathcal{C}_4} } \approx  \frac{\Delta ^{6.5} +510 \Delta ^{2}}{1327}\label{qq:45}    
\ee
The  Fig.~\ref{M.eigen} shows how the data fits by above formula. 
From  Eq.(\ref{qq:40}) and  Eq.(\ref{eq:4}), we have
\be
\frac{\rho}{r_h^3}=\lambda_4\left( 1+\frac{\mathcal{C}_4 g^2\left< \mathcal{O}_{\Delta}\right>^2}{4r_h^{2\Delta}}\right)
\label{qq:46}
\ee
Putting $T=\frac{1}{\pi} r_h  $ with $\lambda_4=\frac{\rho}{r_c^3}$ into Eq.(\ref{qq:46}), we obtain the condensate near $T_c$:
\be
g\frac{\left< \mathcal{O}_{\Delta}\right>}{T_c^{\Delta}}\approx  \mathcal{M}_4 \;  \sqrt{1-\frac{T}{T_c}} \quad
\hbox{  for  } \mathcal{M}_4 = 2 \pi^{\Delta}\sqrt{\frac{3}{\mathcal{C}_4}},
\label{qq:47}
\ee
and the plot is in the FIG. \ref{Odel_crit}.

As we substitute  eq. (\ref{si:1}) into  eq. (\ref{qq:47}), we obtain
\be
g\frac{\left< \mathcal{O}_{\Delta}\right>}{(g \rho)^{\frac{\Delta}{3}} }\approx 2\lambda_{g,4}^{-\frac{\Delta}{3}} \sqrt{\frac{3}{\mathcal{C}_4}} \;  \sqrt{1-\frac{T}{T_c}}  
\label{www:2}
\ee
Fig.~\ref{Odel_crit1} is the plot of eq. (\ref{www:2}).
%
%
%
%
\subsection{Condensate at near the zero temperature}
\subsubsection{Analytic calculation of $g^{\frac{1}{ \Delta}}\frac{\left< \mathcal{O}_{\Delta}\right>^{\frac{1}{\Delta}}}{T_c}$ at $1.5\leq  \Delta<4$}\label{smalla}
The dominant contribution comes from the neighborhood of the boundary $z=0$. So near the $T=0$ we can simplify two coupled equations Eq.(\ref{eq:3}) and Eq.(\ref{eq:3}) with Eq.(\ref{qq:11}) by letting $z\rightarrow0$:
\begin{subequations}
	\begin{equation}
		\frac{d^2 F }{d z^2} +  \frac{2 \Delta-3}{z} \frac{d F}{d z} + \frac{g^2 \Phi^2}{r_h^2 }F =0
		\label{qq:48a}
	\end{equation}
	\begin{equation}
		\frac{d^2 \Phi}{d x^2} -\frac{ g^2\left< \mathcal{O}_{\Delta}\right>^2}{4r_h^{2\Delta}}x^{ \Delta-2}F^2 \Phi =0
		\label{qq:48b}
	\end{equation}
\end{subequations}
where $x=z^2$. Also, we use a boundary condition at the horizon, and
Eq.(\ref{eq:3}) with Eq.(\ref{qq:11}) is rewritten as
\begin{equation}
	-\frac{d^2 F}{d z^2} +\left( \frac{4}{z(1-z^4)}-\frac{2\Delta+1}{z}\right)\frac{d F}{d z}+\left(\frac{\Delta^2 z^2}{1-z^4}- \frac{g^2
		\Phi^2}{r_h^2(1-z^4)^2}\right)F=0.
	\label{qq:49}
\end{equation}
It provides us the following boundary condition at the horizon with Eq.(\ref{eq:3}),  
$\Phi(1)=0 $ 
{and}   $\Psi(1)<\infty$:
\begin{equation}
	4F^{'}(1) +\Delta^2 F(1)=0
	\label{qq:50}
\end{equation}
By multiplying  $z$ to the eq. (\ref{qq:49}) and then taking the limit of $z\to 0$, we get $ F^ {'}(0)=0$.    
Note that $F(0)=1$ should be considered  as the normalization condition of $\vev{{\cal O}_{\Delta}}$ rather than as a boundary condition.  
Also for canonical system,  we regard the 
$\frac{d\Phi(0)}{d x} =-\frac{\rho}{r_h^2} $ as BC 
and $\Phi(0)=\mu$ is not a BC but a value that  should be determined by $\rho$  from the horizon regularity condition  $\Phi(1)=0$. 
In Grand canonical system $\Phi(0)=\mu$ is the boundary condition and $\rho$ should be determined from it by the  $\Phi(1)=0$.  	Here we consider $\rho$ as the given parameter.

If we introduce $b$ by for $b^{\Delta}= \frac{ g\left< \mathcal{O}_{\Delta}\right>}{\Delta r_h^{\Delta}}$, the solution to Eq.(\ref{qq:48b}) for $\Phi$ with $F\approx 1$ is
\begin{equation}
	\Phi(z) =\mathcal{A} r_h b z K_{\frac{1}{ \Delta }}\left(b^{\Delta } z^{\Delta }\right)  
	\label{qq:53b}
\end{equation}
At the horizon $\Phi(1)\varpropto \exp(-b^{\Delta})\rightarrow 0$ because $b\rightarrow\infty$ as $r_h \rightarrow 0$, which takes care  the boundary condition $\Phi(1)=0 $. 
Substituting Eq.(\ref{qq:53b}) into Eq.(\ref{qq:48a}), $F$ becomes
\begin{equation}
	\frac{d^2 F }{d z^2} +  \frac{2 \Delta-3}{z} \frac{d F}{d z} +g^2 b^2 \mathcal{A}^2 z^2 \left( K_{\frac{1}{ \Delta }}\left(b^{\Delta } z^{\Delta
	}\right)\right)^2 F=0
	\label{qq:54}
\end{equation} 
$F(z)$ can be obtained iteratively starting from $F=1$. 
The result is  
\begin{subequations}
	\begin{equation}
		F(z)= 1- g^2 b^2 \mathcal{A}^2 \int_{0}^{z} d\acute{z} \; \acute{z}^{3-2\Delta } \int_{0}^{\acute{z}} d\fH{z} \; \fH{z}^{2\Delta-1} \left( K_{\frac{1}{ \Delta }}\left(b^{\Delta } \fH{z}^{\Delta }\right)\right)^2
		\label{qq:55a}
	\end{equation}
	\begin{equation}
		F^{'}(z)=- g^2 b^2 \mathcal{A}^2 z^{3-2\Delta } \int_{0}^{z} d\fH{z}\;  \fH{z}^{2\Delta-1} \left( K_{\frac{1}{ \Delta }}\left(b^{\Delta } \fH{z}^{\Delta}\right)\right)^2.
		\label{qq:55b}
	\end{equation}
\end{subequations}
with the boundary condition $ F^{'}(0)=0$ and normalized $F(0)=1$. Applying  Eq.(\ref{qq:50}),  we have
\begin{equation}
	g^2\mathcal{A}^2=\frac{\Delta^2 b^2}{4 F_{\Delta}^{'}(b)+ \Delta^2 F_{\Delta}(b)}
	\label{qq:56}
\end{equation}
where
\begin{subequations}
	\begin{equation}
		F_{\Delta}(b)=  \int_{0}^{b} dz \; z^{3-2\Delta} \int_{0}^{z} d\tilde{z} \; \tilde{z}^{2\Delta-1} \left( K_{\frac{1}{ \Delta }}\left(
		\tilde{z}^{\Delta }\right)\right)^2
		\label{qq:57a}
	\end{equation}
	\begin{equation}
		F_{\Delta}^{'}(b)= b^{4-2\Delta} \int_{0}^{b} dz\;  z^{2\Delta-1} \left( K_{\frac{1}{ \Delta }}\left( z^{\Delta }\right)\right)^2
		\label{qq:57b}
	\end{equation}
\end{subequations}
Letting $x=z^{\Delta}$, Eq.(\ref{qq:57b}) is simplified as
\begin{equation}
	F_{\Delta}^{'}(b)= \frac{b^{4-2\Delta}}{\Delta} \int_{0}^{b^{\Delta}} dx\;  x \left( K_{\frac{1}{ \Delta }}\left( x\right)\right)^2
	= \frac{b^{4-2\Delta}}{\Delta} \int_{0}^{\infty} dx\;  x \left( K_{\frac{1}{ \Delta }}\left( x\right)\right)^2.
	\label{qq:58}
\end{equation}
Using Eq.(\ref{eq:59}), Eq.(\ref{qq:58}) becomes
\begin{equation}
	F_{\Delta}^{'}(b)= \frac{\pi b^{4-2\Delta}}{2\Delta^2} \csc{\left( \frac{\pi}{\Delta}\right)}
	\label{qq:60}
\end{equation}
Letting $x=\tilde{z}^{\Delta}$, Eq.(\ref{qq:57a}) is also simplified as
\begin{equation}
	F_{\Delta}(b)= \frac{1}{\Delta} \int_{0}^{b} dz \; z^{3-2\Delta} \int_{0}^{z^{\Delta}} dx \; x \left( K_{\frac{1}{ \Delta }}\left(  x\right)\right)^2
	\label{qq:61}
\end{equation}
As we apply Eq.(\ref{eq:59}), Eq.(\ref{eq:62}) and Eq.(\ref{eq:63}) into Eq.(\ref{qq:61}), we obtain 
\begin{footnotesize}
	\begin{eqnarray}
		F_{\Delta}(b) &=&  \frac{\pi b^{4-2\Delta}}{4 \Delta^2 (2-\Delta)} \csc{\left( \frac{\pi}{\Delta}\right)} - \frac{\pi \epsilon^{4-2\Delta}}{4 \Delta^2 (2-\Delta)} \csc{\left( \frac{\pi}{\Delta}\right)}\nonumber\\ 
		&+&  \frac{1}{2\Delta^2}\lim_{\epsilon\rightarrow 0}\int_{\epsilon^{\Delta}}^{b^{\Delta}}dx x^{\frac{4-\Delta}{\Delta}}  K_{\frac{1}{\Delta}}\left( x\right)^2
		-\frac{1}{2\Delta^2}\lim_{\epsilon\rightarrow 0}\int_{\epsilon^{\Delta}}^{b^{\Delta}}dx x^{\frac{4-\Delta}{\Delta}}  K_{\frac{1}{\Delta}-1}\left( x\right) K_{\frac{1}{\Delta}+1}\left( x\right)\nonumber\\ 
		&\approx&  \frac{\pi b^{4-2\Delta}}{4 \Delta^2 (2-\Delta)} \csc{\left( \frac{\pi}{\Delta}\right)} - \frac{\pi \epsilon^{4-2\Delta}}{4 \Delta^2 (2-\Delta)} \csc{\left( \frac{\pi}{\Delta}\right)}\nonumber\\ 
		&+&  \frac{1}{2\Delta^2} \int_{0}^{\infty}dx x^{\frac{4-\Delta}{\Delta}}  K_{\frac{1}{\Delta}}\left( x\right)^2
		-\frac{1}{2\Delta^2}\lim_{\epsilon\rightarrow 0}\int_{\epsilon^{\Delta}}^{b^{\Delta}}dx x^{\frac{4-\Delta}{\Delta}}  K_{\frac{1}{\Delta}-1}\left( x\right) K_{\frac{1}{\Delta}+1}\left( x\right)\nonumber\\ 
		&=&   \frac{\pi b^{4-2\Delta}}{4 \Delta^2 (2-\Delta)} \csc{\left( \frac{\pi}{\Delta}\right)} - \frac{\pi \epsilon^{4-2\Delta}}{4 \Delta^2 (2-\Delta)} \csc{\left( \frac{\pi}{\Delta}\right)}\nonumber\\ 
		&+&  \frac{\sqrt{\pi}}{8\Delta^2}\frac{\Gamma(1/\Delta)\Gamma(2/\Delta)\Gamma(3/\Delta)}{\Gamma(1/2+2/\Delta)}  
		-\frac{1}{2\Delta^2}\lim_{\epsilon\rightarrow 0}\int_{\epsilon^{\Delta}}^{b^{\Delta}}dx x^{\frac{4-\Delta}{\Delta}}  K_{\frac{1}{\Delta}-1}\left( x\right) K_{\frac{1}{\Delta}+1}\left( x\right) 
		\label{uu:64}
	\end{eqnarray} 
\end{footnotesize}
here, we introduce small $\epsilon$, and take zero at the end of calculations. 
After some long but simple calculations using the properties Eq.(\ref{eq:75}), Eq.(\ref{eq:77}) and Eq.(\ref{eq:79}), an integral in Eq.(\ref{uu:64}) is shows
\begin{footnotesize}
	\begin{equation} 
		-\frac{1}{2\Delta^2}\lim_{\epsilon\rightarrow 0}\int_{\epsilon^{\Delta}}^{b^{\Delta}}dx x^{\frac{4-\Delta}{\Delta}}  K_{\frac{1}{\Delta}-1}\left( x\right) K_{\frac{1}{\Delta}+1}\left( x\right)  =  \frac{ \sqrt{\pi } \Gamma \left( \frac{1}{\Delta }\right) \Gamma \left(\frac{2}{\Delta } \right) \Gamma \left(\frac{3}{\Delta }\right)}{4\Delta^2 (\Delta-2) \Gamma \left( \frac{1}{2}+\frac{2}{\Delta}\right)} 	+\frac{\pi \epsilon^{4-2\Delta}}{4\Delta^2 (2-\Delta)} \csc{\left( \frac{\pi}{2\Delta}\right)}	
		\label{uu:77}
	\end{equation} 
\end{footnotesize}
with $b\rightarrow \infty$. Substitute Eq.(\ref{uu:77}) into Eq.(\ref{uu:64}), and we have 
\begin{equation}
	F_{\Delta}(b) = \frac{\pi b^{4-2\Delta}}{4 \Delta^2 (2-\Delta)} \csc{\left( \frac{\pi}{\Delta}\right)}
	-\frac{\sqrt{\pi} \Gamma \left(-1+\frac{2}{\Delta }\right)\Gamma \left(\frac{1}{\Delta }\right) \Gamma \left(\frac{3}{\Delta }\right) }{8\Delta ^2 \Gamma \left(\frac{1}{2}+\frac{2}{\Delta }\right)} 
	\label{qq:64}
\end{equation} 
Putting  Eq.(\ref{qq:60}) and Eq.(\ref{qq:64}) into Eq.(\ref{qq:56}), we have
\begin{equation}
	g^2\mathcal{A}^2= \frac{b^2}{\frac{\pi(4-\Delta)^2 \csc\left(\frac{\pi}{\Delta} \right)}{4\Delta^4 (2-\Delta)} b^{4-2\Delta}-\frac{\sqrt{\pi}\Gamma\left(\frac{1}{\Delta} \right)\Gamma\left(\frac{3}{\Delta} \right)\Gamma\left(-1+\frac{2}{\Delta}\right)}{8\Delta^2 \Gamma\left( \frac{1}{2}+\frac{2}{\Delta}\right)}}
	\label{qq:65}
\end{equation}
Apply  Eq.(\ref{eq:63}) into Eq.(\ref{qq:53b}) using Eq.(\ref{eq:4}), we deduce
\begin{equation}
	\frac{\rho}{r_h^3}=-\frac{\Gamma \left(\frac{-1}{\Delta }\right)}{2^{1+\frac{1}{\Delta}}}\mathcal{A} b^2
	\label{qq:66}
\end{equation} 
As we combine $
T_c =\frac{1}{\pi}r_c=\frac{1}{\pi}\left(\frac{\rho}{\lambda_4}\right)^{\frac{1}{3}}
$, Eq.(\ref{qq:36}), Eq.(\ref{qq:65}) and Eq.(\ref{qq:66}) with $ b= \left(\frac{ g\left< \mathcal{O}_{\Delta}\right>}{\Delta r_h^{\Delta}}\right)^{\frac{1}{\Delta}}$ in Eq.(\ref{qq:53b}) in the form of
$X$; here, $X:=$	$\left< \mathcal{O}_{\Delta}\right>^{\frac{1}{\Delta}}/{T_c}$ for simple notation, we describec the condensate at $T\approx 0$: 
\begin{equation}
	X^{6}
	= G_4^{6} \left(\alpha_{4} +\beta_4 \tau_{4}^{4-2\Delta} X^{4-2\Delta} \right)
	\label{qq:67}
\end{equation} 
\begin{eqnarray}
	\hbox{where }\quad 		G_4 &=& \pi \Delta^{1/\Delta}\left(\frac{-2^{1+\frac{1}{\Delta}} \lambda_{g,4}}{\Gamma(-\frac{1}{\Delta})}\right)^{\frac{1}{3}} \nonumber\\  
	\alpha_{4} &=&-\frac{\sqrt{\pi}\Gamma\left(\frac{2-\Delta}{\Delta}\right)\Gamma\left(\frac{1}{\Delta}\right)\Gamma\left(\frac{3}{\Delta}\right)}{8\Delta^2 \Gamma\left(\frac{4+\Delta}{2\Delta}\right)} \label{qqq1}\\  
	\beta_4 &=&\frac{\nu \pi  (4-\Delta)^2 \csc(\nu \pi)}{4\Delta^3 \left(2-\Delta\right)}
	\nonumber
\end{eqnarray}  	
with    $\nu=\frac{1}{\Delta}$ and $\tau_{4}=\frac{1}{\pi \Delta^{1/\Delta}}\frac{T_c}{T} $.  
\subsubsection{Analytic calculation of $g^{\frac{1}{\Delta}}\frac{\left< \mathcal{O}_{\Delta}\right>^{\frac{1}{\Delta}}}{T_c}$   at $\Delta=2$}\label{biga}
\begin{figure}[ht!]
	\centering
	\subfigure[ $X$  vs $T/T_c$ at $\Delta=2$]
	{\includegraphics[width=0.45\linewidth]{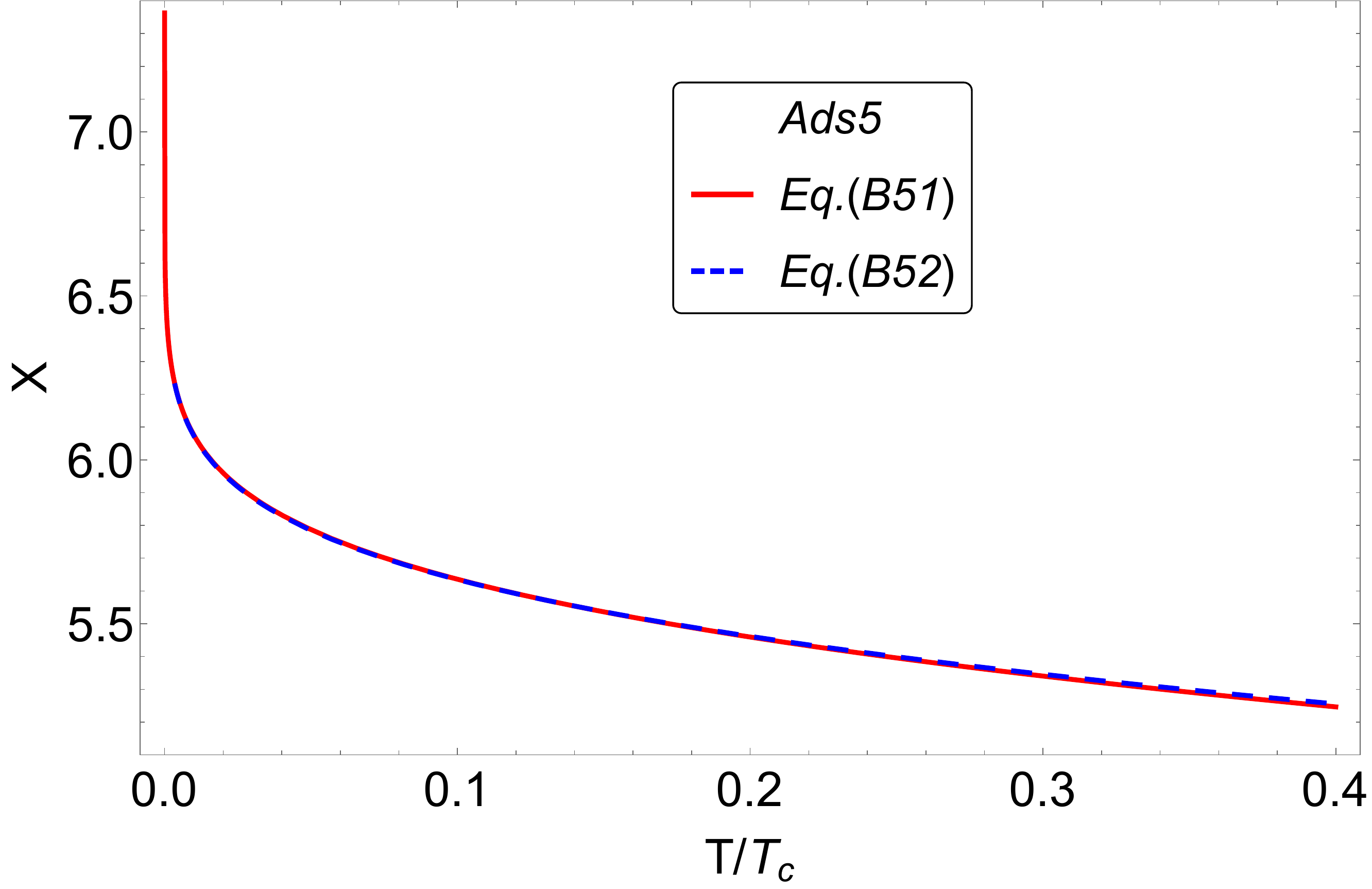}}
	\subfigure[$X^{6}$ vs $\log(T/T_c)$   at $\Delta=2$. ]
	{\includegraphics[width=0.45\linewidth]{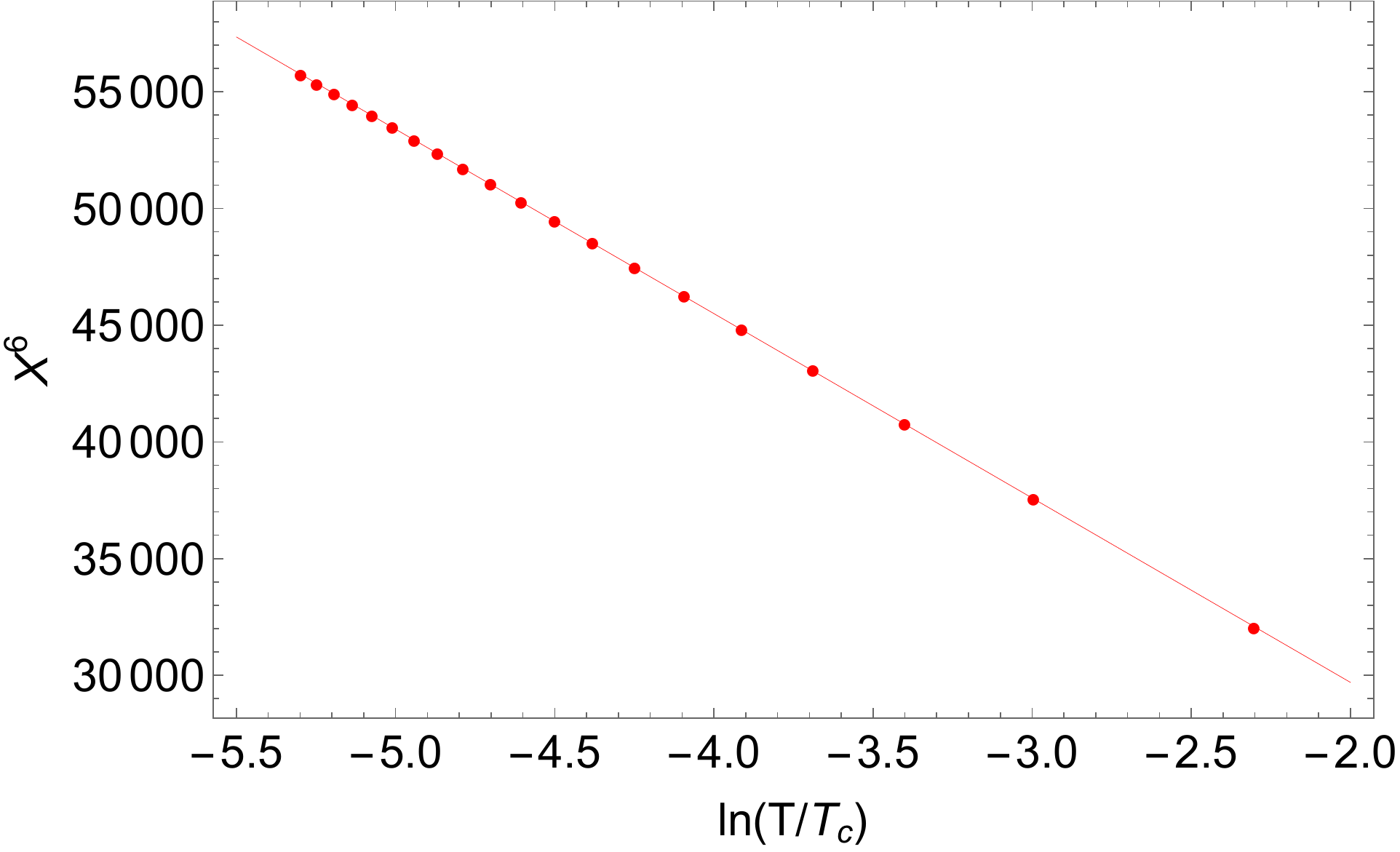}}
	\caption{\small
		(a) $X$  vs $T/T_c$: red  colored curves for $X$ is a plot of Eq.(\ref{qqq:4}). And blue curves is a plot of  Eq.(\ref{ll:62}). These two curves are almost indentical for low temperature.  
		(b)$X^{6}$ vs $\log(T/T_c)$   at $\Delta=2$: The slope of red  dotted line  for $X^{6}$ is $-7900$. }
	\label{logplota} 
\end{figure}

$\alpha_{4}$ and $\beta_4 $	in Eq.(\ref{qqq1}) have series expansions at $\Delta=2$:
\begin{footnotesize}
	\begin{eqnarray}   
		\alpha_{4} &=&\frac{\pi   \csc \left(\frac{ \pi }{2}\right)}{  4^2 \left(\Delta -2\right)} +\frac{ 2\pi   \csc \left(\frac{ \pi }{2}\right) \left(4 -4\log (4) -6 \psi\left(3/2\right)-2 \psi\left(1/2\right)\right)}{4^4}+\mathcal{O} \left(\Delta -2\right) \label{qqq:1}\\  
		\beta_4 &=& -\frac{ \pi \csc \left(\frac{ \pi }{2}\right)}{ 4^2 \left(\Delta -2\right)}+ \frac{2\pi \csc \left(\frac{ \pi }{2}\right) \left(24 +2\pi  \cot \left(\frac{ \pi }{2}\right) \right)  }{4^4}+\mathcal{O}\left(\Delta -2\right). 
		\label{qqq:2}
	\end{eqnarray}  
\end{footnotesize}
As Eq.(\ref{qqq:1}) and Eq.(\ref{qqq:2}) are substituted into Eq.(\ref{qq:67}) with taking the limit $\Delta \rightarrow 2$, we obtain
\begin{equation}
	X^{6}
	= G_4^{6} \left(\rho_{4} +\frac{\sigma_4}{2} \left(\frac{1-\tau_{4}^{4-2\Delta} X^{4-2\Delta}}{\Delta-2} \right)\right)
	\label{qqq:3}
\end{equation} 	
where
\begin{eqnarray}   
	\sigma_4 &=&\frac{ \pi \csc \left(\frac{ \pi }{2}\right)}{8} \nonumber\\  
	\rho_4 &=& \frac{\sigma_4}{4}\left(28-4\ln(4)+2\pi \cot\left(\frac{ \pi}{2} \right)-6\psi(3/2)-2\psi(1/2) \right) \nonumber\\
	&=& \frac{\sigma_4}{4}\left(4-\pi\cot\left(\frac{ \pi}{2} \right)-\ln(4)-2\psi(1/2)\right)  
	\nonumber
\end{eqnarray}  	
By using L'Hopital's rule,  Eq.(\ref{qqq:3}) becomes
\begin{eqnarray}   
	X^{6} &=& G_4^{6} \left(\rho_{4} -\frac{\sigma_4}{2} \frac{\partial }{\partial \Delta}\left(\tau_{4}^{4-2\Delta} X^{4-2\Delta} \right)\right) \nonumber\\
	&=&  G_4^{6}  \left(\rho_{4} +\sigma_4 \ln\left(\tau_{4}X \right)\right) \label{qqq:4} 
\end{eqnarray} 	 	
Fig.~\ref{logplota} (b) tells us that $X\sim \ln(T_c/T)^{1/6}$ for low temperature; 
Numerical result tells us that $X^{6}$-$\log(T/T_c)$ plot  demonstrates the validity of our result  with high precision: 	
$X$ is numerically 
\begin{small}
	\bea 
	X &&\approx  4.9 \left( 1+0.57 \ln\left(\frac{T_c}{T}\right)\right)^{1/6} 
	\label{ll:62}
	\eea
\end{small}  
\subsubsection{Analytic calculation of $g^{\frac{1}{\Delta}}\frac{\left< \mathcal{O}_{\Delta}\right>^{\frac{1}{\Delta}}} {(g \rho)^{1/3}}$   at  $1<\Delta<4$}\label{smalla1}
Apply  Eq.(\ref{qq:65}) into Eq.(\ref{qq:66}) with $T  =\frac{1}{\pi}r_h$ with $ b= \left(\frac{ g\left< \mathcal{O}_{\Delta}\right>}{\Delta r_h^{\Delta}}\right)^{\frac{1}{\Delta}}$   in the form of $Y$; here, $Y:=$	$\frac{g^{1/\Delta}\left< \mathcal{O}_{\Delta}\right>^{1/\Delta}}{(g \rho)^{1/3}}$ for simple notation, we obtain  the condensate at $T\approx 0$: 		
\begin{equation}
	Y^{6}
	= \widetilde{G_4}^{6}   \left(\alpha_{4} +\beta_4 \widetilde{\tau_{4}}^{4-2\Delta}  Y^{4-2\Delta} \right)
	\label{t4:1}
\end{equation} 
\begin{eqnarray}
	\hbox{where }\quad 	 \widetilde{G_4} &=& \Delta^{1/\Delta} \left(\frac{-2^{1+\nu}  }{\Gamma(-\nu)}\right)^{\frac{1}{3}}  \label{t4:2}\\  
	\widetilde{\tau_{4}} &=&\frac{1}{ \pi \Delta^{1/\Delta}}\frac{(g \rho)^{1/3}}{T}   \label{t4:3} 
\end{eqnarray} 
with    $\nu=\frac{1}{\Delta}$. Here, $\alpha_{4}$ and $\beta_4$ are in Eq.(\ref{qqq1}).   
\subsubsection{Analytic calculation of  $g^{\frac{1}{\Delta}}\frac{\left< \mathcal{O}_{\Delta}\right>^{\frac{1}{\Delta}}} {(g \rho)^{1/3}}$    at $\Delta=2$}\label{biga1}
\begin{figure}[ht!]
	\centering
	\subfigure[ $Y$  vs $T/(g \rho)^{1/3}$ at $\Delta=2$]
	{\includegraphics[width=0.45\linewidth]{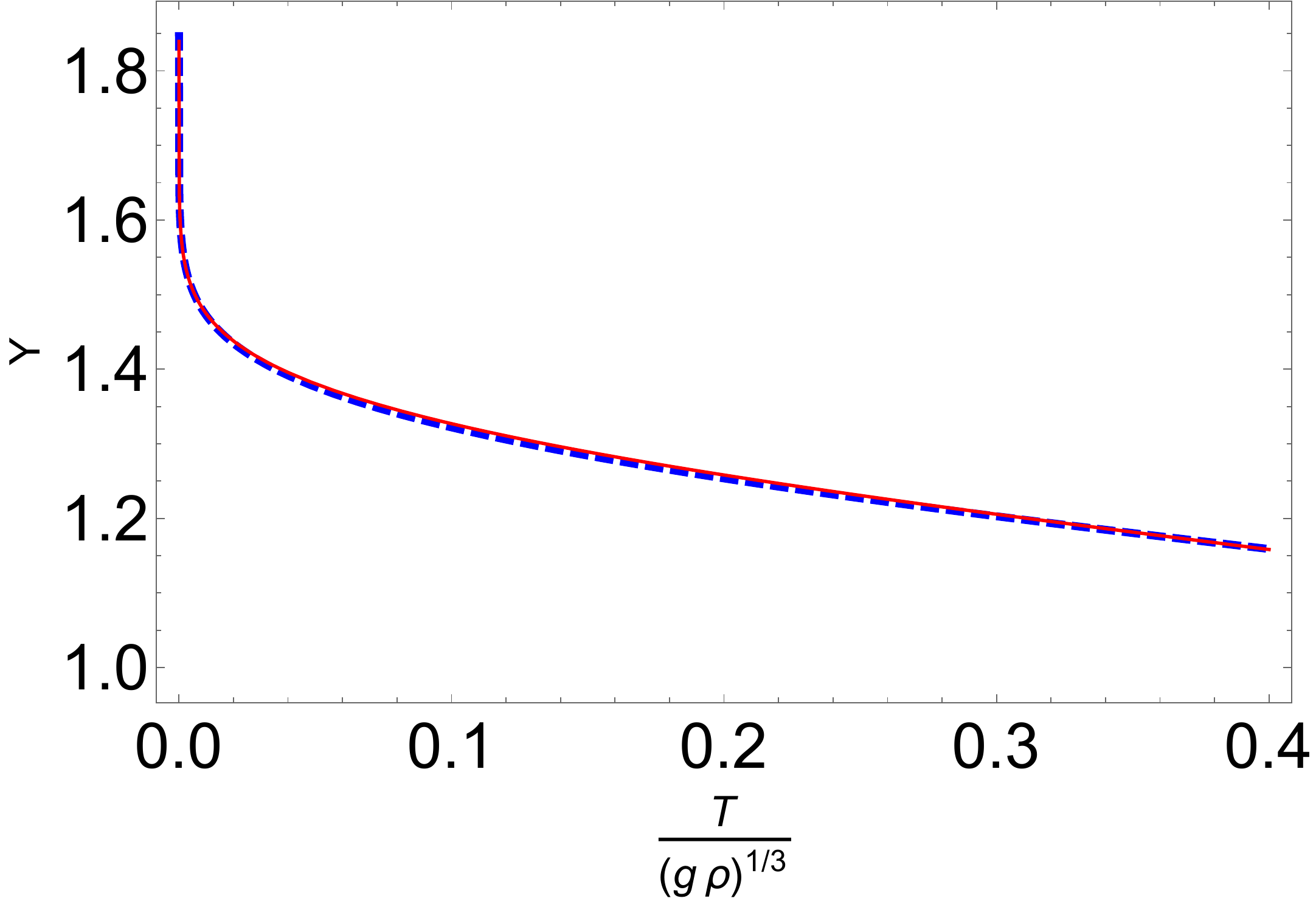}}
	\subfigure[$Y^{4}$-$\log(T/(g \rho)^{1/3})$ graph at $\Delta=2$. ]
	{\includegraphics[width=0.45\linewidth]{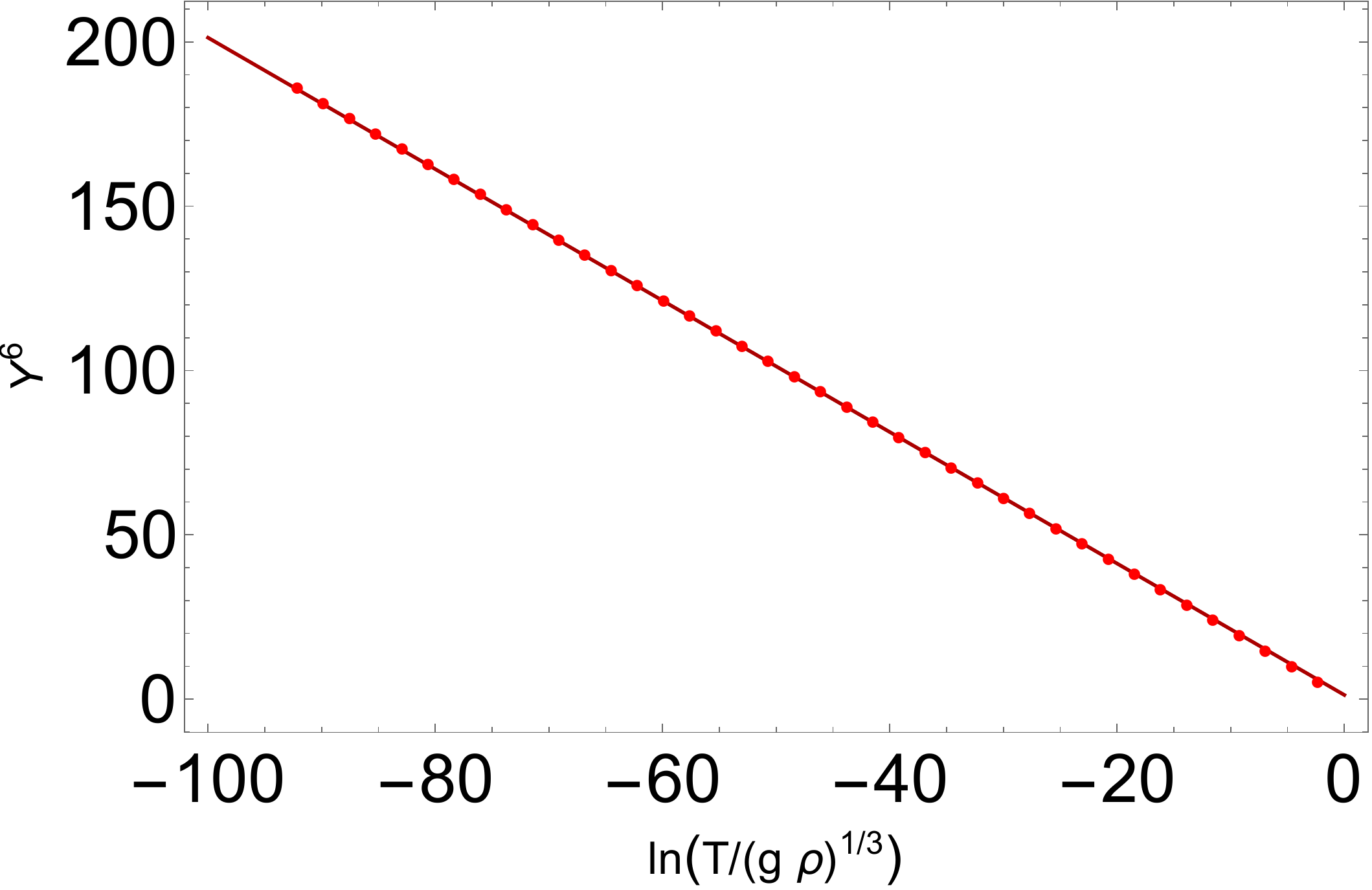}}
	\caption{\small
		(a) $Y$  vs $T/(g \rho)^{1/3}$: red  colored curves for $Y$ is a plot of Eq.(\ref{t4:4}). And blue dashed curves is a plot of  Eq.(\ref{t4:5}). These two curves are almost indentical for low temperature.  \\
		(b)$Y^{6}$-$\log(T/(g \rho)^{1/3})$ graph at $\Delta=2$: The slope of  red  dotted line  for $Y^{6}$ is $-2$. }
	\label{logplot11}
\end{figure}

As Eq.(\ref{qqq:1}) and Eq.(\ref{qqq:2}) are substituted into Eq.(\ref{t4:1}) with taking the limit $\Delta \rightarrow 2$, we obtain
\begin{equation}      
	Y^{6}
	= \widetilde{G_4}^{6}  \left(\rho_{4} +\frac{\sigma_4}{2} \left(\frac{1-\widetilde{\tau_{4}}^{4-2\Delta} Y^{4-2\Delta}}{\Delta-2} \right)\right).
	\label{t4:3}
\end{equation} 		
By using L'Hopital's rule,  Eq.(\ref{t4:3}) becomes
\begin{eqnarray}   
	Y^{6} &=& \widetilde{G_4}^{6} \left(\rho_{4} -\frac{\sigma_4}{2} \frac{\partial }{\partial \Delta}\left(\widetilde{\tau_{4}}^{4-2\Delta} Y^{4-2\Delta} \right)\right) \nonumber\\
	&=&  \widetilde{G_4}^{6}  \left(\rho_{4} +\sigma_4 \ln\left(\widetilde{\tau_{4}}Y \right)\right) \label{t4:4} 
\end{eqnarray} 	 
Fig.~\ref{logplot11} (b) tells us that $Y\sim \ln((g \rho)^{1/3}/T)^{1/6}$ for low temperature; 
Numerical result tells us that $Y^{6}$-$\log(T/(g \rho)^{1/3})$ plot  demonstrates our arguements with high precision. 

And $Y$ is numerically 
\begin{small}
	\bea
	Y &&\approx 0.89 \left( 1+ 4.23 \ln \left(\frac{\sqrt(g \rho)^{1/3}}{T}\right)\right)^{1/6} 
	\label{t4:5}
	\eea    
\end{small}  

\section{Discussion}

In this paper,  we   calculated  the physical observables $T_{c}, \vev{{\cal O}_{\Delta}}, \sigma({\omega}), \omega_{g},  \omega_{i}, n_{s}, $  as  functions of ${\mathcal O_{\Delta}}, T, \rho$.   
 Here we describe the  main  differences so that the readers understand the source of the differences in the results. 
\begin{enumerate}
	\item
	We use matrix algorithm by applying Pincherle’s Theorem to obtain the smallest value $\lambda_{g,3}$.   On the other hand, 
	The authors of  ref. \cite{Siop2010} obtained the minimum value of $\lambda_{g,3}$'s by using variational  method (see Eq.(\ref{eq:37})) and they used  the trial function $F(z)=1-\alpha
	z^2$.  $F(z)$ does not converge on the boundary of the disc of convergence at $z=1$ in general. However, Pincherle’s Theorem tells us that $F(z)$ converges at $z=1$  for  some quantized value of $\lambda_{g,3}$.  As a consequence,   our methodology works straightforwardly without ambiguity caused by the divergences and effective to get an eigenvalue when a power series  is made up of three or more term recurrence relation. Notice also that  our method shows that there is no well defined solution for $1/2<\Delta< 1$ because of three branches of $\lambda_{g,3}$. But variational method do not show this phenomemon:  Moreover, it  tell us that there is $\lambda_{g,3} =0$ at $\Delta =1/2$ which is unphysical, since  $\lambda_{g,3} =0$ means that $T_c$ is infinite.
	
	\item The authors of  ref. \cite{Siop2010}    applied perturbation theory  to obtain the condensate near $T_c$. It   leads  to the integral   such as ${\cal C}_{3}=\int_{0}^{1}dz\;\frac{  z^{2(\Delta-1)}F^2(z)}{z^2+z+1}$.   Instead, we first obtain  the analytic solution   given by $F(z)=\left(z^2+z+1\right)^{-\frac{\lambda_{g,3}}{\sqrt{3}}} \sum_{n=0}^{15}d_n z^n$ (see Eq.(\ref{eq:43})). Then  we used it   to evaluate  Eq.(\ref{eq:42}).   The result gives dramatic  differences: For $\Delta=3$ in $d=3$, we have a finite result for ${\cal C}_{3}$,  while they claimed they  got   ${\cal C}_{3}=0$. As a consequence,  $g^{\frac{1}{\Delta}}\frac{\left< \mathcal{O}_{\Delta}\right>^{\frac{1}{\Delta}}}{T_c} $  is finite at 
	$\Delta=3$ in our result, while  they have divergent result.  
 
 
	\item  Our the boundary condition of $F(z)$ and $\Phi(z)$ in AdS$_{4}$ is given  in the following table. 
	\begin{table}[!htb]   
		\centering
		\begin{tabular}{|l|} 
			\hline
			\hspace{0.8cm} Over all the regime we consider, i.e,  $\frac{1}{2}< \Delta <3$   \\ \hline
			$(i)$  $F(0)=1$,\;   $\Phi^{'}(0)= -{\rho}/{r_h}$ \\   
			$(ii)$  $\Phi(1)=0 $  ,  $3F^{'}(1)+\Delta^2 F(1)=0 $   \\ 
			\hline
		\end{tabular}
		\caption{Boundary condition of $F(z)$ and $\Phi(z)$ at the origin and the unity  } \label{boundary}
	\end{table}   
	
	On the other hand, the authors of ref.  \cite{Siop2010}    used different boundary condition and different trial wave function according the regimes: 		
	$\frac{1}{2}< \Delta <\frac{3}{2}$ and 
	$\frac{3}{2}< \Delta <{3}$. 
	To compute  Eq.(\ref{eq:57a}) and Eq.(\ref{eq:57b}), they applied $K_{\nu}(z)\sim \frac{\Gamma(\nu)}{2}\left(\frac{2}{z}\right)^{\nu}$ as $z\rightarrow 0$ into them. Because modified Bessel   function  $K$    is exponentially suppressed in large $z$. So they thought the dominant contribution comes from near $z=0$ region. Unfortunately, we cannot use  near zero expression of $K_{\nu}$  inside the non-local double integral. In fact,   using $K_{\nu}(z)\sim \frac{1}{z^{\nu}} $  in  Eq.(\ref{eq:57a}) gives completely different result from using the full expression of   $K_{\nu}(z) \sim \frac{e^{-z}}{\sqrt{z}}$, which we did here.  
	
	Unlike in the case of  $\frac{1}{2}< \Delta <\frac{3}{2}$,  they used  variational method without condition   $3F^{'}(1)+\Delta^2 F(1)=0 $ to compute $\mathcal{A}^2$ in Eq.(\ref{eq:56})  in the regime  $\frac{3}{2}< \Delta <3 $.  They used different boundary condition at different region of $\Delta$. 
	We believe that this is not necessary. 		
\end{enumerate}

\acknowledgments
We thank   Ki-Seok Kim for  the useful discussion. 
This  work is supported by Mid-career Researcher Program through the National Research Foundation of Korea grant No. NRF-2021R1A2B5B0200260. 
We also  thank the APCTP for the hospitality during the focus program, “Quantum Matter and   Entanglement with Holography”, where part of this work was discussed.

\bibliographystyle{JHEP}
\bibliography{Refs_SC}

\end{document}